\documentclass[reprint, amssymb, amsmath, aip,cha,floatfix]{revtex4-1}
\usepackage{bm}%
\usepackage{booktabs,tabularx,array}
\usepackage{graphicx,color}
\usepackage{hyperref}%
\usepackage{enumerate}
\usepackage{amsmath}
\newcommand*\mathinhead[2]{\texorpdfstring{$\boldsymbol{#1}$}{#2}}
\begin{document}
  
\title{Mechanisms and Rates of Nucleation of Amyloid Fibrils}
\author{Cheng-Tai Lee and Eugene M. Terentjev}
\affiliation{Cavendish Laboratory, University of Cambridge,  J.J. Thomson Avenue, Cambridge, CB3 0HE, U.K.}

\begin{abstract}
The classical nucleation theory finds the rate of nucleation proportional to the monomer concentration raised to the power, which is the `critical nucleaus size', ${n_c}$. The implicit assumption, that amyloids nucleate in the same way, has been recently challenged by an alternative two-step mechanism, when the soluble monomers first form a metastable aggregate (micelle), and then undergo conversion into the conformation rich in ${\beta}$-strands that are able to form a stable growing nucleus for the protofilament. Here we put together the elements of extensive knowledge about aggregation and nucleation kinetics, using a specific case of A${\beta_{1\mathrm{-}42}}$ amyloidogenic peptide for illustration, to find theoretical expressions for the effective rate of amyloid nucleation. We find that at low monomer concentration in solution, and also at low interaction energy between two peptide conformations in the micelle, the nucleation occurs via the classical route. At higher monomer concentration, and a range of other interaction parameters between peptides, the two-step `aggregation-conversion' mechanism of nucleation takes over. In this regime, the effective rate of the process can be interpreted as a power of monomer concentration in a certain range of parameters, however, the exponent is determined by a complicated interplay of interaction parameters and is not related to the minimum size of the growing nucleus (which we find to be ${\sim}$ 7-8 for  A${\beta_{1-42}}$). 
\end{abstract}

\maketitle

\section*{Introduction}

Amyloid fibrils are insoluble linear ordered aggregates of particularly misfolded proteins or peptides, which are closely connected with neurodegenerative disorders \cite{Murphy2010,Edison2008,DiFiglia1997,Prusiner1991}. As more evidence emerges that oligomers produced at the early stages of amyloid aggregation could be the most toxic species \cite{Pham2011,Shankar2008,Nimmrich2008}, researchers have been keen on understanding the details of the nucleation mechanism of fibrils, in particular, determining the critical nucleus size $n_{c}$ of primary nucleation: the minimum size that enables the extension of amyloid fibrils. Yet, due to the transient nature of critical nuclei and the low concentration of nuclei over the whole aggregation time course, there have been no direct experimental methods of observing amyloid nucleation process \cite{Cohen2013}.
 
\begin{figure}[b]  
\centering	\includegraphics[width=0.7\columnwidth]{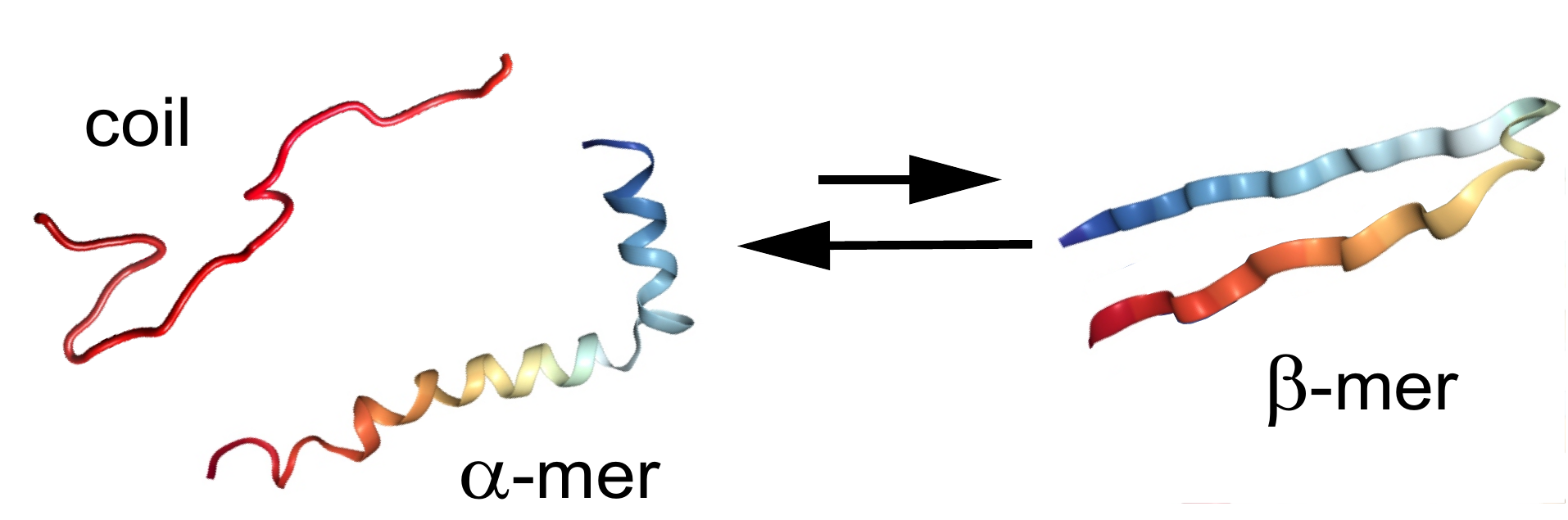}
	\caption{Key monomeric states of A$\beta_{1-42}$. The soluble monomeric state is likely to be the random coil, although the  $\alpha$-mer from Protein Data Bank (PDB: 1YIT) is also soluble and only slightly higher in free energy \cite{Granata2015}. The insoluble $\beta$-mer (PDB: 2BEG) is the standard building block of amyloid fibril. }
	\label{fig:monotype}
\end{figure}

For now, experimental studies monitor the total fibril mass in real time (e.g. by optical experiments\cite{Meehan2007,Wang2010,Lee2009c,Nilsson2004,Kusumoto1998}), obtaining kinetic plots with a characteristic sigmoidal shape \cite{Nielsen2001,Evans1995,Ionescu-Zanetti1999}. An important quantity, called the lag time $t_\mathrm{lag}$ \cite{Knowles2009a,Arosio2015}, can then be extracted; it is defined as the waiting time before a sharp increase of fibril mass appears in the sigmoidal plot. This lag time approximately holds a power-law relationship to the initial monomer concentration $C_{1}$, as originally suggested by the Oosawa model \cite{Oosawa1962}, which considers only primary nucleation and irreversible elongation in aggregation kinetics. Other more advanced models that further incorporate fragmentation and annealing of filaments \cite{Knowles2009a,Cohen2011,Michaels2014}, all retain this characteristic relationship. Without secondary nucleation, the power-law exponent in $t_\mathrm{lag} \sim C_1^\gamma $ is approximately $\gamma = -n_{c}/2$, which then should enables experimental determination of $n_{c}$ {{by plotting $\ln{t_\mathrm{lag}}$ against $\ln{C_1}$ or through a global fitting scheme of the total fibril mass plots against time with different initial monomer concentrations as used in ref. \cite{Cohen2012}}}. However, the validity of such methods of obtaining the critical nucleus size $n_c$ depends on how closely the assumed microscopic aggregation mechanisms and kinetic equations match the actual ones in experiments. 

Before forming an amyloid, monomeric subunits in solution have to switch from their native soluble structure into a partially unfolded intermediate state, which has a higher free energy in solution \cite{Eakin2006, Chiti1999, Chiti2009}. There are many possible configurations a soluble peptide can exist in solution: the recent simulation study \cite{Granata2015} finds the whole hierarchy, from compact conformations rich in $\alpha$-helix to a fully unfolded random coil -- definitively finding the random coil having a lower free energy. This challenges the earlier assumption that the soluble monomer state of A$\beta_{1-42}$ is $\alpha$-helical \cite{Granata2015,Baumketner2006},  see Fig.~\ref{fig:monotype}. However, we find it is sufficient to use a two-state simplification to capture the essence of amyloid aggregation mechanism, as has been suggested in molecular simulations \cite{Bieler2012,Saric2014}: we denote the soluble monomer as `$\alpha$-mer' (for its assumed increasing content of $\alpha$ helix when forming micelles), while the $\beta$-mer is the monomeric unit of mature amyloid fibrils. We later use this two-state simplification to schematically show two different nucleation mechanisms as an aid to point out the weakness of previous theoretical kinetic models on determining the critical nucleus size.

\begin{figure}
\centering	\includegraphics[width=1\columnwidth]{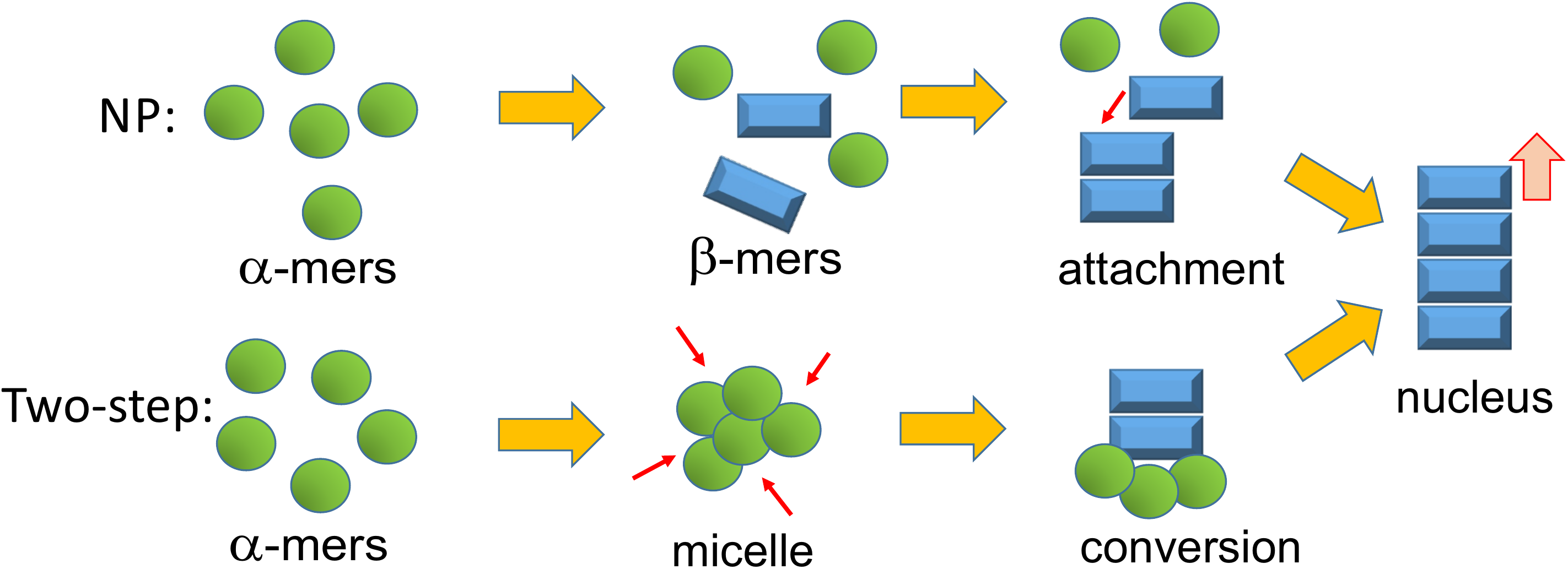}
	\caption{Two mechanisms of amyloid nucleation}
	\label{fig:2nucl}
\end{figure}

Conventional theoretical models used to find the critical nucleus size, are usually based on the `nucleated polymerization' (NP) concept (see Fig.~\ref{fig:2nucl}). Other more complicated models that add secondary nucleation, fragmentation and annealing processes to this NP model \cite{Knowles2009a,Michaels2014,Cohen2011}, were formulated based on the classical nucleation theory, where the primary nucleation rate is proportional to $C_{1}^{n_c}$, and further assumed a fixed $n_c$ value throughout all monomer concentration regimes. However, already in 1984, Ferrone has pointed out  that the critical nucleus size $n_c$ should change with varied monomer concentration to account for observations of the sickle hemoglobin polymerization rate \cite{Ferrone1985}. The assumption of a fixed $n_c$ value with varied monomer concentration therefore must be re-examined.    

In recent single-molecule experiments \cite{Iii2010,Cremades2012,Lee2011,Vitalis2011}, an alternative two-step nucleation mechanism has been suggested: micellation of $\alpha$-mers (monomers in the native soluble state) followed by a gradual conversion into $\beta$-mers within the dense micelle (Fig.~\ref{fig:2nucl}). Under this scenario, the nucleation rate may or may not take the power-law scaling of the initial monomer concentration as would follow from  NP and its derivative models. This may in turn change the previously predicted power-law relationship of $t_\mathrm{lag}$. Even if the power-law scaling is found to exist in practice, the physical meaning of its exponent and its relation to the actual nucleus or micelle size become unclear. All these questions challenge the validity of employing these NP models in all monomer concentration regimes. 

Several theoretical works have been proposed for the two-step nucleation mechanism. The work by Lomakin et al. in 1997 considered the process of producing critical nuclei from pre-formed micelles, under the assumption of a fast thermal equilibrium between monomers and micelles before any nucleation events happen \cite{Lomakin1997}. A more recent theoretical framework of two-step mechanism (the term `nucleation-conversion-polymerization' model is used there) considered not only nucleation of micelles but also the step-by-step conversion of this micelle into its fibrillar form \cite{Garcia2014}. However, both these studies make an assumption of a single micelle size, which was questioned by the observed presence of multi-size micelles in the coarse-grained molecular simulations of the A$\beta_{1-42}$ system \cite{Saric2014}.  Later, Auer et al. derived theoretical expressions for nucleation rates of both NP and two-step nucleation mechanisms, predicting the monomer concentration at crossover between these two mechanisms \cite{Auer2012}.  However, it was not clear what is the micelle size that optimizes the nucleation rate of the two-step mechanism.

The main of our work is to first derive the nucleation rates, then to estimate the critical nucleus size in the NP mechanism, and the micelle size that maximizes the nucleation rate of the two-step mechanism -- and finally compare nucleation rates of NP and two-step mechanisms to estimate the crossover monomer concentration (similar to the one in ref. \cite{Auer2012}). In order to fully test the capability of the two-step model, we deliberately stay with the basic mechanism, without considering secondary pathways (secondary nucleation and fragmentation): these are not expected to contribute significantly at the early nucleation stage.

We choose the free energy approach, originally developed by Ferrone et al. \cite{Ferrone2006}, to obtain free energy functions of intermediate states, final products of amyloid aggregation, and compare the free energy landscape of both nucleation mechanisms. This not only allows the analytical calculation of nucleation rates of variable micelle and nucleus size, but also makes possible a simplified kinetic analysis of nucleation rates. Although this generic scheme of using free energy landscape of nucleation to find nucleation kinetics is formally similar to the work by Auer and Kashchiev \cite{Auer2012,Kashchiev2010}, our work does investigate how different micelle sizes can facilitate the conversion process, which was not addressed before.

A$\beta_{1-42}$ peptide is used as our model system, since it has been studied more than any other amyloidogenic protein; molecular simulations are more reliable due to its affordable small size, in turn allowing more information on thermodynamic parameters to build free energy functions. However, our model is generic, and only the binding energy and geometric parameters would differ for other amyloid systems. For simplicity and clarity of the main text, many of the derivations, and the justification of parameter values in our free energy calculation, are removed to Appendices.

\section*{Nucleated polymerization model}
\subsection*{Fibril structures of A\mathinhead{\beta_{1-42}}{\beta1-42} peptides} \label{sec:abstruc}
     Within this direct nucleation route, monomeric A$\beta_{1-42}$ peptides first undergo structural conversion from the $\alpha$-state into $\beta$-state, and spontaneously stack along one direction to construct a nucleus of a protofilament, or fibrils with more than one protofilaments (typically this number lies between 2 to 6) \cite{Jimenez2002,Sachse2006,Jimenez2001,Arimon2012,Goldsbury2005}. These fibrils have a twisted linear structure along the fibril axis, making a complete pitch every 33 $\beta$-mers  \cite{Paravastu2008,Yu2010,Knowles2012}. Since our aim  is to investigate the critical nucleus size, which will be much less than 33 subunits, the twisted fibril structure and its effect on the later free energy calculations can be neglected.

\begin{figure}[h]
\centering	\includegraphics[width=0.7\columnwidth]{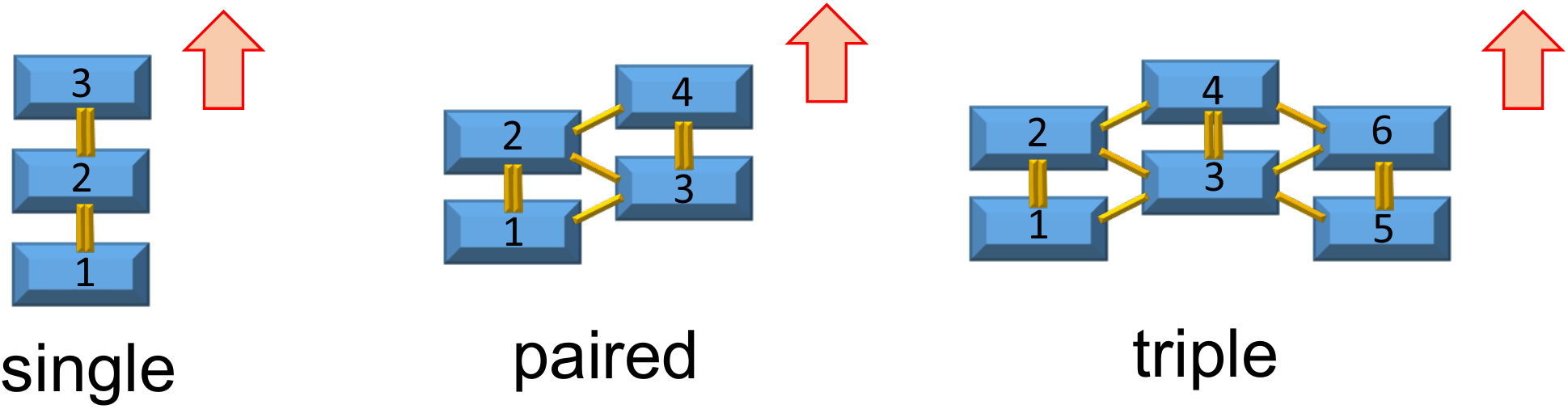}
	\caption{$\beta$-aggregate structure: the number in each $\beta$-mer refers to the order of attachment in the nucleation process. Double orange sticks represent $\beta$ bonds, while single orange ones are side bonds of $\beta$-mers.}
	\label{fig:bstruc}
\end{figure}

A cut of the short length in a fibril from the results of simulation and cryo-electron microscopy further indicated that each protofilament of this fibril was aligned  on the same plane \cite{Luhrs2005}. To portray a coarse-grained fibril structure, it is necessary to further clarify the relative position of one protofilament with respect to another, when they are associated together. Although most of the detailed structural studies \cite{Jimenez2002,Otzen2013} are not often resolving this longitudinal aspect of monomer packing in protofilament pairs, there are clear indications for the period shift. As concluded in a computation study of periods of helical twisting of single and paired protofilaments \cite{Correia2006}, the second protofilament is shifted by half a period. That is, the lateral monomer binds in the middle between two subunits of the existing filament, making equal-strength diagonal bonds with both. Based on these facts, we construct the coarse-grained structures of single, paired, and multiple protofilaments, as illustrated in Fig.~\ref{fig:bstruc}. Two types of bonds are then involved in these fibril structures: the end-to-end bond between two $\beta$-mers, with a binding energy $\Delta_\beta$, and the lateral (side) bond for two neighboring monomers from two different protofilaments, with a binding energy $\Delta_s$ each.  

\subsection*{Free energy functions of fibrils} \label{sec:benergy}
To construct the free energy of a protofilament, we use the approach originally used by Ferrone \cite{Ferrone1985,Ferrone2006}. This method follows the free energy change of a given species along the reaction path, relative to the reference state of pure $\alpha$-monomers in solution. The formation of an $N$-size aggregate is a process where $N$ monomers undergo an internal transition into $\beta$-state, with a conversion free energy penalty $\Delta_c$, and then flock together to form bonds within the $N$-size aggregate. In this process, the translational and rotational free energies of free monomers in solution are lost, but this loss is offset by the translational and rotational free energy of an $N$-size aggregate as a whole. For simplification, the difference in rotational and standard translational free energies between a monomer and the $N$-size aggregate is neglected, as this difference is of the magnitude $\ln{N}$ with coefficient of 3 $k_B T$ only \citep{Ferrone2006}, while all other interaction free energies have a higher dependence on $N$ and with coefficients of tens of $k_BT$ (see Appendix~\ref{app:abpara}). 

From the bond scheme in Fig.~\ref{fig:bstruc}, the free energy function of an $N$-size aggregate of a single protofilament takes the following form:
\begin{equation} \label{eq:s-proto}
F_{\beta,1}(N)=(1-N)\left(\mu^{0}+k_B T\ln{{C_1}}\right) +(N-1)\Delta_{\beta}+N\Delta_{c} .
\end{equation}
Here $\mu^{0}$ is the sum of translational and rotational free energies of a single $\alpha$-mer at a standard concentration of 1~mM. Strictly, $\mu^{0}$ includes many internal rotational degrees of freedom of a peptide subunit in solution, most of which become frozen when the monomer adopts the closely-packed $\beta$-sheet configuration in the filament. For this reason, it is hard to accept any theoretical estimate for $\mu^{0}$ based on the ideal-gas statistics. Therefore, we shall use an estimate for  $\mu^{0}$ based on experimental measurement of elongation free energy (see Appendix~\ref{app:abpara} for details on all material parameters). The estimate gives $\mu^{0} = -34.5\, k_BT$, for the room temperature $T=25^{\circ}$C, and we will use this definition in the remainder of this work. The initial monomer concentration in solution $C_{1}$, in the units of mM (with the reference monomer concentration of 1 mM that we shall use throughout this work). Values of the free energy of the longitudinal $\beta$-bond, and the conversion free energy in solution are also discussed  in Appendix~\ref{app:abpara}: $\Delta_{\beta}=-44 \, k_BT$ \cite{Davis2010}, and  $\Delta_c=20 \, k_BT$ \cite{Bieler2012}. 

Similarly, the free energy for a paired protofilament will be:
\begin{align} \label{eq:d-proto}
F_{\beta,2}(N)&=(1-N)\left(\mu^{0}+k_B T\ln{{C_1}}\right) \notag \\
&+N_s(N) \Delta_s + N_{\beta}(N)\Delta_{\beta}   +N\Delta_{c}  \ .
\end{align}
Here $N_s(N)$ is the number of lateral (side) bonds in an $N$-size protofilament pair: it is zero for $N \leq 2$, and we assume a symmetric $N_s=(N-1)$ for larger  $N$ vales. The free energy of a side bond $\Delta_s = -22 \, k_BT$ \cite{GarciCuesta2014}. Finally, $N_{\beta}(N)$ is the number of longitudinal $\beta$-bonds of energy $\Delta_\beta$  in a paired protofilament of $N$-length, which is zero at $N=1$, equal to 1 at $N=2$, and then $(N-2)$ for other $N$ values (cf. Fig. \ref{fig:bstruc}).

It is noticeable that lateral addition of one $\beta$-mer to a pre-formed protofilament would give an interaction energy of $2\Delta_{s}$: almost exactly the same magnitude as $\Delta_{\beta}$. It immediately suggests that lateral addition is equally likely as addition along the fibril axis, and shall be experimentally observed in fibril formation. This indication is indirectly supported by the fact that no mature fibrils have frayed ends in amyloid aggregation, which means that a protofilament pair is likely to form at the phase of amyloid nucleation, and is further reinforced by observations of lateral addition at the early stage of fibril formation of human amylin \cite{Arimon2012,Green2004}.

Obviously, multi-protofilaments will have more protofilaments joined through lateral addition. To initiate the growth of an extra protofilament by side addition of one $\beta$-mer to the pre-existing $\beta$-aggregate contributes the binding energy of $2\Delta_s$ only, which is weaker than $(\Delta_{\beta}+\Delta_s)$ of an alternative addition of one $\beta$-mer on the fibril end to keep the original protofilament number. Therefore, multi-protofilament cases will not have a more favorable bond free energy, and thus the aggregation free energy would not favor the multi-protofilament over the paired case at the nucleation stage. The difference between the protofilament pair and the multi-protofilament lies in that a larger critical nucleus and a higher nucleation free energy barrier are required for multi-protofilament nucleation. The presence of multi-protofilament cases in mature A$\beta_{1-42}$ fibrils is due to other free energy contributions that originate from the twisted structure of fibrils, and will become significant as fibrils grow longer. But this effect does not contribute at the nucleation stage where the aggregate size is small, and its analysis is outside the scope of this work. Consequently, multi-protofilament cases will not be discussed further.

\begin{figure}
\centering	\includegraphics[width=0.9\columnwidth]{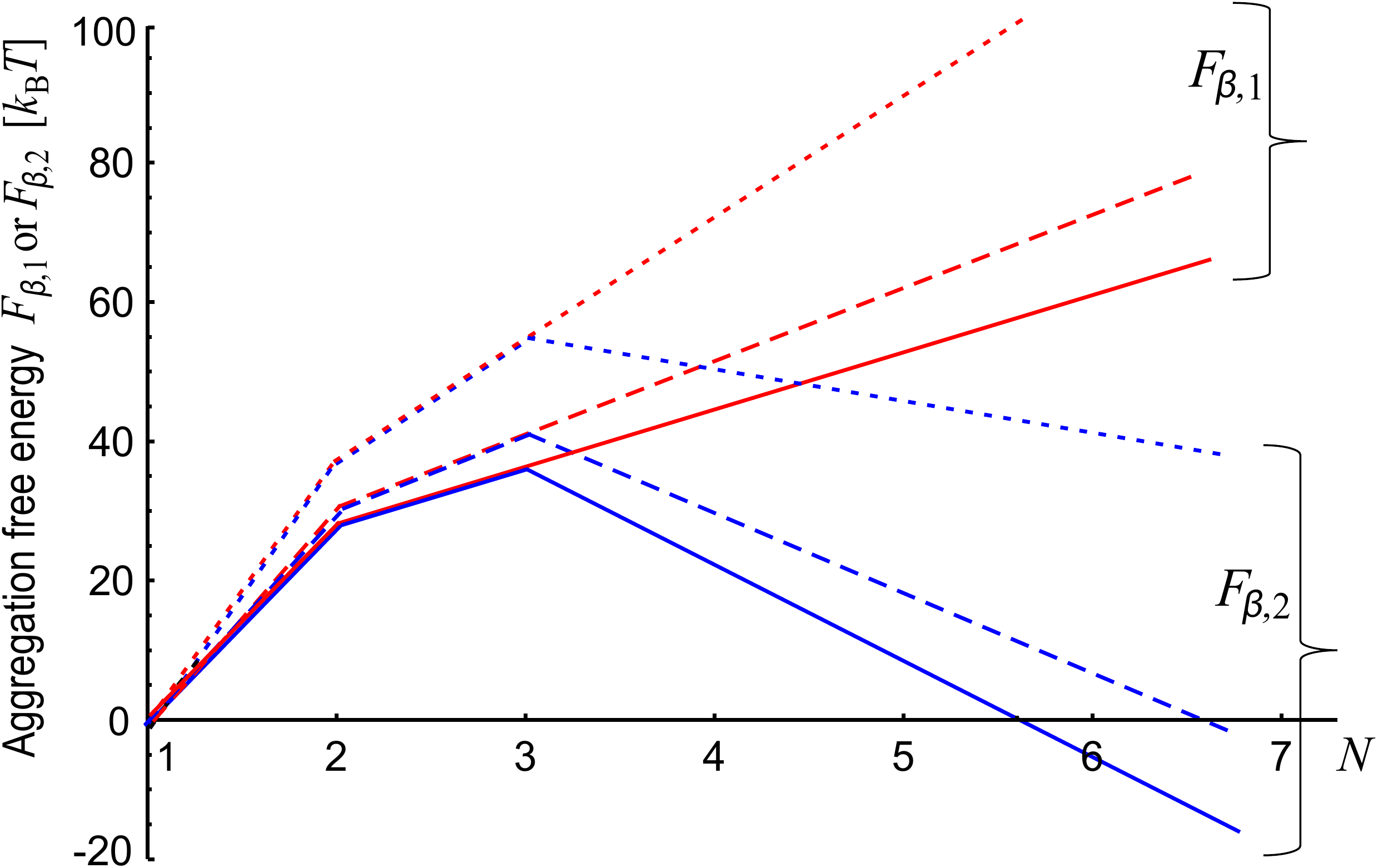}
	\caption{Aggregation free energie of a single protofilament and a protofilament pair, $F_{\beta,1}(N,C_1)$ and $F_{\beta,2}(N,C_1)$, as functions of the number of aggregated peptides $N$ with three initial monomer concentrations in solution: $C_1=$10 mM (solid lines), 1 mM (dashed) and 1 $\mu$M (dotted). The units of the $y$-axis are $k_BT$ at room temperature.}
	\label{fig:benerg}
\end{figure}

The comparison of free energies of a single and a paired protofilaments is given in Fig.~\ref{fig:benerg}. Due to the increase of the number of bond sites per monomer addition in the protofilament pair when $N$ exceeds 3 (one at the end and the other on the side of the protofilament), the protofilament pair has a lower free energy than a single protofilament with the same number of units. A thermally stable aggregate is the one that should have a larger population than monomers, when thermal equilibrium is reached, i.e. a negative free energy with respect to the reference state chosen to be $\alpha$-monomers in solution. Single protofilaments cannot be thermally stable and must transform into protofilament pairs through lateral addition at the early stage of nucleation. We therefore conclude that the critical nucleus size is always $n_c=3$ in the paired protofilament. The free energy barrier of aggregation in the NP model is then equal to $F_{\beta,2}(3,C_1)$. 

The $n_c$ value we predict here is not the same as, albeit close to $n_c=2$ obtained by recent experimental studies of A$\beta_{1-42}$ \cite{Cohen2013,Meisl2014}. This inconsistency may be due to the denaturant used in experiments (i.e. sodium azide) to initiate amyloid aggregation, which can significantly change the interactions between monomers, and our interaction parameters do not cover this effect. Kashchiev and Auer also analyzed the nucleation free energy of $\beta$ strands in a 2D nucleus model \cite{Kashchiev2010}, yet they concluded a variable critical nucleus size, which we do not observe in the examined concentration regime. In their work, the $n_c$ value starts from a rather large size (over 40 $\beta$-strands) at low concentration, and then shrinks as the function of $1/(\ln{C_1})^2$. In this sense, our result as well as the experimental work by Knowles \cite{Meisl2014} are likely to fall within the high concentration/saturation regime, where further increase in concentration can cause little effect on the critical nucleus size than at a rather low concentration regime Kashchiev and Auer were interested in. However, later in this paper we will demonstrate that the two-step mechanism of nucleation becomes prevalent at higher monomer concentrations. 

Our conclusion that critical nuclei are exclusively in the form of paired protofilaments is supported by the fact that only multi-protofilament aggregates are observed at neutral pH and high ionic concentration \cite{Arimon2012}. It is arguable that our conclusion cannot be applied at low pH and ionic strength, where single protofilaments do appear \cite{Arimon2012}, because molecular simulations giving bond free energy parameters, $\Delta_{\beta}$ and $\Delta_s$, were only implemented under neutral pH values so far (and so we do not know the values of these energy parameters in other situations). We shall not worry about this limitation of our model since it is under physiological conditions, at neutral pH, that we intend to investigate the problem.

\subsection{Nucleation rate of the NP mechanism} \label{sec:NPrate}
Since the rate-limiting (slow) process is actually the nucleation itself, while the subsequent elongation is fast, we could take the rate of producing $(n_c + 1)$-mers, i.e. tetramers here, as the measure of nucleation rate in the NP model. We assume the pre-thermal equilibrium for the population of the critical nucleus size, $C_1 \exp{[-F_{\beta,2}(n_c,C_1 )/k_B T]}$, with $n_c=3$ here, and then use the theoretical expression for the elongation rate for the pre-existing fibril obtained in ref. \cite{Buell2010}. 

Two consecutive processes are involved in elongation: the diffusive arrival of a monomer at the fibril end, and the attempt to cross an additional free energy barrier to achieve internal conversion, leading to a frequency factor: $\exp{(-\Delta F_{el}/k_B T)}/(\tau_D + \tau_I)$ \cite{Buell2010}, see Appendix~\ref{app:k1} for derivation. Here $\Delta F_{el}$ is the barrier to overcome in elongation and is roughly 3.42 $k_BT$ for A$\beta_{1-42}$ peptide \cite{Buell2012,Vendruscolo2007}. $\tau_I$ is the time-scale for internal $\alpha$ to $\beta$ rearrangement of amyloidogenic species (the value of $\tau_I$ is estimated as $10^{-5}$ s in Appendix~\ref{app:abpara}), while $\tau_D$ is the arrival time to the fibril end. The rate of producing tetramers, $k_1 C_1$, is the product of this frequency factor and the concentration of trimers:

\begin{equation} \label{eq:elong}
k_1 C_1=\frac{C_1 }{\tau_D + \tau_I} e^{-\left[F_{\beta,2}(3,C_1 )+\Delta F_{el}\right]/k_B T} 
\end{equation}
  
The rate constant $k_1$ is interpreted as the nucleation rate constant for the NP mechanism. The arrival time $\tau_D$ is the Smoluchowski diffusion rate in solution (assuming no crowding effects). Taking into account the non-spherical shape of critical nuclei, one can modify the Smoluchowski theory for bimolecular reaction rate constants \cite{Richter1974}, as well as the Stokes-Einstein equation for diffusion coefficients, producing the estimate: $1/\tau_D=4k_B T C_1 f_{geo}/3\eta r_1$, where $\eta$ is the solvent viscosity, $r_1$ is the hydrodynamic radius of one $\alpha$-mer in solution. Note that $4k_B T C_1/3\eta r_1$ is the classical Smoluchowski result, modified here by the geometrical factor $f_{geo}$ that accounts for the shape of the $\beta$-aggregate (this geometry factor makes only a small correction and will not influence the conclusions we make in this paper). With this expression for the arrival time, the rate constant of the NP mechanism, $k_1$ in \eqref{eq:elong}, is expressed as:
\begin{equation} \label{eq:k1}
k_1=\frac{f_{geo}C_1}{ f_{geo}\tau_IC_1+3\eta r_1/4k_BT} e^{-\left[F_{\beta,2}(3,C_1 )+\Delta F_{el}\right]/k_B T} 
\end{equation}
Equation~(\ref{eq:k1}) shows that $k_1$ is proportional to $C_1^3$ when the monomer concentration $C_1$ is low and the term $f_{geo}C_1 \tau_I$ can be neglected in the denominator (remember that $F_{\beta,2}(3,C_1)$ is proportional to $-2\ln C_1$).  In contrast, when the term $f_{geo}C_1 \tau_I$ dominates in the denominator at high monomer concentration, the rate constant  only has the square of the $C_1$ concentration left: $k_1 \propto C_1^2$.

\section*{Two-step nucleation mechanism}
\subsection*{Micellation of soluble peptides} \label{sec:mic}
Several experiments have reported that amyloidogenic proteins or peptides in their native state can coalesce into a single micelle (some papers may use the term oligomer in this context) utilizing solvent-exposed hydrophobic patches on the surface of monomers \cite{Bleiholder2011,Yong2002,Sabate2005,Frare2009}. For example, taking lysozymes (a type of amyloidogenic proteins), micelles have a lower content of $\alpha$-helix and a higher content of $\beta$-sheet yet with the majority being disordered loops compared with native monomeric units \citep{Frare2009}. Therefore, aggregated micelles are often classified as amorphous. It is suggested that these amorphous micelles serve as intermediates, or initiation states for amyloid nucleation \cite{Serio2000a}. This generic hypothesis of the role of micelles is indirectly hinted by the increase in $\beta$-sheet structural content while losing their original $\alpha$-helix content \citep{Frare2009}, and is further supported by molecular simulation of coarse-grained peptides \cite{Auer2008a,Saric2014}. These facts imply an alternative aggregation pathway for amyloid fibrils: the two-step nucleation mechanism as described in Fig.~\ref{fig:2nucl}. 

In our work, micelles are defined as composed of a few $\alpha$-mers aggregated from solution. Though micelles could have a variety of structures \cite{Uversky2010}, we will assume them to be amorphous globular aggregates of densely packed monomers for simplicity. The driving force for micellation of A$\beta_{1-42}$ peptides originates mainly from hydrophobic interaction between $\alpha$-mers \cite{Saric2014,Urbanc2010}. We define one $\alpha$-bond as the bond formed between a pair of $\alpha$-mers inside a micelle. The total number of such $\alpha$ bonds, $N_{\alpha}(N)$, for a spherical $N$-size micelle can be assumed to have the bulk and surface terms. Since the expression $N_{\alpha}(N)$ has to reduce to $N_{\alpha}(2)=1$, it is easy to obtain: $N_{\alpha}(N)=AN+(1-2A)2^{-2/3}N^{2/3}$, with just one unknown parameter $A$. 

With only hydrophobic attraction accounted for, an infinite micelle size could be expected in equilibrium. However, aggregation is constrained by other unfavorable free energy factors, e.g. the electrostatic repulsion due to accumulation of negative charge on the surface \cite{Guo2005}, and the entropic loss from the compact packing of several $\alpha$-mers into a micelle \cite{Maibaum2004}. Therefore, a large micelle size is never observed in experiment.

A crude but convenient way to estimate the electrostatic repulsion is to assume that electrostatic charges distribute evenly on the spherical surface of the packed micelle. Then the repulsion energy is $(N q_e)^2/8\pi \epsilon \epsilon_0 r_N$ \cite{Groenewold2001}, where $q_e$ is the effective charge on a single $\alpha$-mer, $\epsilon$ and $\epsilon_0$ are the relative dielectric constant and the permittivity in vacuum, $r_N$ is the radius of the $N$-size micelle, proportional to $N^{1/3}$ if we assume it is roughly spherical. Accordingly, the electrostatic potential energy has the overall scaling of $N^{5/3}$. 

The screening effect from counter ions in solution may challenge the validity of using the expression $(N q_e)^2/8\pi \epsilon \epsilon_0 r_N$ for the potential energy of electrostatic repulsion. However, this Debye screening effect requires mobile charges in thermal motion. Without a doubt it would exist between any two charged micelles or monomers in solution. Yet the potential energy quoted above refers to the energy to confine charges in a small volume when starting from a well-separated distance (i.e. the interaction of immobile surface charges across the packed micelle itself). In this case, no counter charges exist inside micelles, and even if there were some, they would have very little mobility due to the compactness of internal structure. Therefore, the Debye screening effect cannot play a role. This enables the use of electrostatic potential in an effective dielectric medium in our case. 

In addition, there is an entropic cost of forcing polar amino acid groups to the micelle surface, an effect well-studied in the formation of micelles of polar surfactants \cite{Maibaum2004}. The simplest expression of the free energy expressing this reduction in conformational freedom turns out to scale as $N^{5/3}$, in an analogy to the electrostatic repulsive energy of a sphere with evenly distributed surface charge \cite{Woo1996}. Although the actual entropic repulsion term can be of more complex forms, this $N^{5/3}$ scaling can be the leading term, and helps to correctly predict the experimentally determined concentration threshold where monomers starts aggregating into micelles \cite{Maibaum2004}. Therefore, we account for these two repulsive free energy contributions (electrostatic and entropic) as a single term $h N^{5/3}$, with its parameter $h$ to be determined. 

Assembling together the free energy of $\alpha$ bonds and the repulsive free energy terms, the micellation free energy, $F_\mathrm{mic}(N,C_1)$, can be written as:
\begin{align} \label{eq:mic}
F_\mathrm{mic}(N,C_1)=&(1-N)\left(\mu^0 + k_B T \ln{ C_1 } \right)+h N^{5/3}\notag\\&
+\Delta_{\alpha}\left[AN+(1-2A)2^{-2/3}N^{2/3}\right] 
\end{align}
Note a close resemblance of this expression, and Eqs. (\ref{eq:s-proto}) and (\ref{eq:d-proto}) for protofilaments. Here the important parameter is $\Delta_{\alpha}$, the free energy of an attractive $\alpha$-bond, which is approximately $-17 k_B T$ estimated from the work of Hills and Brooks \cite{Hills2007} (see Appendix~\ref{app:abpara}). $F_\mathrm{mic}(N,C_1)$ is analogous to free energy of micelle formation in surfactant solutions, or flocculation in colloids. At a certain threshold monomer concentration, monomers and micelles of a specific size are equally favored at thermal equilibrium; this is the `critical micelle concentration' ($cmc$), and the corresponding micelle size is the `critical micelle size' ($cms$). In fact, the presence of the critical micelles in the A$\beta$ peptide system with different solvents and pH values has been already reported \cite{Soreghan1994,Lomakin1997,Nuallain2005,Kim2004}.

The micellation free energy, $F_\mathrm{mic}(N,C_1)$, and the slope of $F_\mathrm{mic}(N,C_1)$, are both zero at $N=cms$, as in any coexisting equilibrium. These two independent conditions let us evaluate the parameters $A$ and $h$ from experimentally determined values of $cmc=17.6$ $\mathrm{\mu}$M, and $cms=25$, which were measured in the solvent system that more closely reproduces the physiological conditions \cite{Sabate2005}. In this way we obtain the parameters to be used in the rest of this work: $A=4.86$ (dimensionless) and $h=1.6$ $k_BT$. Both values make good physical sense, although we shall not spend any more time on this discussion. 

We can now plot $F_\mathrm{mic}(N,C_1)$, at $cmc$ and several other values of monomer concentration, in Fig.~\ref{fig:mic}. {At low monomer concentrations (below approximately 1 $\mathrm{\mu}$M for our chosen set of parameters), $F_\mathrm{mic}(N,C_1)$ is a monotonically increasing function of $N$, and no metastable micelles can exist. This threshold concentration can be easily derived from \eqref{eq:mic}. It is therefore impossible to have the two-step nucleation mechanism for amyloid aggregation below 1 $\mathrm{\mu}$M in the A$\beta_{1-42}$ system. 
When monomer concentration exceeds 1 $\mathrm{\mu}$M, $F_\mathrm{mic}(N,C_1)$ has a metastable state, and a free energy barrier to cross to reach a micelle. We define the micelle size that gives the  barrier position as $N_h$, and the size that sits at the lowest point of the free energy trap as $N_l$, both labelled in Fig.~\ref{fig:mic}.  Only micelles with the size between $N_h$ and $N_l$ are metastable and can follow the two-step nucleation mechanism. Since the barrier occurs at relatively low $N$, the stabilizing effect of the electrostatic $h$-term is not yet important and we can find an approximate expression for the barrier position, useful in the subsequent analysis ($N_h$ varies between $\sim 4.5$ at the lowest $C_1$ end, to $\sim 2$ at the highest):
\begin{equation}
N_h \approx \frac{2}{27}\left( \frac{ (2A-1)\Delta_\alpha /k_BT }{A \Delta_\alpha /k_BT - \mu_0/k_BT - \ln C_1} \right)^3 . \nonumber
\end{equation}

\begin{figure}[htbp]
\centering	\includegraphics[width=1\columnwidth]{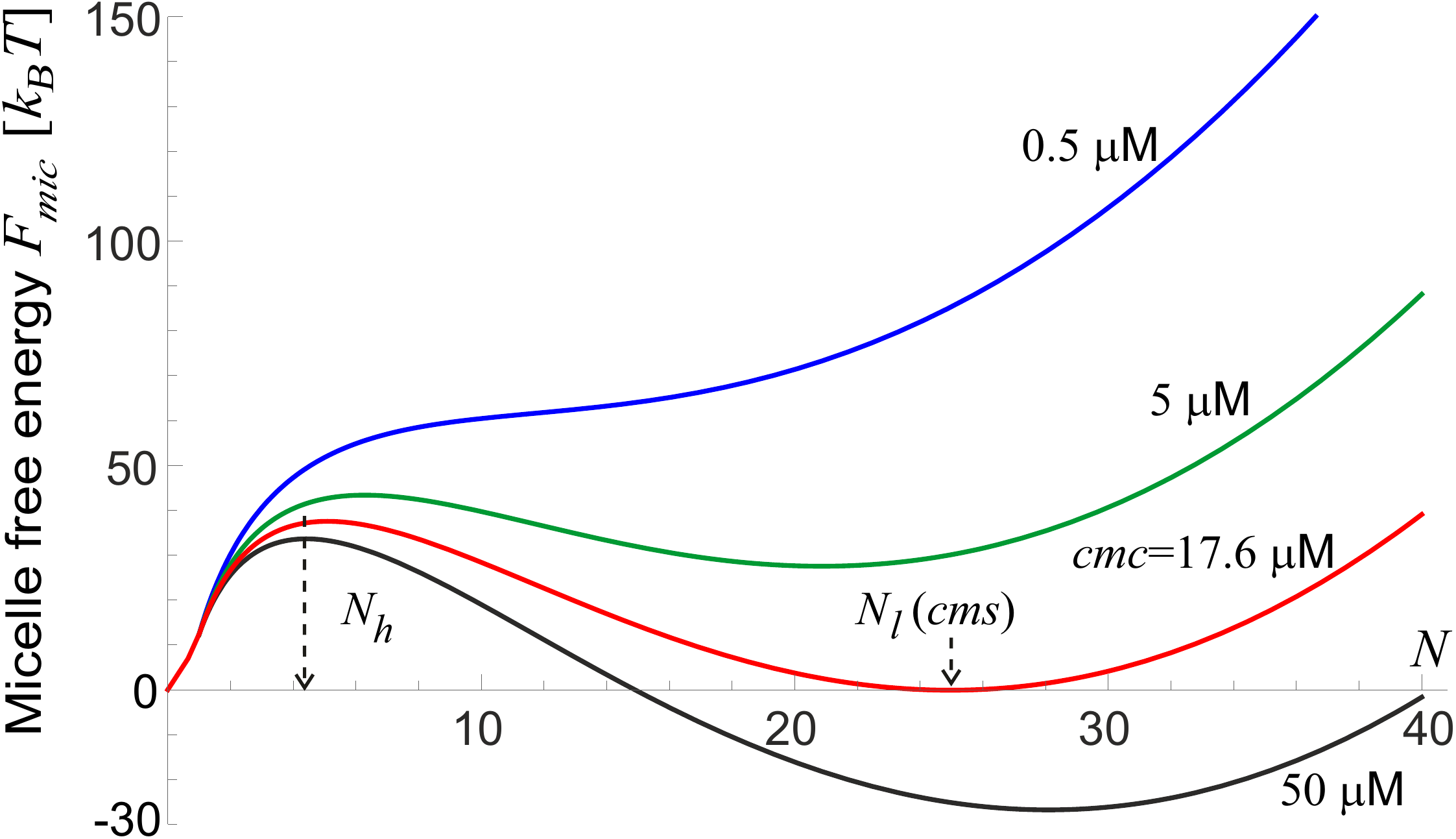}
	\caption{Micellation free energy $F_\mathrm{mic}(N,C_1)$ (in units of $k_BT$ at room temperature) vs. the micelle size $N$. From top to bottom: $C_1$ lower than the threshold monomer concentration to have metastable micelles (0.5 $\mathrm{\mu}$M), $C_1=cmc$ (17.6 $\mathrm{\mu}$M), and $C_1$ well above $cmc$ (50 $\mathrm{\mu}$M). $N_h$ and $N_l$ are the local maximum and minimum of $F_\mathrm{mic}(N,C_1)$, respectively. }
	\label{fig:mic}
\end{figure}

\subsection*{Conversion of micelles} \label{sec:conv}
Based on two structural facts about fibrils, we can assume that $\alpha$-$\beta$ conversion happens on the surface of the remaining $\alpha$-mers in the packed micells as schematically illustrated in Fig.~\ref{fig:2nucl}. Firstly, due to the twisted structure of $\beta$-aggregates, it would cost more free energy to twist inside a dense amorphous micelle than being unconstrained on the surface of this micelle \cite{Knowles2012}. Besides, $\alpha$-mers interact with the pre-formed $\beta$-aggregate mainly at the end of the emerging fibril. This is because the region of exposed hydrophobic groups designed to attach to the $\beta$-aggregate is on the end of a $\beta$-stack. Accordingly, $\alpha$-mers tend to gather near the end area of the fibril to gain more interaction energy. 

We choose $x \in (0,N)$, the number of converted $\alpha$-mers, as the reaction coordinate in the micelle of size $N$. The free energy along the conversion path, $F_{c} (N,x,C_1 )$, is composed of four contributions: the free energy of a remaining $(N-x)$-size micelle, Eq. \eqref{eq:mic}; the free energy of the emerging $x$-size $\beta$-aggregate, Eq. \eqref{eq:d-proto}; the bonding free energy of an interface between the micelle and the aggregate, which will be detailed later; and an additional free energy loss of $-(\mu^{\circ}+k_BT\ln {C_1})$ in translational and rotational motions, which is to compensate for one degree of center-of-mass freedom that is present in separate expressions for the $\alpha$-micelle and the $\beta$-aggregate, but is removed when they are bound and move as one entity. Organizing these separate terms, the conversion free energy $F_{c} (N,x,C_1 )$ takes the form:
\begin{align} \label{eq:con}
F_{c}(N,x,C_1)&=(1-N)\left(\mu^{\circ}+k_B T\ln {C_1} \right)+
h (N-x)^{5/3}\notag\\&
+\Delta_{\alpha}N_{\alpha}+\Delta_{\alpha\beta}N_{\alpha\beta}(N,x)\notag\\& +x\Delta_{c}+\Delta_{s}N_{s}(x)
+\Delta_{\beta}N_{\beta}(x) \  , 
\end{align} 
where $N_{\alpha} = A(N-x)+(1-2A)2^{-2/3}(N-x)^{2/3}$ for the remaining micelle, and  $\Delta_{\alpha\beta}$ is the free energy per $\alpha\beta$ bond, which is defined as the bond formed between an $\alpha$-mer and a $\beta$-mer. This is a parameter we know the least about; the range of reasonable  $\Delta_{\alpha\beta}$ values is given in Appendix~\ref{app:abpara}. We now proceed to find the expression for $N_{\alpha\beta}(N,x)$, the number of total $\alpha\beta$ bonds in this intermediate aggregate.

Under the assumption that $\alpha\beta$ bonds originate from the replacement of pre-existing $\alpha$ bonds, and that the emerging $\beta$-aggregate is located on the surface of the micelle,  we can compare two expressions for $N_\alpha$ -- one including the $\beta$-mers at the contact interface as yet non-converted $\alpha$-mers, and the other with an actual remaining $(N-x)$ $\alpha$-mers. 
The resulting expression is a non-linear function of $N$ due to the surface term present in the definition of $N_\alpha$. 
 When $x$ is smaller than 3 (giving only a single protofilament geometry), the number of $\alpha\beta$ bonds is $N_{\alpha\beta}(N,x) = N_{\alpha}(N-x+1)-N_{\alpha}(N-x)$. On the other hand, when $x \geq 3$ (paired protofilament starting to form), $N_{\alpha\beta} = N_{\alpha}(N-x+2)-N_{\alpha}(N-x)-1$, where the additional $-1$ is due to the replacement of one original $\alpha$ bond with the $\beta$ bond that bridges two protofilaments on the micelle surface instead of contributing to $N_{\alpha\beta}(N-x)$.   

\begin{figure}[b]
\centering	\includegraphics[width=0.9\columnwidth]{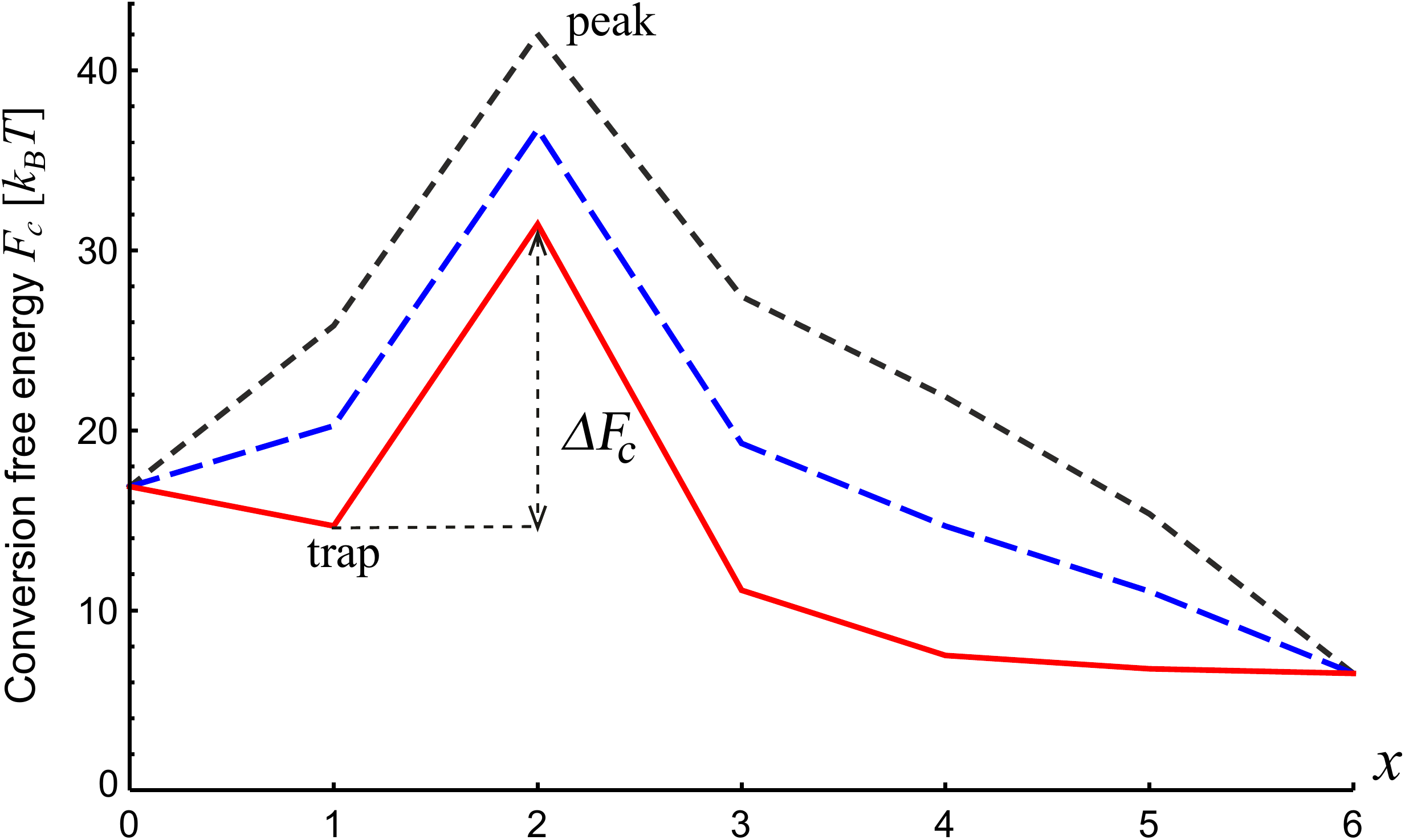}
	\caption{Free energy of conversion $F_c$ of a hexamer ($N=6$) at monomer concentration of 1 mM, as a function of the conversion fraction $x$, for three $\Delta_{\alpha\beta}$ values: $-18$ (solid line), $-20$ (dashed) and $-22$ (dotted) (in units of $k_BT$). $\Delta F_{c}$ is the conversion barrier. The peak of $F_c$ occurs at $x=2$, while the `trap', if present, is at $x=1$.}
	\label{fig:conv}
\end{figure}

In  Fig.~\ref{fig:conv} we examine the conversion free energy of a hexamer ($N=6$) as an illustration of two key features of free energy evolution during conversion: the `peak' and the ‘trap’ in $F_c(N,x)$. The peak of the conversion free energy is always at $x=2$ no matter how strong the coupling $\Delta_{\alpha\beta}$ is. This is a generic characteristic for the micelle size larger than 4 due to the presence of the second protofilament at $x=3$, almost doubling the number of $\alpha\beta$ bonds and causing an enormous stabilizing effect to drag down the value of $F_c$. A local free energy minimum, which we call  a `trap' could exist, just before this barrier, at $x=1$ (a minimum appears at $\Delta_{\alpha\beta}=-22$ $k_B T$, in Fig.~\ref{fig:conv}). The difference $N_{\alpha\beta}(N,2)-N_{\alpha\beta}(N,1)$ is negative, while the difference in  $N_{\alpha\beta}$ is positive between $x=1$ and $x=0$; this assures that the trap can only be at $x=1$. It should be noted that even though the local minimum appears only at $\Delta_{\alpha\beta}=-22$ $k_B T$ in the hexamer case of Fig. \ref{fig:conv}, it can be present at other $\Delta_{\alpha\beta}$ values with a larger micelle size, when the value of $N_{\alpha\beta} (N,1)-N_{\alpha\beta} (N,0)$ and the negative value of the difference $F_c (N,1,C_1 )-F_c (N,0,C_1 )$ increase.

The conversion barrier for an $N$-size micelle, $\Delta F_{c}$, is defined as the free energy difference between the peak and the minimum (the ‘trap’ or $x=0$) along the reaction coordinate of conversion. In other words, the barrier $\Delta F_{c}$ is evaluated either as the difference $F_c (N,2)-F_c (N,0)$, or instead as $F_c (N,2)-F_c (N,1)$ when a trap is present, as eventually always happens at a sufficiently large $N$. As a result, $\Delta F_c$ is not a simple single-valued  function.

\subsection*{Nucleation rate of the two-step nucleation mechanism} \label{sec:2rate}
Each individual nucleation path in the two-step mechanism is characterized by the final size $N$ of the micelle that forms along the pathway. The nucleation free energy landscape is plotted with combined $F_\mathrm{mic}$ and $F_c$. For the micelle size between $N_h$ and $N_l$ (see Fig. \ref{fig:mic}), the nucleation free energy landscape has two barriers: one for the micellation formation and the other for the $\alpha$-$\beta$ conversion with a metastable (intermediate) state in between. This pathway is illustrated in Fig.~\ref{fig:nuclfree} for the case of a hexamer micelle ($N=6$) and $C_1=1$\,mM. The plot starts with the expression for the micelle free energy $F_\mathrm{mic}(N)$, which passes over the micelle nucleation barrier $N_h \approx 3$ and then starts decreasing towards the micelle minimum (which for this concentration will be an $N_l=35$). However, at the micelle size $N=6$ the conversion starts, and the remaining part of the plot gives the free energy $F_c(6,x)$. This part is not a continuous but a piece-wise function of $x$, same as in Fig. \ref{fig:conv}, because the expressions for the bond counts $N_s$, $N_\beta$, and $N_{\alpha \beta}$ are all piecewise. This, however, does not affect the conclusions on the barrier height or kinetic parameters calculated below. This overall free energy profile allows the use of a three-state kinetic model.

Several ways can be used to approach this three-state kinetics problem, a most direct being the application of the general Kramers-style analysis of steady-state flux \cite{Kramers1940,Hanggi1990}. However, even a simpler problem, with just one variable (degree of freedom) and a potential with two maxima surrounding the intermediate state, cannot be solved analytically except in limiting cases when one barrier is much higher than the other. In this work, this is further complicated by the fact that we have two separate variables, $N$ for the micelle size and $x$ for the degree of conversion, and in most cases the free energy barriers in Fig.~\ref{fig:nuclfree} are not high, when viewed from the intermediate state. Therefore, no formal analytical solution is anticipated in this diffusion-type scheme.

Instead, we choose a different method of evaluating the overall nucleation rate. We separate the nucleation process into three distinct elements: the transition rate of association from the monomer into the intermediate state  (written as $k_{+}C_1$), the reverse transition rate of dissociation from the intermediates back to monomers (as $k_{-}C_\mathrm{int}$), and the transition rate of conversion from the intermediate state into the final $\beta$-aggregate state ($k_c C_\mathrm{int}$), where $C_\mathrm{int}$ is the concentration of micelles (the intermediate species). Then we derive the effective nucleation rate from monomers to the final aggregate, defined as $k_2C_1$, and its rate constant $k_2$.

\begin{figure}
\centering	\includegraphics[width=0.95\columnwidth]{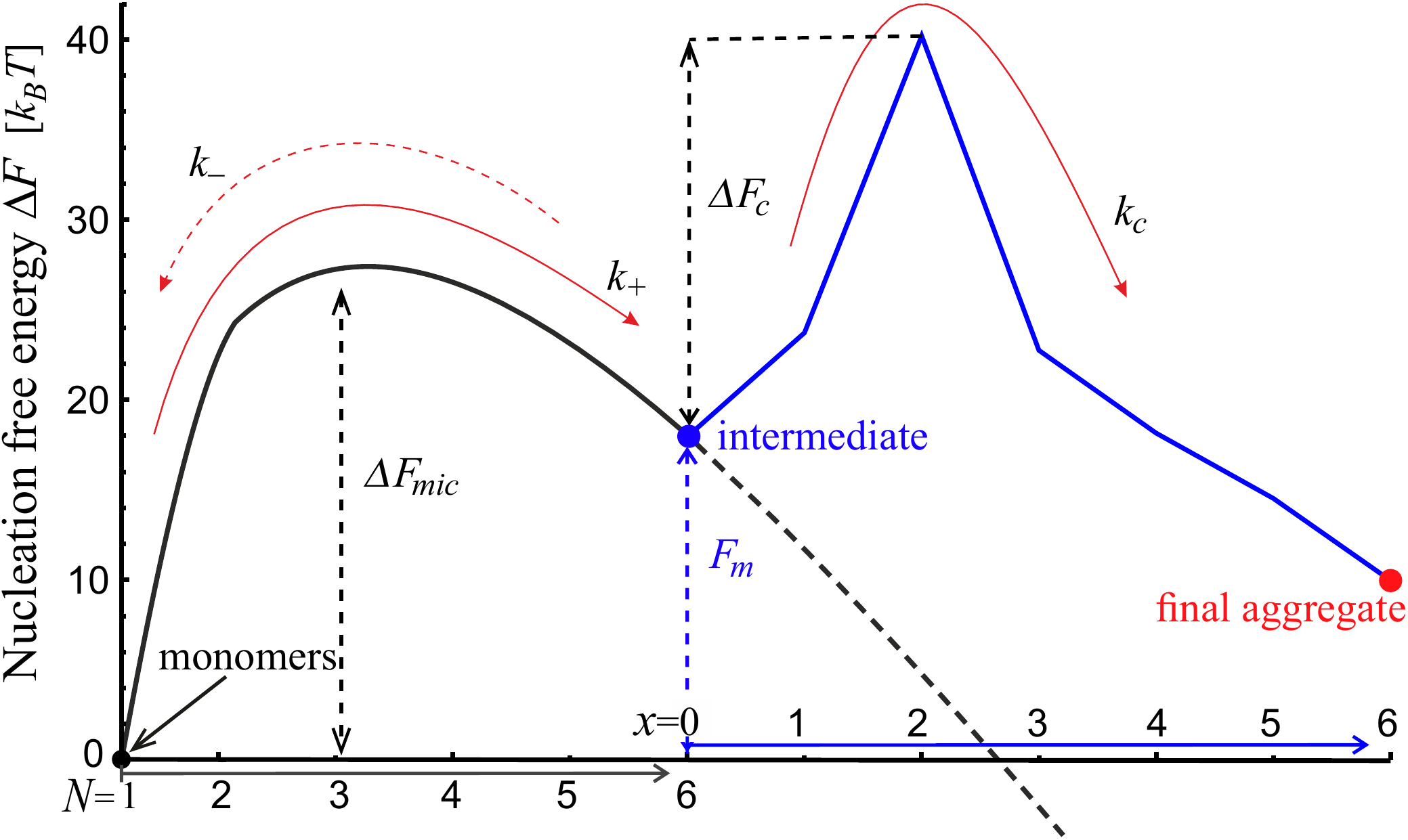}
	\caption{
Nucleation free energy landscape of an example hexamer ($N=6$) for $C_1=1$\,mM. Note the ‘composite’ nature of the $x$-axis: The left half of the plot is in the reaction coordinate for micellation, $N$, and the free energy function here is $F_\mathrm{mic}(N)$. The right half is plotted along the conversion coordinate, $x$, and producing $F_c(x)$ values. Two barriers($\Delta F_\mathrm{mic}$ and $\Delta F_c$) and three key states are labelled in the plot. $F_m$ is the free energy of the intermediates (in this example: the $N=6$ micelle before conversion).}
	\label{fig:nuclfree}
\end{figure}

Strictly, there has to be a reverse process from $\beta$-aggregate back to the intermediate micelle state, and also a process of ‘explosive’ instant dissociation of $\beta$-aggregate into monomers. These two processes are necessary if one is to find an equilibrium steady-state solution to this problem. However, our interest is to study the amyloid nucleation from solution, which is a very 
non-equilibrium process: the first stage of a fibril growth process leading to much lower free energy states where one would need to investigate the possible equilibration \cite{Michaels2014}. Besides, the subsequent elongation of the aggregate larger than the critical nucleus size is usually rather rapid, and therefore the reverse conversion from the final aggregate state back to the metastable state is slow and can be ignored. Consequently, we ignore the two very low-probability processes and regard the final $\beta$-aggregate state as irreversible.
 
We use a strategy similar to the probabilistic single-molecular approach by King and Altman tackling the Michaelis-Menten enzymatic reaction \cite{Michel2013}. Evaluation of the average time to reach the final state of nucleation, made up of an infinite number of paths with varied times of repeated micellation and dissociation processes, gives the effective nucleation rate constant in one two-step mechanism path, $k_2$ is expressed as follows (see Appendix~\ref{app:k2} for derivation):
\begin{equation} \label{eq:3-state}
k_2=\frac{k_{+} k_{c}}{2k_{+}+k_{-}+k_c} 
\end{equation}
Although this effective rate constant $k_2$ is slightly different from the traditionally used expression (see ref.~\cite{Raines1988} for instance),  Eq. \eqref{eq:3-state} recovers the average rate constant of the steady-state approximation, where the assumption that $k_c+k_{-}\gg k_{+}$ is used, giving the more familiar expression: $k_2 = k_{+} k_c/(k_c+k_{-})$. 
Two limiting cases further validate Eq. \eqref{eq:3-state}. When $k_c \gg k_{-}, \, k_+$, then the rate limiting process is the micelle formation, and $k_2=k_+$. This result is reasonable since the metastable species quickly converts into the aggregate. The other limiting case is when $k_- \gg k_+, \, k_c$. In this case $k_2 = k_+ (k_c/k_-) \ll k_+$, which reflects the alternating between the monomeric and intermediate states: completion of the reaction becomes a rare event. Both are also limiting cases of the steady-state kinetics. 

We now proceed to find explicit expressions for the rate constants $k_{+}$, $k_{-}$ and $k_c$. A convenient way to estimate the micellation rate of $N$ monomers is to separate the formation of an $N$-size micelle into the formation of one $(N-1)$-size pre-micelle followed by one $\alpha$-mer attachment. In this scheme, we use the Smoluchowski rate of forming a spherical micelle of size $N$: $4\pi D_m (r_1+r_{N-1}) C_{N-1} C_1$, where $D_m$ is the mutual diffusion coefficient, $C_{N-1}$ the concentration of $(N-1)$-size micelles, and $r_1$ and $r_{N-1}$ the radii of one monomer and an $(N-1)$-size micelle \cite{Berg1985a}. To form an $(N-1)$-size micelle, there is a free energy barrier $\Delta F_\mathrm{mic}$ to cross, since $(N-1)$ is always larger than $N_h$ in the three-state kinetic model. To find the concentration $C_{N-1}$, we separate it into two parts: a pre-thermal equilibrium concentration of $N_h$-size  micelles is first assumed to be reached, and then the simultaneous adsorption of $(N-N_h-1)$ monomers to this pre-micelle takes place, which is assumed to act as a deep adsorbing sink  (see Appendix~\ref{app:k+} for derivation). Assembling the terms, $k_+$ takes the form: 
\begin{align} \label{eq:k+}
&k_+=\omega_+ e^{-{\Delta F_\mathrm{mic}}/{k_B T}}\ , \  \\ \ \mathrm{where} \ 
&\omega_+= \frac{2k_BT}{3\eta}C_1^{N-N_h}\left(2+\sqrt[3]{N-1}+\frac{1}{\sqrt[3]{N-1}}\right) \notag \\
& \qquad \cdot \left[\frac{4\pi r_1^3}{3}(1+\sqrt[3]{N_h})^3 \right]^{N-N_h-1} . \notag
\end{align} 
From Eq. \eqref{eq:mic}, it is clear that $e^{-\Delta F_\mathrm{mic}/k_B T}$ is proportional to $C_1^{N_h -1}$. {Consequently, the product $k_+ C_1$ implicitly has the factor $C_1^N$ in it, and therefore corresponds to the kinetic expression for the $N$-particle collision, which validates our formulation for the constant $k_+$.}

In a system where the monomer concentration is maintained constant (either by re-supplying the depleted molecules from a reservoir, or simply ignoring the small change at the nucleation stage), the micelle dissociation rate constant $k_-$ can be derived from an argument based on the thermal equilibrium condition that requires the dissociation rate equal to the micellation rate, namely the equation $k_-=k_+ \exp(F_m/k_B T)$. The expression for $k_-$ is then:
\begin{equation} \label{eq:k-}
k_-=\omega_+e^{-[{\Delta F_\mathrm{mic}-F_m}]/{k_B T}} . 
\end{equation}	
Here $F_m$ is the free energy of the metastable state: the micelle of size $N$, see Fig.~\ref{fig:nuclfree}. {The expression for $F_m$ has the term of $(1-N)  k_B T \ln {C_1}$ arising from Eqs. \eqref{eq:mic} or (\ref{eq:con}), and therefore the product of $C_1^{N-N_h}$ in $\omega_+$ and $\exp [-(\Delta F_\mathrm{mic}-F_m)/k_B T]$ does not have the explicit power-law dependence on $C_1$. The rate constant $k_-$ depends on $C_1$ only implicitly, and weakly, through $N_h$ that appears in $\Delta F_\mathrm{mic}$.} At a fixed monomer concentration, an increase in $N$ decreases the value of $k_-$ by having a larger negative $\alpha$ and $\alpha\beta$ bond free energies in $\Delta F_\mathrm{mic}-F_m$, which results in a smaller probability to break all the bonds into monomers.

The rate constant of the conversion reaction $k_c$ is found directly from the Kramers escape theory, and is the product of the activation factor over the barrier $\Delta F_{c}$ and the attempt frequency from thermal fluctuations of positions of the A$\beta_{1-42}$ segments, $1/\tau_I$, which has first appeared in the derivation of $k_1$, Eq. \eqref{eq:elong}:
\begin{equation} \label{eq:kc}
k_c=\frac{1}{\tau_I} e^{-{\Delta F_{c}}/{k_B T}} 
\end{equation}

With the three rate constants determined, we can write down the overall rate of amyloid nucleation $k_2$ for a given size $N$. The full expression is cumbersome, but let us examine $k_2$ in two limiting cases. One is $k_2\approx k_+$ when $k_c \gg k_- , \, k_+$ (micelle formation from solution is the rate-limiting process); the other is $k_2\approx k_+ (k_c/k_-)$ when $k_- \gg k_+$ and $k_c$ (conversion of a micelle is the rate-limiting process). These two extreme cases give the expressions of 
$\omega_+ e^{-\Delta F_\mathrm{mic}/k_B T}$ and 
$e^{-(F_m+\Delta F_{c})/k_B T}/\tau_I $, respectively. When conversion is the major barrier, $k_2$ depends on the highest free energy, i.e. $F_m+\Delta F_c$, cf. Fig.~\ref{fig:nuclfree}, and is the product of a quasiequilibrium population for the metastable state and the probability to cross the conversion barrier. On the other hand, when the micellation process is slow, $k_2$ depends on the $\Delta F_\mathrm{mic}$ barrier only. In general, Eq. \eqref{eq:3-state} is a mixture of these three kinetic processes.

\subsection*{The fastest growing nuclei} \label{sec:optsize}
Since the monomer concentration in experiments mostly is within the range from $\mu$M to a few mM, and using the standard concentration of 1~mM, {we let $\ln{C_1}$ vary from $-6.5$ (i.e. $C_1 \approx 1.5\, \mu$M, approximately the threshold monomer concentration to initiate the two-step mechanism, discussed in the section of micellation of soluble peptides, Fig. \ref{fig:mic}), up to $\ln{C_1}=2$ (corresponding to $C_1 \approx 7.4$\,mM).} Values of ${C_1}$ outside of this common experimental range will not be discussed further.  {From Eqs.~(\ref{eq:k1}) and (\ref{eq:3-state})-(\ref{eq:kc}), we evaluate $\ln{(k_2/k_{1}^{\circ})}$ of different micelle sizes at several monomer concentrations, where the reference rate constant $k_{1}^{\circ}$ is the  value of the rate  $k_1$ of the NP mechanism at $\ln{C_1}=-6.5$, referring to the lowest monomer concentration case we investigate.} In this way, the fastest rate of two-step nucleation can be detected, identifying the dominant micelle species $N^*$ in this nucleation mechanism. 

The $x$-axis in the $\ln{k_2}$ plot varies from $N_h$ to $N_l$, which is the range of the metastable micelle sizes that can adopt the two-step mechanism for each chosen monomer concentration $C_1$. For clarity and convenience, we normalize the plots of $\ln{k_2}$ to extend between 0 and 1. The width of the peak of nucleation rate, if such a peak exists, is defined as the size range of micelles with the fraction of 0.5 in the $y$-axis. All the parameters involved in the estimation of $k_2$ have been listed in Appendix~\ref{app:abpara} with the room temperature for $T$ and the viscosity of water at room temperature for $\eta$. The results for $\ln{k_2 (N)}$, at $\Delta_{\alpha\beta}=-18\,k_B T$ as an illustrative example, are shown in Fig.~\ref{fig:micpeak}. The plots in Fig.~\ref{fig:micpeak} show two main features: the peak of the nucleation rate that occurs at a certain micelle size $N^*(C_1)$, and the width of this peak increasing as the monomer concentration increases. 

\begin{figure} 
\centering	\includegraphics[width=1\columnwidth]{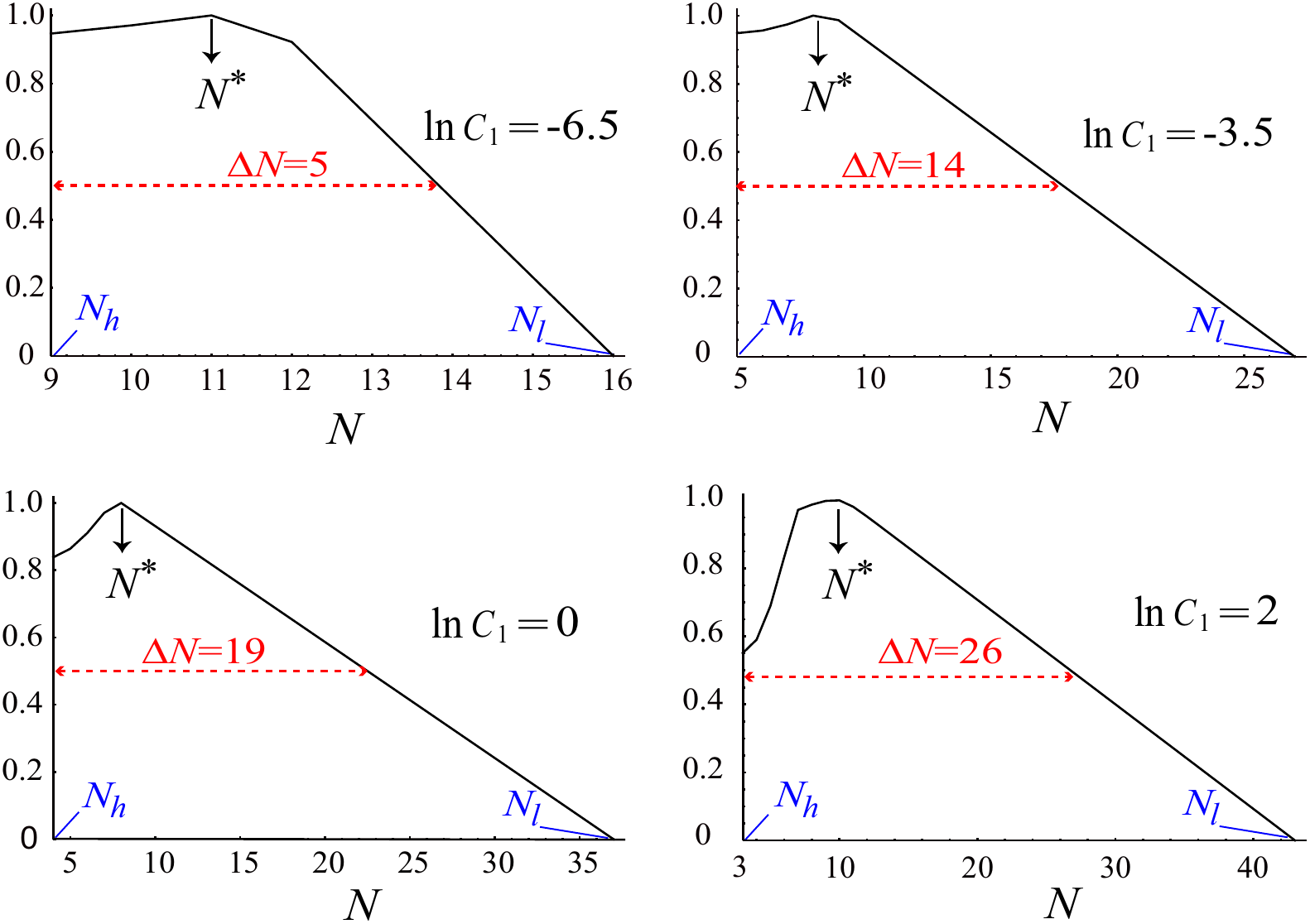}
	\caption{Plots of $\ln{k_2}$ (normalized between 0 and 1) against the micelle size $N$ for several $C_1$ values (using an example of $\Delta_{\alpha\beta}=-18\, k_B T$).  The $N$-axis range  in each plot is between $N_h(C_1)$ and $N_l(C_1)$, where the relevant range of two-step nucleation occurs. The width of each peak is indicated in the plots.}
	\label{fig:micpeak}
\end{figure}

{To understand the fastest growing rate, let us examine the contributing effects of micelle formation at a rate $k_+$, and its conversion at a rate $k_c$. For the parameters corresponding to A$\beta_{1-42}$ (Appendix~\ref{app:abpara}), the conversion barrier $\Delta F_c$ is always between 10 and 25 $k_B T$ for the micelle size below 40 (which is clearly the maximum micelle size value that we will encounter).} On the other hand, the barrier to form a micelle, $\Delta F_\mathrm{mic}$,  starts from roughly 50 $k_B T$ at $\ln{C_1}=-6.5$, and decreases to 20 $k_B T$ at $\ln{C_1}=2$, meaning that the rate $k_+$ at $N=N_h$ is very small -- and then sharply increases as $C_1$ grows to finally become comparable with $k_c$ at high concentrations. 
The value of $k_-$ at $N=N_h$ is very high, which is easy to see by inserting $F_m=\Delta F_\mathrm{mic}$ at $N=N_h$ in Eq. \eqref{eq:k-}. Then $k_-$ decreases dramatically as $N$ grows from $N_h$, due to the $C_1^N e^{F_m/k_B T}$ factor in \eqref{eq:k-}. Hence there must exist a particular micelle size $N^*$ when $k_- \approx k_c$. When $N<N^*$, we have the nucleation rate constant $k_2 \approx k_ck_+/k_- = k_c e^{-F_m/k_B T}$, which a is growing function of $N$. On the other side, at $N>N^*$,  the $k_-$ is small and $k_2\approx k_+$, which is a decreasing function of $N$. Between these regimes we will always find the point $N=N^*$ of the fastest rate of two-step nucleation. 

To put it more physically, when micelle size is too small, it suffers a rather large dissociation rate and cannot undergo the full subsequent conversion into fibrils. As micelle size grows bigger, they become less prone to dissociation and will have adequate time to complete conversion process. However, a further growth of micelles will decrease the micellation rate. In this case, micellation is the rate-limiting process and the total time of the two-step nucleation will therefore increase with micelle size. Basically, this $N^*$ value is the size where the effect of slow micellation process starts to take over other two kinetic processes.

{Figure~\ref{fig:optsize} illustrates this effective size of the critical nucleus $N^*$ (the `peak' in Fig.~\ref{fig:micpeak}) against the monomer concentration ${C_1}$, using  three values of the $\alpha$-$\beta$ interaction parameter $\Delta_{\alpha\beta}$ from its plausible range. At very low monomer concentrations, this nucleation size $N^*$ starts from a larger value, gradually decreasing to  finally reach a plateau that spans over a broad range of concentrations. (There is an additional feature: the re-increase of $N^*$ for a low $\Delta_{\alpha\beta}=-18$ $k_B T$ at very high concentrations, which we do not fully understand and will not be discussing further). These values of `critical nucleus' size  are within the range obtained by the coarse-grained molecular simulations of Sari\'c et al. \cite{Saric2014}: from 2 to 14; they also fall within the experimentally observed values (2 to roughly 25) \cite{Sakono2010}. Notably, the fact that $N^*$ is not a constant value in Fig.~\ref{fig:optsize} indicates the weakness of the assumption about a fixed, concentration-independent micelle size before conversion, which was employed in the previous two-step kinetic model \cite{Garcia2014}.} Nevertheless, a typical prediction of the critical nucleus size $N^* \simeq $ 7-8 over a broad range of monomer concentrations seems to be appropriate. We believe this $N^*$ value is reasonable, and is close to the previous work of the two-step nucleation kinetics of amyloid peptides by Lomakin et al.  \cite{Lomakin1997}, which gave the nucleus size of 10,  although they did not consider the conversion process from micelles into fibrils. 
 
{To qualitatively understand the dependence of  $N^*$ on concentration $C_1$, we can use the approximate condition for the  peak of the total rate constant $k_2$, namely $k_-\approx k_c$, where we take $k_c$ to be a constant (it has only a slow variation with $N$ compared with $k_-$). From Eq. \eqref{eq:k-} we know that $k_-$ does not have a power-law dependence on $C_1$, and we further ignore the electrostatic/entropic repulsion term in $F_\mathrm{mic}$ since the micelle size is quite small and this repulsion free energy is relatively low in this region. 
   The difference $N-N_h$, required for this ratio to reach a constant value, depends on the choice of $N_h$ due to the nonlinear term in $N_{\alpha}$. A small $N_h$ requires a larger difference compared with higher $N_h$ values. One should further notice that $N_h$ depends on $C_1$ and gradually decreases as $C_1$ increases (see Fig.~\ref{fig:mic}). Therefore, $N^{*}$ is controlled by two opposite effects: the shift of $N_h$ to a lower value with an increased $\ln{C_1}$, and the growing $N-N_h$ difference required for $\ln{k_-}$ to reach a specific constant as $\ln{C_1}$ increases. At a small $C_1$, the shift of $N_h$ wins, causing a larger $N^*$ value. As $C_1$ increases, these two effects may offset each other, explaining the plateau region in Fig.~\ref{fig:optsize}. The re-increase of the $N^*$ value with $\Delta_{\alpha\beta}=-18 k_B T$ at $\ln{C_1} \geq 2$ is likely due to the other peak condition, that $k_+\approx k_c$. At such a high $C_1$, the micelle formation rate $k_+$ decreases slowly with $N$, which allows the micelle size $N$ to grow more significantly before $k_+$ reaches the required $k_c$ value, producing a noticeable increase in $N^*$.}

\begin{figure}
\centering	\includegraphics[width=0.9\columnwidth]{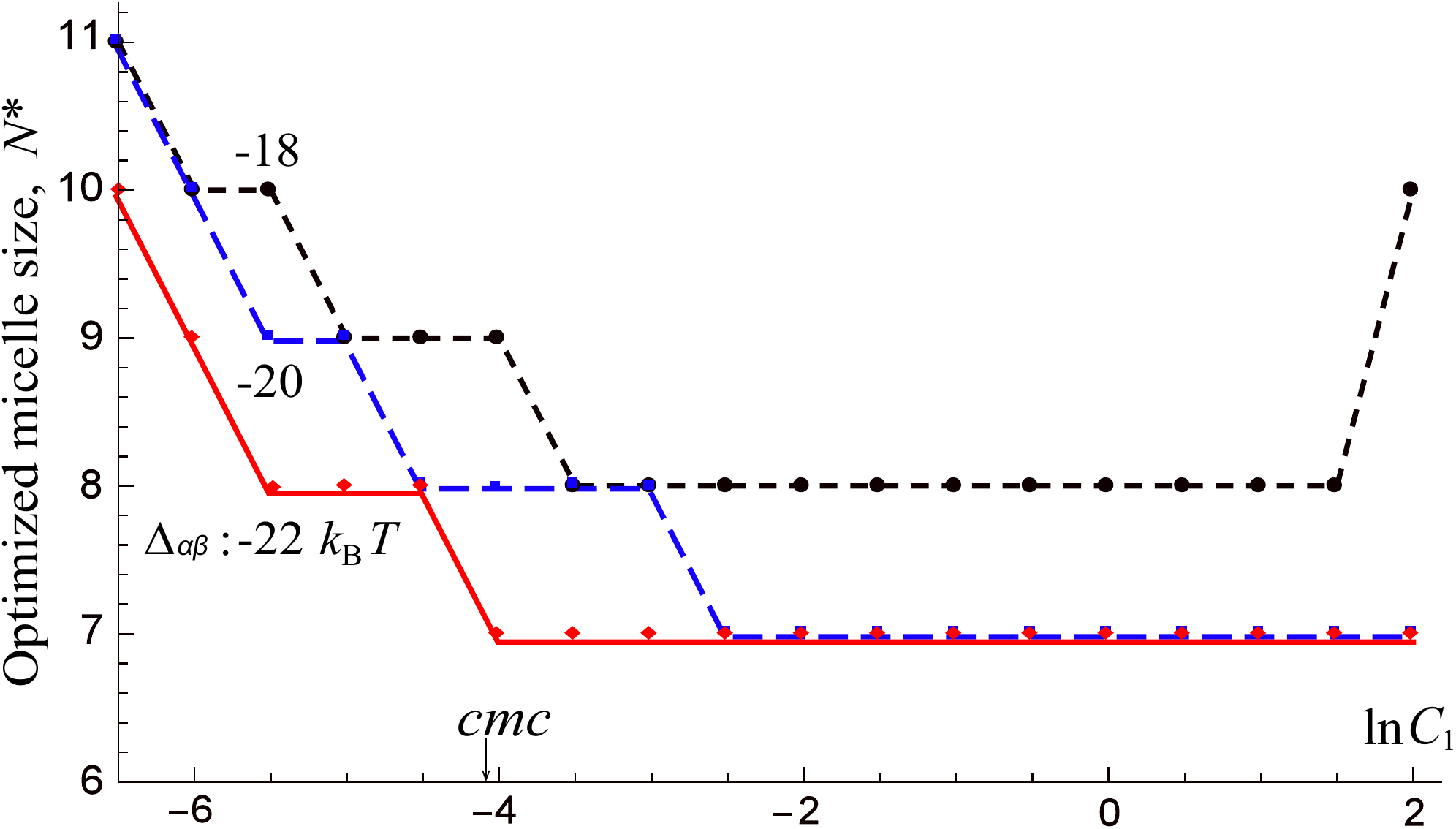}
	\caption{
The plot of the fastest-growing nucleus size $N^*$ against $\ln  C_1$, for three values of $\Delta_{\alpha\beta}/k_BT$: $-18$ (dotted line, circles), $-20$ (dashed line, squares) and $-22$ (solid line, diamonds) from top to bottom. We conclude that a safe assumption is that $N^*\approx 7$ for A$\beta_{1-42}$, for a broad range of monomer concentrations. }
	\label{fig:optsize}
\end{figure}

Strictly, the total nucleation rate in the two-step mechanism, $k_2^{'}$ is the sum of all $k_2$ values of different individual pathways labelled by the micelle size $N$, at a given monomer concentration $C_1$. However, due to the presence of the sharp peak in $k_2(N)$, we may ignore the spread of different nucleus sizes and only represent nucleation by the peak rate  $k_2(N^*)$. A comparison of these two values is given in the inset of Fig.~\ref{fig:k1k2}, proving that the error in ignoring nucleation pathways other than $N^*$ is very small.  This allows us to investigate how $k_2^{*}$ changes with increasing  monomer concentration, and propose possible explanations for the features present in Fig.~\ref{fig:k1k2}.

\section*{Comparison of nucleation rates}\label{sec:compare}
With Eqs. \eqref{eq:k1} and \eqref{eq:3-state}, a comparison of rate constants of the two competing mechanisms, $k_1 (C_1)$ and $k_2(C_1)$, is plotted in Fig.~\ref{fig:k1k2}.  Both rate constants are normalized by the same factor: the $k_1$ value at $\ln{C_1}=-6.5$, referring to the lowest monomer concentration case we investigate.  The $\ln{k_1}$ curve is not perfectly linear on the log-log plot of  Fig.~\ref{fig:k1k2} due to the two-competing time-scales in \eqref{eq:k1}. For the concentration regime we investigate, the diffusive arrival time $\tau_D$, which is proportional to $C_1$, is smaller than the internal re-arrangement time of peptides, $\tau_I$. The $\tau_D$ factor is hence screened off, giving roughly a dependence  $k_1 \propto C_1^2$ in Eq. \eqref{eq:k1}. {However, in a traditional approach to nucleation, with the critical nucleus size in the NP mechanism at $n_c=4$ (since we used the rate of producing tetramers as the measure for our nucleation rate), one expects a cubic dependence:  $k_1 \propto C_1^{N_c}$;} this will be observed when the diffusion time is dominant -- but not in the relevant concentration regime. 

For the three versions of $k_2$, for three values of the conversion penalty $\Delta_{\alpha\beta}$, there are two features to observe. First, a linear trend line could be used to fit $\ln{k_2}$ in the range of $\ln{C_1}$ between -6.5 and 0, which suggests an effective power-law dependence $k_2 \propto C_1^n$. However, the exponent of this power law does not match the value of the critical nucleus size for the two-step mechanism, $N^* \approx 7$. Moreover, this slope is not universal and clearly varies  with the values of $\Delta_{\alpha\beta}$ parameter.  

We first focus on how $\Delta_{\alpha\beta}$ comes into play. At low monomer concentration, we found that $k_2$ is first determined by the relation $k_2=k_+ k_c/k_-$, which then changes to $k_2=k_+$ (the crossover at $k_-=k_c$ is what determines the fastest growing size $N^*$). Accordingly, an increase in $\Delta_{\alpha\beta}$ has a strong effect on the rate $k_2$ by decreasing the conversion rate $k_c$ at low $C_1$. In contrast, at higher $C_1$ we have $k_2 = k_+$ which does not depend on $\Delta_{\alpha\beta}$: different $k_2$ lines overlapping in the high-concentration regime in Fig.~\ref{fig:k1k2}.

The crossover concentration when the two-step mechanism starts to overtake the direct NP mechanism of amyloid nucleation is also strongly dependent on the $\alpha$-$\beta$ interaction energy $\Delta_{\alpha\beta}$, which is why it is important to have better estimates of its value. But in all cases this happens in the low- to medium-concentration regime. From Fig.~\ref{fig:k1k2}, we find that the two-step mechanism takes over from $\ln{C_1}=-6$ (2.5 $\mathrm{\mu}$M)  for $\Delta_{\alpha\beta}=-22 \,$ $k_B T$, this crossover concentration increasing to $\ln{C_1}=-3$  (50 $\mathrm{\mu}$M) for a lower bound $\Delta_{\alpha\beta}=-18\,$ $k_B T$. The crossover concentration was given as approximately 8 $\mathrm{\mu}$M by Auer et al. \citep{Auer2012}, where the two-step mechanism was also concluded to take place at a higher concentration than the NP mechanism.

\begin{figure} 
\centering	\includegraphics[width=1\columnwidth]{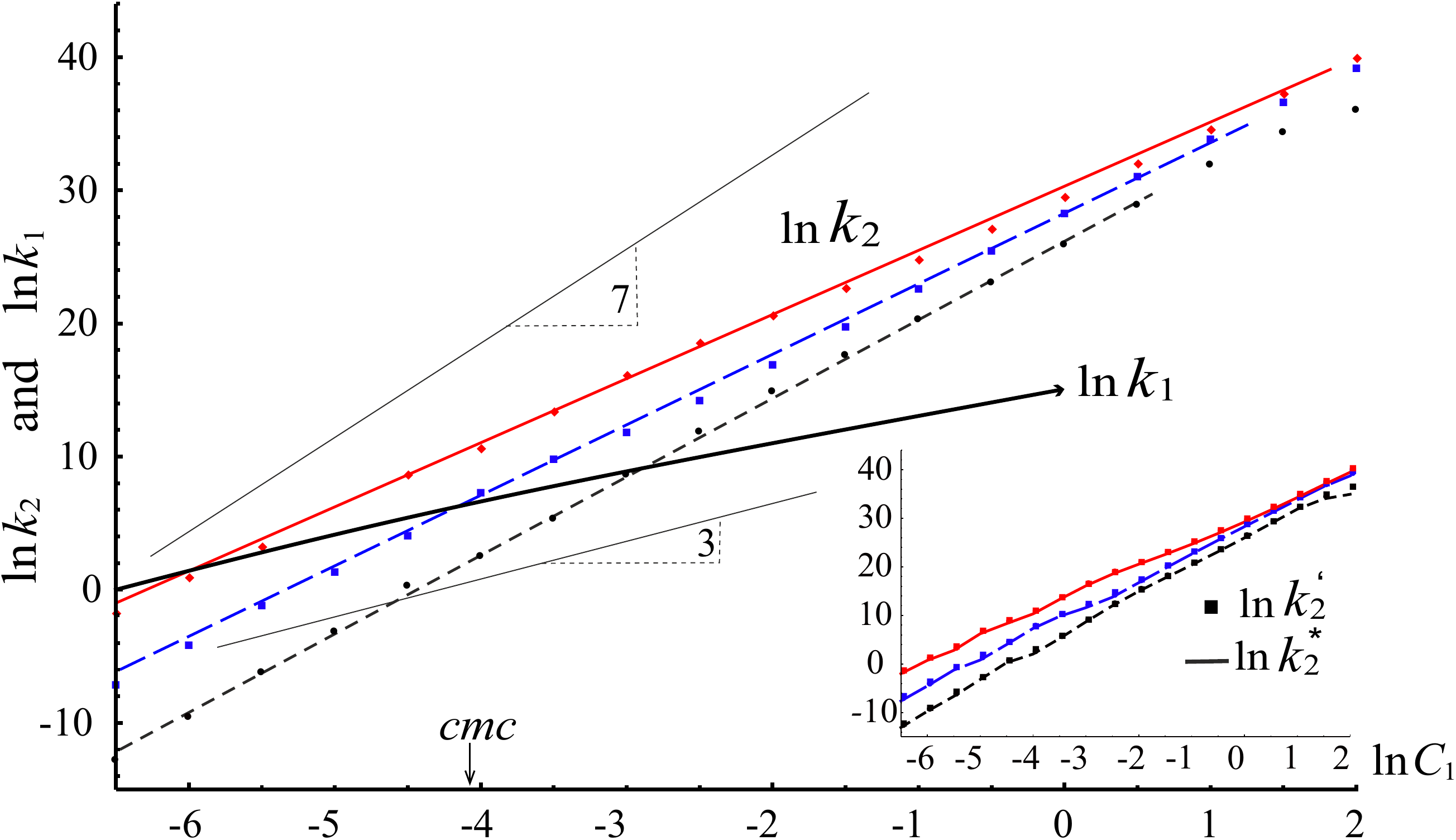}
	\caption{The log-log plot of the normalized effective nucleation rate in the two-step mechanism, $\ln{k_2}$, and in the NP mechanism, $\ln{k_1}$, against the monomer concentration $C_1$, at three values of $\Delta_{\alpha\beta}=-18$ (dotted line), $-20$ (dashed line) and $-22$ (solid line). The lines are fitted linear trends of $\ln{k_2}$, whose slopes are given as approximately 5.9 (black), 5.3 (blue) and 4.8 (red). The bold solid line of $\ln{k_1}$ is not perfectly linear, but has the slope exponent of approximately 2 at higher $C_1$. For comparison, we show the slopes of $C_1^7$ and $C_1^3$, where the exponents are the critical nucleus sizes in the two mechanisms. The inset shows the comparison of the total rate $\ln{k_2^{'}}$ (squares) and the peak rate $\ln{k_2^{*}}$ (lines), demonstrating that the difference is very small.}
	\label{fig:k1k2}
\end{figure}

\section*{Conclusions}
We elucidate the kinetics of two alternative amyloid nucleation mechanisms, the direct (NP) nucleation and the two-step nucleation via an aggregation of $\alpha$-micelle and its subsequent conversion into the $\beta$ filament. We used an equilibrium free energy approach to consider the possibilities that the critical nucleus and favorable micelle size change with monomer concentration. This analysis of the composite free energy landscape in the two-step mechanism allows the use of a simplified three-state kinetic model and thus the analytical derivation of the effective nucleation rate constant. We define the critical nucleation size by the fastest growing rate criterion, having verified that this rate dominates a practically measured total rate of nucleation. 

We find that theoretically predicted variation of nucleation rate with monomer concentration could be easily interpreted as a power law in the analysis of experimental data, even though the actual expression is not.  For the NP mechanism, the critical nucleus size is determined to be the trimer of the protofilament pair, and this does not change within the monomer concentration range we investigate. On the other hand, in the two-step mechanism, the micelle size with the largest nucleation rate is dependent not only on monomer concentration, but also on the strength of the interaction between $\alpha$- and $\beta$-mers ($\Delta_{\alpha\beta}$). Generally speaking, the optimized micelle size decreases from around 10 to around 7, and then remains at this value for a wide concentration range, pointing out the inappropriateness of the pre-assumption of the fixed micelle size in the previous kinetic approaches. 
 
Although a nearly power-law relationship of both nucleation rates on monomer concentration is observed, the exponent of the `apparent power law' in the two-step mechanism have no simple relationship on the critical nucleus size, as would be expected from the kinetic formalism of  classical nucleation theory. The exponent of $k_2 \propto C_1^n$ also depend on $\Delta_{\alpha\beta}$, and is determined by a complex interplay between micellation, conversion and dissociation processes. Unfortunately, such complexity makes the rigorous analysis of these effective power laws hard to carry out. But it is certain that the critical nucleus size obtained by fitting the Oosawa model, or other kinetic models that all assume the classical nucleation formula in amyloid aggregation is not actually the real critical nucleus size. One may think of some direct experimental methods of determining the critical nucleus size that could extract this important nucleation size (in both mechanisms), and thus compare with our theoretical predictions.  Unfortunately, no such experiments are yet available, and instead molecular simulations would be a more plausible way to understand the amyloid nucleation.

It is also important that the theoretical prediction for the crossover  point of concentration where the two-step mechanism takes over from the direct filament  nucleation depends on the strength of a not well-known interaction strength $\Delta_{\alpha\beta}$. Experiment or simulation are also needed to focus on finding a more accurate value of this parameter, in order for our predictions to become more quantitative.  

\subsection*{Acknowledgements}
The authors acknowledge a significant input from T.P.J. Knowles, A. \v{S}ari\'{c}, D. Frenkel,  I. Ford, M. Vendruscolo and P. Sormanni in discussing the general concepts, details, presentation, as well as experimental and simulation data. This work has been supported by the Theory of Condensed Matter Critical Mass Grant from
EPSRC (EP/J017639).

\subsection*{References}

\begin{thebibliography}{84}%
\makeatletter
\providecommand \@ifxundefined [1]{%
 \@ifx{#1\undefined}
}%
\providecommand \@ifnum [1]{%
 \ifnum #1\expandafter \@firstoftwo
 \else \expandafter \@secondoftwo
 \fi
}%
\providecommand \@ifx [1]{%
 \ifx #1\expandafter \@firstoftwo
 \else \expandafter \@secondoftwo
 \fi
}%
\providecommand \natexlab [1]{#1}%
\providecommand \enquote  [1]{``#1''}%
\providecommand \bibnamefont  [1]{#1}%
\providecommand \bibfnamefont [1]{#1}%
\providecommand \citenamefont [1]{#1}%
\providecommand \href@noop [0]{\@secondoftwo}%
\providecommand \href [0]{\begingroup \@sanitize@url \@href}%
\providecommand \@href[1]{\@@startlink{#1}\@@href}%
\providecommand \@@href[1]{\endgroup#1\@@endlink}%
\providecommand \@sanitize@url [0]{\catcode `\\12\catcode `\$12\catcode
  `\&12\catcode `\#12\catcode `\^12\catcode `\_12\catcode `\%12\relax}%
\providecommand \@@startlink[1]{}%
\providecommand \@@endlink[0]{}%
\providecommand \url  [0]{\begingroup\@sanitize@url \@url }%
\providecommand \@url [1]{\endgroup\@href {#1}{\urlprefix }}%
\providecommand \urlprefix  [0]{URL }%
\providecommand \Eprint [0]{\href }%
\providecommand \doibase [0]{http://dx.doi.org/}%
\providecommand \selectlanguage [0]{\@gobble}%
\providecommand \bibinfo  [0]{\@secondoftwo}%
\providecommand \bibfield  [0]{\@secondoftwo}%
\providecommand \translation [1]{[#1]}%
\providecommand \BibitemOpen [0]{}%
\providecommand \bibitemStop [0]{}%
\providecommand \bibitemNoStop [0]{.\EOS\space}%
\providecommand \EOS [0]{\spacefactor3000\relax}%
\providecommand \BibitemShut  [1]{\csname bibitem#1\endcsname}%
\let\auto@bib@innerbib\@empty
\bibitem [{\citenamefont {Murphy}\ and\ \citenamefont
  {Iii}(2010)}]{Murphy2010}%
  \BibitemOpen
  \bibfield  {author} {\bibinfo {author} {\bibfnamefont {M.~P.}\ \bibnamefont
  {Murphy}}\ and\ \bibinfo {author} {\bibfnamefont {H.~L.}\ \bibnamefont
  {Iii}},\ }\href {\doibase 10.3233/JAD-2010-1221.Alzheimer} {\bibfield
  {journal} {\bibinfo  {journal} {J. Alzheimer's Dis.}\ }\textbf {\bibinfo
  {volume} {19}},\ \bibinfo {pages} {1} (\bibinfo {year} {2010})}\BibitemShut
  {NoStop}%
\bibitem [{\citenamefont {Edison}\ \emph {et~al.}(2008)\citenamefont {Edison},
  \citenamefont {Rowe}, \citenamefont {Rinne}, \citenamefont {Ng},
  \citenamefont {Ahmed}, \citenamefont {Kemppainen}, \citenamefont
  {Villemagne}, \citenamefont {O'Keefe}, \citenamefont {N{\aa}gren},
  \citenamefont {Chaudhury}, \citenamefont {Masters},\ and\ \citenamefont
  {Brooks}}]{Edison2008}%
  \BibitemOpen
  \bibfield  {author} {\bibinfo {author} {\bibfnamefont {P.}~\bibnamefont
  {Edison}}, \bibinfo {author} {\bibfnamefont {C.~C.}\ \bibnamefont {Rowe}},
  \bibinfo {author} {\bibfnamefont {J.~O.}\ \bibnamefont {Rinne}}, \bibinfo
  {author} {\bibfnamefont {S.}~\bibnamefont {Ng}}, \bibinfo {author}
  {\bibfnamefont {I.}~\bibnamefont {Ahmed}}, \bibinfo {author} {\bibfnamefont
  {N.}~\bibnamefont {Kemppainen}}, \bibinfo {author} {\bibfnamefont {V.~L.}\
  \bibnamefont {Villemagne}}, \bibinfo {author} {\bibfnamefont
  {G.}~\bibnamefont {O'Keefe}}, \bibinfo {author} {\bibfnamefont
  {K.}~\bibnamefont {N{\aa}gren}}, \bibinfo {author} {\bibfnamefont {K.~R.}\
  \bibnamefont {Chaudhury}}, \bibinfo {author} {\bibfnamefont {C.~L.}\
  \bibnamefont {Masters}}, \ and\ \bibinfo {author} {\bibfnamefont {D.~J.}\
  \bibnamefont {Brooks}},\ }\href {\doibase 10.1136/jnnp.2007.127878}
  {\bibfield  {journal} {\bibinfo  {journal} {J. Neurol. Neurosurg.
  Psychiatry}\ }\textbf {\bibinfo {volume} {79}},\ \bibinfo {pages} {1331}
  (\bibinfo {year} {2008})}\BibitemShut {NoStop}%
\bibitem [{\citenamefont {DiFiglia}\ \emph {et~al.}(1997)\citenamefont
  {DiFiglia}, \citenamefont {Sapp}, \citenamefont {Chase}, \citenamefont
  {Davies}, \citenamefont {Bates}, \citenamefont {Vonsattel},\ and\
  \citenamefont {Aronin}}]{DiFiglia1997}%
  \BibitemOpen
  \bibfield  {author} {\bibinfo {author} {\bibfnamefont {M.}~\bibnamefont
  {DiFiglia}}, \bibinfo {author} {\bibfnamefont {E.}~\bibnamefont {Sapp}},
  \bibinfo {author} {\bibfnamefont {K.~O.}\ \bibnamefont {Chase}}, \bibinfo
  {author} {\bibfnamefont {S.~W.}\ \bibnamefont {Davies}}, \bibinfo {author}
  {\bibfnamefont {G.~P.}\ \bibnamefont {Bates}}, \bibinfo {author}
  {\bibfnamefont {J.~P.}\ \bibnamefont {Vonsattel}}, \ and\ \bibinfo {author}
  {\bibfnamefont {N.}~\bibnamefont {Aronin}},\ }\href {\doibase
  10.1126/science.277.5334.1990} {\bibfield  {journal} {\bibinfo  {journal}
  {Science}\ }\textbf {\bibinfo {volume} {277}},\ \bibinfo {pages} {1990}
  (\bibinfo {year} {1997})}\BibitemShut {NoStop}%
\bibitem [{\citenamefont {Prusiner}(1991)}]{Prusiner1991}%
  \BibitemOpen
  \bibfield  {author} {\bibinfo {author} {\bibfnamefont {S.~B.}\ \bibnamefont
  {Prusiner}},\ }\href {\doibase 10.1126/science.1675487} {\bibfield  {journal}
  {\bibinfo  {journal} {Science}\ }\textbf {\bibinfo {volume} {252}},\ \bibinfo
  {pages} {1515} (\bibinfo {year} {1991})}\BibitemShut {NoStop}%
\bibitem [{\citenamefont {Pham}\ \emph {et~al.}(2011)\citenamefont {Pham},
  \citenamefont {Crews}, \citenamefont {Ubhi}, \citenamefont {Hansen},
  \citenamefont {Adame}, \citenamefont {Salmon}, \citenamefont {Galasko},
  \citenamefont {Michael}, \citenamefont {Savas},\ and\ \citenamefont
  {John}}]{Pham2011}%
  \BibitemOpen
  \bibfield  {author} {\bibinfo {author} {\bibfnamefont {E.}~\bibnamefont
  {Pham}}, \bibinfo {author} {\bibfnamefont {L.}~\bibnamefont {Crews}},
  \bibinfo {author} {\bibfnamefont {K.}~\bibnamefont {Ubhi}}, \bibinfo {author}
  {\bibfnamefont {L.}~\bibnamefont {Hansen}}, \bibinfo {author} {\bibfnamefont
  {A.}~\bibnamefont {Adame}}, \bibinfo {author} {\bibfnamefont
  {D.}~\bibnamefont {Salmon}}, \bibinfo {author} {\bibfnamefont
  {D.}~\bibnamefont {Galasko}}, \bibinfo {author} {\bibfnamefont
  {S.}~\bibnamefont {Michael}}, \bibinfo {author} {\bibfnamefont {J.~N.}\
  \bibnamefont {Savas}}, \ and\ \bibinfo {author} {\bibfnamefont
  {R.}~\bibnamefont {John}},\ }\href {\doibase
  10.1111/j.1742-4658.2010.07719.x.Progressive} {\bibfield  {journal} {\bibinfo
   {journal} {FEBS J.}\ }\textbf {\bibinfo {volume} {277}},\ \bibinfo {pages}
  {3051} (\bibinfo {year} {2011})}\BibitemShut {NoStop}%
\bibitem [{\citenamefont {Shankar}\ \emph {et~al.}(2008)\citenamefont
  {Shankar}, \citenamefont {Li}, \citenamefont {Mehta}, \citenamefont
  {Garcia-munoz}, \citenamefont {Nina}, \citenamefont {Smith}, \citenamefont
  {Brett}, \citenamefont {Farrell}, \citenamefont {Rowan}, \citenamefont
  {Lemere}, \citenamefont {Regan}, \citenamefont {Walsh}, \citenamefont
  {Sabatini},\ and\ \citenamefont {Selkoe}}]{Shankar2008}%
  \BibitemOpen
  \bibfield  {author} {\bibinfo {author} {\bibfnamefont {G.~M.}\ \bibnamefont
  {Shankar}}, \bibinfo {author} {\bibfnamefont {S.}~\bibnamefont {Li}},
  \bibinfo {author} {\bibfnamefont {T.~H.}\ \bibnamefont {Mehta}}, \bibinfo
  {author} {\bibfnamefont {A.}~\bibnamefont {Garcia-munoz}}, \bibinfo {author}
  {\bibfnamefont {E.}~\bibnamefont {Nina}}, \bibinfo {author} {\bibfnamefont
  {I.}~\bibnamefont {Smith}}, \bibinfo {author} {\bibfnamefont {F.~M.}\
  \bibnamefont {Brett}}, \bibinfo {author} {\bibfnamefont {M.~A.}\ \bibnamefont
  {Farrell}}, \bibinfo {author} {\bibfnamefont {M.~J.}\ \bibnamefont {Rowan}},
  \bibinfo {author} {\bibfnamefont {C.~A.}\ \bibnamefont {Lemere}}, \bibinfo
  {author} {\bibfnamefont {C.~M.}\ \bibnamefont {Regan}}, \bibinfo {author}
  {\bibfnamefont {D.~M.}\ \bibnamefont {Walsh}}, \bibinfo {author}
  {\bibfnamefont {B.~L.}\ \bibnamefont {Sabatini}}, \ and\ \bibinfo {author}
  {\bibfnamefont {D.~J.}\ \bibnamefont {Selkoe}},\ }\href@noop {} {\bibfield
  {journal} {\bibinfo  {journal} {Nat. Med.}\ }\textbf {\bibinfo {volume}
  {14}},\ \bibinfo {pages} {837} (\bibinfo {year} {2008})}\BibitemShut
  {NoStop}%
\bibitem [{\citenamefont {Nimmrich}\ \emph {et~al.}(2008)\citenamefont
  {Nimmrich}, \citenamefont {Grimm}, \citenamefont {Draguhn}, \citenamefont
  {Barghorn}, \citenamefont {Lehmann}, \citenamefont {Schoemaker},
  \citenamefont {Hillen}, \citenamefont {Gross}, \citenamefont {Ebert},\ and\
  \citenamefont {Bruehl}}]{Nimmrich2008}%
  \BibitemOpen
  \bibfield  {author} {\bibinfo {author} {\bibfnamefont {V.}~\bibnamefont
  {Nimmrich}}, \bibinfo {author} {\bibfnamefont {C.}~\bibnamefont {Grimm}},
  \bibinfo {author} {\bibfnamefont {A.}~\bibnamefont {Draguhn}}, \bibinfo
  {author} {\bibfnamefont {S.}~\bibnamefont {Barghorn}}, \bibinfo {author}
  {\bibfnamefont {A.}~\bibnamefont {Lehmann}}, \bibinfo {author} {\bibfnamefont
  {H.}~\bibnamefont {Schoemaker}}, \bibinfo {author} {\bibfnamefont
  {H.}~\bibnamefont {Hillen}}, \bibinfo {author} {\bibfnamefont
  {G.}~\bibnamefont {Gross}}, \bibinfo {author} {\bibfnamefont
  {U.}~\bibnamefont {Ebert}}, \ and\ \bibinfo {author} {\bibfnamefont
  {C.}~\bibnamefont {Bruehl}},\ }\href {\doibase
  10.1523/JNEUROSCI.4771-07.2008} {\bibfield  {journal} {\bibinfo  {journal}
  {J. Neurosci.}\ }\textbf {\bibinfo {volume} {28}},\ \bibinfo {pages} {788}
  (\bibinfo {year} {2008})}\BibitemShut {NoStop}%
\bibitem [{\citenamefont {Cohen}\ \emph {et~al.}(2013)\citenamefont {Cohen},
  \citenamefont {Linse}, \citenamefont {Luheshi}, \citenamefont {Hellstrand},
  \citenamefont {White}, \citenamefont {Rajah}, \citenamefont {Otzen},
  \citenamefont {Vendruscolo}, \citenamefont {Dobson},\ and\ \citenamefont
  {Knowles}}]{Cohen2013}%
  \BibitemOpen
  \bibfield  {author} {\bibinfo {author} {\bibfnamefont {S.~I.~A.}\
  \bibnamefont {Cohen}}, \bibinfo {author} {\bibfnamefont {S.}~\bibnamefont
  {Linse}}, \bibinfo {author} {\bibfnamefont {L.~M.}\ \bibnamefont {Luheshi}},
  \bibinfo {author} {\bibfnamefont {E.}~\bibnamefont {Hellstrand}}, \bibinfo
  {author} {\bibfnamefont {D.~A.}\ \bibnamefont {White}}, \bibinfo {author}
  {\bibfnamefont {L.}~\bibnamefont {Rajah}}, \bibinfo {author} {\bibfnamefont
  {D.~E.}\ \bibnamefont {Otzen}}, \bibinfo {author} {\bibfnamefont
  {M.}~\bibnamefont {Vendruscolo}}, \bibinfo {author} {\bibfnamefont {C.~M.}\
  \bibnamefont {Dobson}}, \ and\ \bibinfo {author} {\bibfnamefont {T.~P.~J.}\
  \bibnamefont {Knowles}},\ }\href@noop {} {\bibfield  {journal} {\bibinfo
  {journal} {Proc. Natl. Acad. Sci. U. S. A.}\ }\textbf {\bibinfo {volume}
  {110}},\ \bibinfo {pages} {9758} (\bibinfo {year} {2013})}\BibitemShut
  {NoStop}%
\bibitem [{\citenamefont {Granata}\ \emph {et~al.}(2015)\citenamefont
  {Granata}, \citenamefont {Baftizadeh}, \citenamefont {Habchi}, \citenamefont
  {Galvagnion}, \citenamefont {Simone}, \citenamefont {Camilloni},
  \citenamefont {Laio},\ and\ \citenamefont {Vendruscolo}}]{Granata2015}%
  \BibitemOpen
  \bibfield  {author} {\bibinfo {author} {\bibfnamefont {D.}~\bibnamefont
  {Granata}}, \bibinfo {author} {\bibfnamefont {F.}~\bibnamefont {Baftizadeh}},
  \bibinfo {author} {\bibfnamefont {J.}~\bibnamefont {Habchi}}, \bibinfo
  {author} {\bibfnamefont {C.}~\bibnamefont {Galvagnion}}, \bibinfo {author}
  {\bibfnamefont {A.~D.}\ \bibnamefont {Simone}}, \bibinfo {author}
  {\bibfnamefont {C.}~\bibnamefont {Camilloni}}, \bibinfo {author}
  {\bibfnamefont {A.}~\bibnamefont {Laio}}, \ and\ \bibinfo {author}
  {\bibfnamefont {M.}~\bibnamefont {Vendruscolo}},\ }\href {\doibase
  10.1038/srep15449} {\bibfield  {journal} {\bibinfo  {journal} {Sci. Rep.}\
  }\textbf {\bibinfo {volume} {5}},\ \bibinfo {pages} {15449} (\bibinfo {year}
  {2015})}\BibitemShut {NoStop}%
\bibitem [{\citenamefont {Meehan}\ \emph {et~al.}(2007)\citenamefont {Meehan},
  \citenamefont {Knowles}, \citenamefont {Baldwin}, \citenamefont {Smith},
  \citenamefont {Squires}, \citenamefont {Clements}, \citenamefont {Treweek},
  \citenamefont {Ecroyd}, \citenamefont {Tartaglia}, \citenamefont
  {Vendruscolo}, \citenamefont {MacPhee}, \citenamefont {Dobson},\ and\
  \citenamefont {Carver}}]{Meehan2007}%
  \BibitemOpen
  \bibfield  {author} {\bibinfo {author} {\bibfnamefont {S.}~\bibnamefont
  {Meehan}}, \bibinfo {author} {\bibfnamefont {T.~P.~J.}\ \bibnamefont
  {Knowles}}, \bibinfo {author} {\bibfnamefont {A.~J.}\ \bibnamefont
  {Baldwin}}, \bibinfo {author} {\bibfnamefont {J.~F.}\ \bibnamefont {Smith}},
  \bibinfo {author} {\bibfnamefont {A.~M.}\ \bibnamefont {Squires}}, \bibinfo
  {author} {\bibfnamefont {P.}~\bibnamefont {Clements}}, \bibinfo {author}
  {\bibfnamefont {T.~M.}\ \bibnamefont {Treweek}}, \bibinfo {author}
  {\bibfnamefont {H.}~\bibnamefont {Ecroyd}}, \bibinfo {author} {\bibfnamefont
  {G.~G.}\ \bibnamefont {Tartaglia}}, \bibinfo {author} {\bibfnamefont
  {M.}~\bibnamefont {Vendruscolo}}, \bibinfo {author} {\bibfnamefont {C.~E.}\
  \bibnamefont {MacPhee}}, \bibinfo {author} {\bibfnamefont {C.~M.}\
  \bibnamefont {Dobson}}, \ and\ \bibinfo {author} {\bibfnamefont {J.~A.}\
  \bibnamefont {Carver}},\ }\href@noop {} {\bibfield  {journal} {\bibinfo
  {journal} {J. Mol. Biol.}\ }\textbf {\bibinfo {volume} {372}},\ \bibinfo
  {pages} {470} (\bibinfo {year} {2007})}\BibitemShut {NoStop}%
\bibitem [{\citenamefont {Wang}\ \emph {et~al.}(2010)\citenamefont {Wang},
  \citenamefont {Petty}, \citenamefont {Trojanowski}, \citenamefont {Knee},
  \citenamefont {Goulet}, \citenamefont {Mukerji},\ and\ \citenamefont
  {King}}]{Wang2010}%
  \BibitemOpen
  \bibfield  {author} {\bibinfo {author} {\bibfnamefont {Y.}~\bibnamefont
  {Wang}}, \bibinfo {author} {\bibfnamefont {S.}~\bibnamefont {Petty}},
  \bibinfo {author} {\bibfnamefont {A.}~\bibnamefont {Trojanowski}}, \bibinfo
  {author} {\bibfnamefont {K.}~\bibnamefont {Knee}}, \bibinfo {author}
  {\bibfnamefont {D.}~\bibnamefont {Goulet}}, \bibinfo {author} {\bibfnamefont
  {I.}~\bibnamefont {Mukerji}}, \ and\ \bibinfo {author} {\bibfnamefont
  {J.}~\bibnamefont {King}},\ }\href {\doibase 10.1167/iovs.09-3987} {\bibfield
   {journal} {\bibinfo  {journal} {Investig. Ophthalmol. Vis. Sci.}\ }\textbf
  {\bibinfo {volume} {51}},\ \bibinfo {pages} {672} (\bibinfo {year}
  {2010})}\BibitemShut {NoStop}%
\bibitem [{\citenamefont {Lee}\ \emph {et~al.}(2009)\citenamefont {Lee},
  \citenamefont {Lee}, \citenamefont {Choe}, \citenamefont {Kang},
  \citenamefont {Kim}, \citenamefont {Gai}, \citenamefont {Hahn},\ and\
  \citenamefont {Paik}}]{Lee2009c}%
  \BibitemOpen
  \bibfield  {author} {\bibinfo {author} {\bibfnamefont {J.-H.}\ \bibnamefont
  {Lee}}, \bibinfo {author} {\bibfnamefont {I.-H.}\ \bibnamefont {Lee}},
  \bibinfo {author} {\bibfnamefont {Y.-J.}\ \bibnamefont {Choe}}, \bibinfo
  {author} {\bibfnamefont {S.}~\bibnamefont {Kang}}, \bibinfo {author}
  {\bibfnamefont {H.~Y.}\ \bibnamefont {Kim}}, \bibinfo {author} {\bibfnamefont
  {W.-P.}\ \bibnamefont {Gai}}, \bibinfo {author} {\bibfnamefont {J.-S.}\
  \bibnamefont {Hahn}}, \ and\ \bibinfo {author} {\bibfnamefont {S.~R.}\
  \bibnamefont {Paik}},\ }\href {\doibase 10.1042/BJ20081572} {\bibfield
  {journal} {\bibinfo  {journal} {Biochem. J.}\ }\textbf {\bibinfo {volume}
  {418}},\ \bibinfo {pages} {311} (\bibinfo {year} {2009})}\BibitemShut
  {NoStop}%
\bibitem [{\citenamefont {Nilsson}(2004)}]{Nilsson2004}%
  \BibitemOpen
  \bibfield  {author} {\bibinfo {author} {\bibfnamefont {M.}~\bibnamefont
  {Nilsson}},\ }\href@noop {} {\bibfield  {journal} {\bibinfo  {journal}
  {Methods}\ }\textbf {\bibinfo {volume} {34}},\ \bibinfo {pages} {151}
  (\bibinfo {year} {2004})}\BibitemShut {NoStop}%
\bibitem [{\citenamefont {Kusumoto}\ \emph {et~al.}(1998)\citenamefont
  {Kusumoto}, \citenamefont {Lomakin}, \citenamefont {Teplow},\ and\
  \citenamefont {Benedek}}]{Kusumoto1998}%
  \BibitemOpen
  \bibfield  {author} {\bibinfo {author} {\bibfnamefont {Y.}~\bibnamefont
  {Kusumoto}}, \bibinfo {author} {\bibfnamefont {A.}~\bibnamefont {Lomakin}},
  \bibinfo {author} {\bibfnamefont {D.~B.}\ \bibnamefont {Teplow}}, \ and\
  \bibinfo {author} {\bibfnamefont {G.~B.}\ \bibnamefont {Benedek}},\
  }\href@noop {} {\bibfield  {journal} {\bibinfo  {journal} {Proc. Natl. Acad.
  Sci. U. S. A.}\ }\textbf {\bibinfo {volume} {95}},\ \bibinfo {pages} {12277}
  (\bibinfo {year} {1998})}\BibitemShut {NoStop}%
\bibitem [{\citenamefont {Nielsen}\ \emph {et~al.}(2001)\citenamefont
  {Nielsen}, \citenamefont {Khurana}, \citenamefont {Coats}, \citenamefont
  {Frokjaer}, \citenamefont {Brange}, \citenamefont {Vyas}, \citenamefont
  {Uversky},\ and\ \citenamefont {Fink}}]{Nielsen2001}%
  \BibitemOpen
  \bibfield  {author} {\bibinfo {author} {\bibfnamefont {L.}~\bibnamefont
  {Nielsen}}, \bibinfo {author} {\bibfnamefont {R.}~\bibnamefont {Khurana}},
  \bibinfo {author} {\bibfnamefont {A.}~\bibnamefont {Coats}}, \bibinfo
  {author} {\bibfnamefont {S.}~\bibnamefont {Frokjaer}}, \bibinfo {author}
  {\bibfnamefont {J.}~\bibnamefont {Brange}}, \bibinfo {author} {\bibfnamefont
  {S.}~\bibnamefont {Vyas}}, \bibinfo {author} {\bibfnamefont {V.~N.}\
  \bibnamefont {Uversky}}, \ and\ \bibinfo {author} {\bibfnamefont {A.~L.}\
  \bibnamefont {Fink}},\ }\href {\doibase 10.1021/bi002555c} {\bibfield
  {journal} {\bibinfo  {journal} {Biochemistry}\ }\textbf {\bibinfo {volume}
  {40}},\ \bibinfo {pages} {6036} (\bibinfo {year} {2001})}\BibitemShut
  {NoStop}%
\bibitem [{\citenamefont {Evans}\ \emph {et~al.}(1995)\citenamefont {Evans},
  \citenamefont {Berger}, \citenamefont {Cho}, \citenamefont {Weisgraber},\
  and\ \citenamefont {Lansbury}}]{Evans1995}%
  \BibitemOpen
  \bibfield  {author} {\bibinfo {author} {\bibfnamefont {K.~C.}\ \bibnamefont
  {Evans}}, \bibinfo {author} {\bibfnamefont {E.~P.}\ \bibnamefont {Berger}},
  \bibinfo {author} {\bibfnamefont {C.~G.}\ \bibnamefont {Cho}}, \bibinfo
  {author} {\bibfnamefont {K.~H.}\ \bibnamefont {Weisgraber}}, \ and\ \bibinfo
  {author} {\bibfnamefont {P.~T.}\ \bibnamefont {Lansbury}},\ }\href {\doibase
  10.1073/pnas.92.3.763} {\bibfield  {journal} {\bibinfo  {journal} {Proc.
  Natl. Acad. Sci. U. S. A.}\ }\textbf {\bibinfo {volume} {92}},\ \bibinfo
  {pages} {763} (\bibinfo {year} {1995})}\BibitemShut {NoStop}%
\bibitem [{\citenamefont {Ionescu-Zanetti}\ \emph {et~al.}(1999)\citenamefont
  {Ionescu-Zanetti}, \citenamefont {Khurana}, \citenamefont {Gillespie},
  \citenamefont {Petrick}, \citenamefont {Trabachino}, \citenamefont {Minert},
  \citenamefont {Carter},\ and\ \citenamefont {Fink}}]{Ionescu-Zanetti1999}%
  \BibitemOpen
  \bibfield  {author} {\bibinfo {author} {\bibfnamefont {C.}~\bibnamefont
  {Ionescu-Zanetti}}, \bibinfo {author} {\bibfnamefont {R.}~\bibnamefont
  {Khurana}}, \bibinfo {author} {\bibfnamefont {J.~R.}\ \bibnamefont
  {Gillespie}}, \bibinfo {author} {\bibfnamefont {J.~S.}\ \bibnamefont
  {Petrick}}, \bibinfo {author} {\bibfnamefont {L.~C.}\ \bibnamefont
  {Trabachino}}, \bibinfo {author} {\bibfnamefont {L.~J.}\ \bibnamefont
  {Minert}}, \bibinfo {author} {\bibfnamefont {S.~A.}\ \bibnamefont {Carter}},
  \ and\ \bibinfo {author} {\bibfnamefont {A.~L.}\ \bibnamefont {Fink}},\
  }\href {\doibase 10.1073/pnas.96.23.13175} {\bibfield  {journal} {\bibinfo
  {journal} {Proc. Natl. Acad. Sci. U. S. A.}\ }\textbf {\bibinfo {volume}
  {96}},\ \bibinfo {pages} {13175} (\bibinfo {year} {1999})}\BibitemShut
  {NoStop}%
\bibitem [{\citenamefont {Knowles}\ \emph {et~al.}(2009)\citenamefont
  {Knowles}, \citenamefont {Waudby}, \citenamefont {Devlin}, \citenamefont
  {Cohen}, \citenamefont {Aguzzi}, \citenamefont {Vendruscolo}, \citenamefont
  {Terentjev}, \citenamefont {Welland},\ and\ \citenamefont
  {Dobson}}]{Knowles2009a}%
  \BibitemOpen
  \bibfield  {author} {\bibinfo {author} {\bibfnamefont {T.~P.~J.}\
  \bibnamefont {Knowles}}, \bibinfo {author} {\bibfnamefont {C.~A.}\
  \bibnamefont {Waudby}}, \bibinfo {author} {\bibfnamefont {G.~L.}\
  \bibnamefont {Devlin}}, \bibinfo {author} {\bibfnamefont {S.~I.~A.}\
  \bibnamefont {Cohen}}, \bibinfo {author} {\bibfnamefont {A.}~\bibnamefont
  {Aguzzi}}, \bibinfo {author} {\bibfnamefont {M.}~\bibnamefont {Vendruscolo}},
  \bibinfo {author} {\bibfnamefont {E.~M.}\ \bibnamefont {Terentjev}}, \bibinfo
  {author} {\bibfnamefont {M.~E.}\ \bibnamefont {Welland}}, \ and\ \bibinfo
  {author} {\bibfnamefont {C.~M.}\ \bibnamefont {Dobson}},\ }\href@noop {}
  {\bibfield  {journal} {\bibinfo  {journal} {Science}\ }\textbf {\bibinfo
  {volume} {326}},\ \bibinfo {pages} {1533} (\bibinfo {year}
  {2009})}\BibitemShut {NoStop}%
\bibitem [{\citenamefont {Arosio}\ \emph {et~al.}(2015)\citenamefont {Arosio},
  \citenamefont {Knowles},\ and\ \citenamefont {Linse}}]{Arosio2015}%
  \BibitemOpen
  \bibfield  {author} {\bibinfo {author} {\bibfnamefont {P.}~\bibnamefont
  {Arosio}}, \bibinfo {author} {\bibfnamefont {T.~P.~J.}\ \bibnamefont
  {Knowles}}, \ and\ \bibinfo {author} {\bibfnamefont {S.}~\bibnamefont
  {Linse}},\ }\href {\doibase 10.1039/C4CP05563B} {\bibfield  {journal}
  {\bibinfo  {journal} {Phys. Chem. Chem. Phys.}\ }\textbf {\bibinfo {volume}
  {17}},\ \bibinfo {pages} {7606} (\bibinfo {year} {2015})}\BibitemShut
  {NoStop}%
\bibitem [{\citenamefont {Oosawa}\ and\ \citenamefont
  {Kasai}(1962)}]{Oosawa1962}%
  \BibitemOpen
  \bibfield  {author} {\bibinfo {author} {\bibfnamefont {F.}~\bibnamefont
  {Oosawa}}\ and\ \bibinfo {author} {\bibfnamefont {M.}~\bibnamefont {Kasai}},\
  }\href {\doibase 10.1016/S0022-2836(62)80112-0} {\bibfield  {journal}
  {\bibinfo  {journal} {J. Mol. Biol.}\ }\textbf {\bibinfo {volume} {4}},\
  \bibinfo {pages} {10} (\bibinfo {year} {1962})}\BibitemShut {NoStop}%
\bibitem [{\citenamefont {Cohen}\ \emph {et~al.}(2011)\citenamefont {Cohen},
  \citenamefont {Vendruscolo}, \citenamefont {Welland}, \citenamefont {Dobson},
  \citenamefont {Terentjev},\ and\ \citenamefont {Knowles}}]{Cohen2011}%
  \BibitemOpen
  \bibfield  {author} {\bibinfo {author} {\bibfnamefont {S.~I.~A.}\
  \bibnamefont {Cohen}}, \bibinfo {author} {\bibfnamefont {M.}~\bibnamefont
  {Vendruscolo}}, \bibinfo {author} {\bibfnamefont {M.~E.}\ \bibnamefont
  {Welland}}, \bibinfo {author} {\bibfnamefont {C.~M.}\ \bibnamefont {Dobson}},
  \bibinfo {author} {\bibfnamefont {E.~M.}\ \bibnamefont {Terentjev}}, \ and\
  \bibinfo {author} {\bibfnamefont {T.~P.~J.}\ \bibnamefont {Knowles}},\
  }\href@noop {} {\bibfield  {journal} {\bibinfo  {journal} {J. Chem. Phys.}\
  }\textbf {\bibinfo {volume} {135}},\ \bibinfo {pages} {065105} (\bibinfo
  {year} {2011})}\BibitemShut {NoStop}%
\bibitem [{\citenamefont {Michaels}\ and\ \citenamefont
  {Knowles}(2014)}]{Michaels2014}%
  \BibitemOpen
  \bibfield  {author} {\bibinfo {author} {\bibfnamefont {T.~C.~T.}\
  \bibnamefont {Michaels}}\ and\ \bibinfo {author} {\bibfnamefont {T.~P.~J.}\
  \bibnamefont {Knowles}},\ }\href@noop {} {\bibfield  {journal} {\bibinfo
  {journal} {J. Chem. Phys.}\ }\textbf {\bibinfo {volume} {140}} (\bibinfo
  {year} {2014})}\BibitemShut {NoStop}%
\bibitem [{\citenamefont {Cohen}\ \emph {et~al.}(2012)\citenamefont {Cohen},
  \citenamefont {Vendruscolo}, \citenamefont {Dobson},\ and\ \citenamefont
  {Knowles}}]{Cohen2012}%
  \BibitemOpen
  \bibfield  {author} {\bibinfo {author} {\bibfnamefont {S.~I.~A.}\
  \bibnamefont {Cohen}}, \bibinfo {author} {\bibfnamefont {M.}~\bibnamefont
  {Vendruscolo}}, \bibinfo {author} {\bibfnamefont {C.~M.}\ \bibnamefont
  {Dobson}}, \ and\ \bibinfo {author} {\bibfnamefont {T.~P.~J.}\ \bibnamefont
  {Knowles}},\ }\href {\doibase 10.1016/j.jmb.2012.02.031} {\bibfield
  {journal} {\bibinfo  {journal} {J. Mol. Biol.}\ }\textbf {\bibinfo {volume}
  {421}},\ \bibinfo {pages} {160} (\bibinfo {year} {2012})}\BibitemShut
  {NoStop}%
\bibitem [{\citenamefont {Eakin}\ \emph {et~al.}(2006)\citenamefont {Eakin},
  \citenamefont {Berman},\ and\ \citenamefont {Miranker}}]{Eakin2006}%
  \BibitemOpen
  \bibfield  {author} {\bibinfo {author} {\bibfnamefont {C.~M.}\ \bibnamefont
  {Eakin}}, \bibinfo {author} {\bibfnamefont {A.~J.}\ \bibnamefont {Berman}}, \
  and\ \bibinfo {author} {\bibfnamefont {A.~D.}\ \bibnamefont {Miranker}},\
  }\href {\doibase 10.1038/nsmb1068} {\bibfield  {journal} {\bibinfo  {journal}
  {Nat. Struct. Mol. Biol.}\ }\textbf {\bibinfo {volume} {13}},\ \bibinfo
  {pages} {202} (\bibinfo {year} {2006})}\BibitemShut {NoStop}%
\bibitem [{\citenamefont {Chiti}\ \emph {et~al.}(1999)\citenamefont {Chiti},
  \citenamefont {Webster}, \citenamefont {Taddei}, \citenamefont {Clark},
  \citenamefont {Stefani}, \citenamefont {Ramponi},\ and\ \citenamefont
  {Dobson}}]{Chiti1999}%
  \BibitemOpen
  \bibfield  {author} {\bibinfo {author} {\bibfnamefont {F.}~\bibnamefont
  {Chiti}}, \bibinfo {author} {\bibfnamefont {P.}~\bibnamefont {Webster}},
  \bibinfo {author} {\bibfnamefont {N.}~\bibnamefont {Taddei}}, \bibinfo
  {author} {\bibfnamefont {A.}~\bibnamefont {Clark}}, \bibinfo {author}
  {\bibfnamefont {M.}~\bibnamefont {Stefani}}, \bibinfo {author} {\bibfnamefont
  {G.}~\bibnamefont {Ramponi}}, \ and\ \bibinfo {author} {\bibfnamefont
  {C.~M.}\ \bibnamefont {Dobson}},\ }\href {\doibase 10.1073/pnas.96.7.3590}
  {\bibfield  {journal} {\bibinfo  {journal} {Proc. Natl. Acad. Sci. U. S. A.}\
  }\textbf {\bibinfo {volume} {96}},\ \bibinfo {pages} {3590} (\bibinfo {year}
  {1999})}\BibitemShut {NoStop}%
\bibitem [{\citenamefont {Chiti}\ and\ \citenamefont
  {Dobson}(2009)}]{Chiti2009}%
  \BibitemOpen
  \bibfield  {author} {\bibinfo {author} {\bibfnamefont {F.}~\bibnamefont
  {Chiti}}\ and\ \bibinfo {author} {\bibfnamefont {C.~M.}\ \bibnamefont
  {Dobson}},\ }\href {\doibase 10.1038/nchembio.131} {\bibfield  {journal}
  {\bibinfo  {journal} {Nat. Chem. Biol.}\ }\textbf {\bibinfo {volume} {5}},\
  \bibinfo {pages} {15} (\bibinfo {year} {2009})}\BibitemShut {NoStop}%
\bibitem [{\citenamefont {Baumketner}\ \emph {et~al.}(2006)\citenamefont
  {Baumketner}, \citenamefont {Bernstein}, \citenamefont {Wyttenbach},
  \citenamefont {Bitan}, \citenamefont {Teplow}, \citenamefont {Bowers},\ and\
  \citenamefont {Shea}}]{Baumketner2006}%
  \BibitemOpen
  \bibfield  {author} {\bibinfo {author} {\bibfnamefont {A.}~\bibnamefont
  {Baumketner}}, \bibinfo {author} {\bibfnamefont {S.~L.}\ \bibnamefont
  {Bernstein}}, \bibinfo {author} {\bibfnamefont {T.}~\bibnamefont
  {Wyttenbach}}, \bibinfo {author} {\bibfnamefont {G.}~\bibnamefont {Bitan}},
  \bibinfo {author} {\bibfnamefont {D.~B.}\ \bibnamefont {Teplow}}, \bibinfo
  {author} {\bibfnamefont {M.~T.}\ \bibnamefont {Bowers}}, \ and\ \bibinfo
  {author} {\bibfnamefont {J.}~\bibnamefont {Shea}},\ }\href {\doibase
  10.1110/ps.051762406} {\bibfield  {journal} {\bibinfo  {journal} {Protein
  Sci.}\ }\textbf {\bibinfo {volume} {15}},\ \bibinfo {pages} {420} (\bibinfo
  {year} {2006})}\BibitemShut {NoStop}%
\bibitem [{\citenamefont {Bieler}\ \emph {et~al.}(2012)\citenamefont {Bieler},
  \citenamefont {Knowles}, \citenamefont {Frenkel},\ and\ \citenamefont
  {V{\'{a}}cha}}]{Bieler2012}%
  \BibitemOpen
  \bibfield  {author} {\bibinfo {author} {\bibfnamefont {N.~S.}\ \bibnamefont
  {Bieler}}, \bibinfo {author} {\bibfnamefont {T.~P.~J.}\ \bibnamefont
  {Knowles}}, \bibinfo {author} {\bibfnamefont {D.}~\bibnamefont {Frenkel}}, \
  and\ \bibinfo {author} {\bibfnamefont {R.}~\bibnamefont {V{\'{a}}cha}},\
  }\href@noop {} {\bibfield  {journal} {\bibinfo  {journal} {PLoS Comput.
  Biol.}\ }\textbf {\bibinfo {volume} {8}},\ \bibinfo {pages} {e1002692}
  (\bibinfo {year} {2012})}\BibitemShut {NoStop}%
\bibitem [{\citenamefont {Sari{\'{c}}}\ \emph {et~al.}(2014)\citenamefont
  {Sari{\'{c}}}, \citenamefont {Chebaro}, \citenamefont {Knowles},\ and\
  \citenamefont {Frenkel}}]{Saric2014}%
  \BibitemOpen
  \bibfield  {author} {\bibinfo {author} {\bibfnamefont {A.}~\bibnamefont
  {Sari{\'{c}}}}, \bibinfo {author} {\bibfnamefont {Y.~C.}\ \bibnamefont
  {Chebaro}}, \bibinfo {author} {\bibfnamefont {T.~P.~J.}\ \bibnamefont
  {Knowles}}, \ and\ \bibinfo {author} {\bibfnamefont {D.}~\bibnamefont
  {Frenkel}},\ }\href {\doibase 10.1073/pnas.1410159111} {\bibfield  {journal}
  {\bibinfo  {journal} {Proc. Natl. Acad. Sci. U. S. A.}\ }\textbf {\bibinfo
  {volume} {111}},\ \bibinfo {pages} {17869} (\bibinfo {year}
  {2014})}\BibitemShut {NoStop}%
\bibitem [{\citenamefont {Ferrone}\ \emph {et~al.}(1985)\citenamefont
  {Ferrone}, \citenamefont {Hofrichter},\ and\ \citenamefont
  {Eaton}}]{Ferrone1985}%
  \BibitemOpen
  \bibfield  {author} {\bibinfo {author} {\bibfnamefont {F.~A.}\ \bibnamefont
  {Ferrone}}, \bibinfo {author} {\bibfnamefont {J.}~\bibnamefont {Hofrichter}},
  \ and\ \bibinfo {author} {\bibfnamefont {W.~A.}\ \bibnamefont {Eaton}},\
  }\href {\doibase 10.1016/0022-2836(85)90175-5} {\bibfield  {journal}
  {\bibinfo  {journal} {J. Mol. Biol.}\ }\textbf {\bibinfo {volume} {183}},\
  \bibinfo {pages} {611} (\bibinfo {year} {1985})}\BibitemShut {NoStop}%
\bibitem [{\citenamefont {Pease}\ \emph {et~al.}(2010)\citenamefont {Pease},
  \citenamefont {Sorci}, \citenamefont {Guha}, \citenamefont {Tsai},
  \citenamefont {Zachariah}, \citenamefont {Tarlov},\ and\ \citenamefont
  {Belfort}}]{Iii2010}%
  \BibitemOpen
  \bibfield  {author} {\bibinfo {author} {\bibfnamefont {L.~F.}\ \bibnamefont
  {Pease}}, \bibinfo {author} {\bibfnamefont {M.}~\bibnamefont {Sorci}},
  \bibinfo {author} {\bibfnamefont {S.}~\bibnamefont {Guha}}, \bibinfo {author}
  {\bibfnamefont {D.-H.}\ \bibnamefont {Tsai}}, \bibinfo {author}
  {\bibfnamefont {M.~R.}\ \bibnamefont {Zachariah}}, \bibinfo {author}
  {\bibfnamefont {M.~J.}\ \bibnamefont {Tarlov}}, \ and\ \bibinfo {author}
  {\bibfnamefont {G.}~\bibnamefont {Belfort}},\ }\href {\doibase
  10.1016/j.bpj.2010.10.010} {\bibfield  {journal} {\bibinfo  {journal}
  {Biophysj}\ }\textbf {\bibinfo {volume} {99}},\ \bibinfo {pages} {3979}
  (\bibinfo {year} {2010})}\BibitemShut {NoStop}%
\bibitem [{\citenamefont {Cremades}\ \emph {et~al.}(2012)\citenamefont
  {Cremades}, \citenamefont {Cohen}, \citenamefont {Deas}, \citenamefont
  {Abramov}, \citenamefont {Chen}, \citenamefont {Orte}, \citenamefont
  {Sandal}, \citenamefont {Clarke}, \citenamefont {Dunne}, \citenamefont
  {Aprile}, \citenamefont {Bertoncini}, \citenamefont {Wood}, \citenamefont
  {Knowles}, \citenamefont {Dobson},\ and\ \citenamefont
  {Klenerman}}]{Cremades2012}%
  \BibitemOpen
  \bibfield  {author} {\bibinfo {author} {\bibfnamefont {N.}~\bibnamefont
  {Cremades}}, \bibinfo {author} {\bibfnamefont {S.~I.~A.}\ \bibnamefont
  {Cohen}}, \bibinfo {author} {\bibfnamefont {E.}~\bibnamefont {Deas}},
  \bibinfo {author} {\bibfnamefont {A.~Y.}\ \bibnamefont {Abramov}}, \bibinfo
  {author} {\bibfnamefont {A.~Y.}\ \bibnamefont {Chen}}, \bibinfo {author}
  {\bibfnamefont {A.}~\bibnamefont {Orte}}, \bibinfo {author} {\bibfnamefont
  {M.}~\bibnamefont {Sandal}}, \bibinfo {author} {\bibfnamefont {R.~W.}\
  \bibnamefont {Clarke}}, \bibinfo {author} {\bibfnamefont {P.}~\bibnamefont
  {Dunne}}, \bibinfo {author} {\bibfnamefont {F.~A.}\ \bibnamefont {Aprile}},
  \bibinfo {author} {\bibfnamefont {C.~W.}\ \bibnamefont {Bertoncini}},
  \bibinfo {author} {\bibfnamefont {N.~W.}\ \bibnamefont {Wood}}, \bibinfo
  {author} {\bibfnamefont {T.~P.~J.}\ \bibnamefont {Knowles}}, \bibinfo
  {author} {\bibfnamefont {C.~M.}\ \bibnamefont {Dobson}}, \ and\ \bibinfo
  {author} {\bibfnamefont {D.}~\bibnamefont {Klenerman}},\ }\href {\doibase
  10.1016/j.cell.2012.03.037} {\bibfield  {journal} {\bibinfo  {journal}
  {Cell}\ }\textbf {\bibinfo {volume} {149}},\ \bibinfo {pages} {1048}
  (\bibinfo {year} {2012})}\BibitemShut {NoStop}%
\bibitem [{\citenamefont {Lee}\ \emph {et~al.}(2011)\citenamefont {Lee},
  \citenamefont {Culyba}, \citenamefont {Powers},\ and\ \citenamefont
  {Kelly}}]{Lee2011}%
  \BibitemOpen
  \bibfield  {author} {\bibinfo {author} {\bibfnamefont {J.}~\bibnamefont
  {Lee}}, \bibinfo {author} {\bibfnamefont {E.~K.}\ \bibnamefont {Culyba}},
  \bibinfo {author} {\bibfnamefont {E.~T.}\ \bibnamefont {Powers}}, \ and\
  \bibinfo {author} {\bibfnamefont {J.~W.}\ \bibnamefont {Kelly}},\ }\href@noop
  {} {\bibfield  {journal} {\bibinfo  {journal} {Nat. Chem. Biol.}\ }\textbf
  {\bibinfo {volume} {7}},\ \bibinfo {pages} {602} (\bibinfo {year}
  {2011})}\BibitemShut {NoStop}%
\bibitem [{\citenamefont {Vitalis}\ and\ \citenamefont
  {Pappu}(2011)}]{Vitalis2011}%
  \BibitemOpen
  \bibfield  {author} {\bibinfo {author} {\bibfnamefont {A.}~\bibnamefont
  {Vitalis}}\ and\ \bibinfo {author} {\bibfnamefont {R.~V.}\ \bibnamefont
  {Pappu}},\ }\href {\doibase 10.1016/j.bpc.2011.04.006} {\bibfield  {journal}
  {\bibinfo  {journal} {Biophys. Chem.}\ }\textbf {\bibinfo {volume} {159}},\
  \bibinfo {pages} {14} (\bibinfo {year} {2011})}\BibitemShut {NoStop}%
\bibitem [{\citenamefont {Lomakin}\ \emph {et~al.}(1997)\citenamefont
  {Lomakin}, \citenamefont {Teplow}, \citenamefont {Kirschner},\ and\
  \citenamefont {Benedek}}]{Lomakin1997}%
  \BibitemOpen
  \bibfield  {author} {\bibinfo {author} {\bibfnamefont {A.}~\bibnamefont
  {Lomakin}}, \bibinfo {author} {\bibfnamefont {D.~B.}\ \bibnamefont {Teplow}},
  \bibinfo {author} {\bibfnamefont {D.~A.}\ \bibnamefont {Kirschner}}, \ and\
  \bibinfo {author} {\bibfnamefont {G.~B.}\ \bibnamefont {Benedek}},\ }\href
  {\doibase 10.1073/pnas.94.15.7942} {\bibfield  {journal} {\bibinfo  {journal}
  {Proc. Natl. Acad. Sci. U. S. A.}\ }\textbf {\bibinfo {volume} {94}},\
  \bibinfo {pages} {7942} (\bibinfo {year} {1997})}\BibitemShut {NoStop}%
\bibitem [{\citenamefont {Garcia}\ \emph {et~al.}(2014)\citenamefont {Garcia},
  \citenamefont {Cohen}, \citenamefont {Dobson},\ and\ \citenamefont
  {Knowles}}]{Garcia2014}%
  \BibitemOpen
  \bibfield  {author} {\bibinfo {author} {\bibfnamefont {G.~A.}\ \bibnamefont
  {Garcia}}, \bibinfo {author} {\bibfnamefont {S.~I.~A.}\ \bibnamefont
  {Cohen}}, \bibinfo {author} {\bibfnamefont {C.~M.}\ \bibnamefont {Dobson}}, \
  and\ \bibinfo {author} {\bibfnamefont {T.~P.~J.}\ \bibnamefont {Knowles}},\
  }\href@noop {} {\bibfield  {journal} {\bibinfo  {journal} {Phys. Rev. E}\
  }\textbf {\bibinfo {volume} {89}},\ \bibinfo {pages} {1} (\bibinfo {year}
  {2014})}\BibitemShut {NoStop}%
\bibitem [{\citenamefont {Auer}\ \emph {et~al.}(2012)\citenamefont {Auer},
  \citenamefont {Ricchiuto},\ and\ \citenamefont {Kashchiev}}]{Auer2012}%
  \BibitemOpen
  \bibfield  {author} {\bibinfo {author} {\bibfnamefont {S.}~\bibnamefont
  {Auer}}, \bibinfo {author} {\bibfnamefont {P.}~\bibnamefont {Ricchiuto}}, \
  and\ \bibinfo {author} {\bibfnamefont {D.}~\bibnamefont {Kashchiev}},\ }\href
  {\doibase 10.1016/j.jmb.2012.06.022} {\bibfield  {journal} {\bibinfo
  {journal} {J. Mol. Biol.}\ }\textbf {\bibinfo {volume} {422}},\ \bibinfo
  {pages} {723} (\bibinfo {year} {2012})}\BibitemShut {NoStop}%
\bibitem [{\citenamefont {Ferrone}(2006)}]{Ferrone2006}%
  \BibitemOpen
  \bibfield  {author} {\bibinfo {author} {\bibfnamefont {F.~A.}\ \bibnamefont
  {Ferrone}},\ }\href {\doibase 10.1016/S0076-6879(06)12017-0} {\bibfield
  {journal} {\bibinfo  {journal} {Methods Enzymol.}\ }\textbf {\bibinfo
  {volume} {412}},\ \bibinfo {pages} {285} (\bibinfo {year}
  {2006})}\BibitemShut {NoStop}%
\bibitem [{\citenamefont {Kashchiev}\ and\ \citenamefont
  {Auer}(2010)}]{Kashchiev2010}%
  \BibitemOpen
  \bibfield  {author} {\bibinfo {author} {\bibfnamefont {D.}~\bibnamefont
  {Kashchiev}}\ and\ \bibinfo {author} {\bibfnamefont {S.}~\bibnamefont
  {Auer}},\ }\href {\doibase 10.1063/1.3447891} {\bibfield  {journal} {\bibinfo
   {journal} {J. Chem. Phys.}\ }\textbf {\bibinfo {volume} {132}},\ \bibinfo
  {pages} {215191} (\bibinfo {year} {2010})}\BibitemShut {NoStop}%
\bibitem [{\citenamefont {Jim{\'{e}}nez}\ \emph {et~al.}(2002)\citenamefont
  {Jim{\'{e}}nez}, \citenamefont {Nettleton}, \citenamefont {Bouchard},
  \citenamefont {Robinson}, \citenamefont {Dobson},\ and\ \citenamefont
  {Saibil}}]{Jimenez2002}%
  \BibitemOpen
  \bibfield  {author} {\bibinfo {author} {\bibfnamefont {J.~L.}\ \bibnamefont
  {Jim{\'{e}}nez}}, \bibinfo {author} {\bibfnamefont {E.~J.}\ \bibnamefont
  {Nettleton}}, \bibinfo {author} {\bibfnamefont {M.}~\bibnamefont {Bouchard}},
  \bibinfo {author} {\bibfnamefont {C.~V.}\ \bibnamefont {Robinson}}, \bibinfo
  {author} {\bibfnamefont {C.~M.}\ \bibnamefont {Dobson}}, \ and\ \bibinfo
  {author} {\bibfnamefont {H.~R.}\ \bibnamefont {Saibil}},\ }\href {\doibase
  10.1073/pnas.142459399} {\bibfield  {journal} {\bibinfo  {journal} {Proc.
  Natl. Acad. Sci. U. S. A.}\ }\textbf {\bibinfo {volume} {99}},\ \bibinfo
  {pages} {9196} (\bibinfo {year} {2002})}\BibitemShut {NoStop}%
\bibitem [{\citenamefont {Sachse}\ \emph {et~al.}(2006)\citenamefont {Sachse},
  \citenamefont {Xu}, \citenamefont {Wieligmann}, \citenamefont {Diekmann},
  \citenamefont {Grigorieff},\ and\ \citenamefont
  {F{\"{a}}ndrich}}]{Sachse2006}%
  \BibitemOpen
  \bibfield  {author} {\bibinfo {author} {\bibfnamefont {C.}~\bibnamefont
  {Sachse}}, \bibinfo {author} {\bibfnamefont {C.}~\bibnamefont {Xu}}, \bibinfo
  {author} {\bibfnamefont {K.}~\bibnamefont {Wieligmann}}, \bibinfo {author}
  {\bibfnamefont {S.}~\bibnamefont {Diekmann}}, \bibinfo {author}
  {\bibfnamefont {N.}~\bibnamefont {Grigorieff}}, \ and\ \bibinfo {author}
  {\bibfnamefont {M.}~\bibnamefont {F{\"{a}}ndrich}},\ }\href {\doibase
  10.1016/j.jmb.2006.07.011} {\bibfield  {journal} {\bibinfo  {journal} {J.
  Mol. Biol.}\ }\textbf {\bibinfo {volume} {362}},\ \bibinfo {pages} {347}
  (\bibinfo {year} {2006})}\BibitemShut {NoStop}%
\bibitem [{\citenamefont {Jim{\'{e}}nez}\ \emph {et~al.}(2001)\citenamefont
  {Jim{\'{e}}nez}, \citenamefont {Tennent}, \citenamefont {Pepys},\ and\
  \citenamefont {Saibil}}]{Jimenez2001}%
  \BibitemOpen
  \bibfield  {author} {\bibinfo {author} {\bibfnamefont {J.~L.}\ \bibnamefont
  {Jim{\'{e}}nez}}, \bibinfo {author} {\bibfnamefont {G.}~\bibnamefont
  {Tennent}}, \bibinfo {author} {\bibfnamefont {M.}~\bibnamefont {Pepys}}, \
  and\ \bibinfo {author} {\bibfnamefont {H.~R.}\ \bibnamefont {Saibil}},\
  }\href {\doibase 10.1006/jmbi.2001.4863} {\bibfield  {journal} {\bibinfo
  {journal} {J. Mol. Biol.}\ }\textbf {\bibinfo {volume} {311}},\ \bibinfo
  {pages} {241} (\bibinfo {year} {2001})}\BibitemShut {NoStop}%
\bibitem [{\citenamefont {Arimon}\ \emph {et~al.}(2012)\citenamefont {Arimon},
  \citenamefont {Sanz}, \citenamefont {Giralt},\ and\ \citenamefont
  {Carulla}}]{Arimon2012}%
  \BibitemOpen
  \bibfield  {author} {\bibinfo {author} {\bibfnamefont {M.}~\bibnamefont
  {Arimon}}, \bibinfo {author} {\bibfnamefont {F.}~\bibnamefont {Sanz}},
  \bibinfo {author} {\bibfnamefont {E.}~\bibnamefont {Giralt}}, \ and\ \bibinfo
  {author} {\bibfnamefont {N.}~\bibnamefont {Carulla}},\ }\href {\doibase
  10.1021/bc200077s} {\bibfield  {journal} {\bibinfo  {journal} {Bioconjug.
  Chem.}\ }\textbf {\bibinfo {volume} {23}},\ \bibinfo {pages} {27} (\bibinfo
  {year} {2012})}\BibitemShut {NoStop}%
\bibitem [{\citenamefont {Goldsbury}\ \emph {et~al.}(2005)\citenamefont
  {Goldsbury}, \citenamefont {Frey}, \citenamefont {Olivieri}, \citenamefont
  {Aebi},\ and\ \citenamefont {M{\"u}ller}}]{Goldsbury2005}%
  \BibitemOpen
  \bibfield  {author} {\bibinfo {author} {\bibfnamefont {C.}~\bibnamefont
  {Goldsbury}}, \bibinfo {author} {\bibfnamefont {P.}~\bibnamefont {Frey}},
  \bibinfo {author} {\bibfnamefont {V.}~\bibnamefont {Olivieri}}, \bibinfo
  {author} {\bibfnamefont {U.}~\bibnamefont {Aebi}}, \ and\ \bibinfo {author}
  {\bibfnamefont {S.~A.}\ \bibnamefont {M{\"u}ller}},\ }\href {\doibase
  10.1016/j.jmb.2005.07.029} {\bibfield  {journal} {\bibinfo  {journal} {J.
  Mol. Biol.}\ }\textbf {\bibinfo {volume} {352}},\ \bibinfo {pages} {282}
  (\bibinfo {year} {2005})}\BibitemShut {NoStop}%
\bibitem [{\citenamefont {Paravastu}\ \emph {et~al.}(2008)\citenamefont
  {Paravastu}, \citenamefont {Leapman}, \citenamefont {Yau},\ and\
  \citenamefont {Tycko}}]{Paravastu2008}%
  \BibitemOpen
  \bibfield  {author} {\bibinfo {author} {\bibfnamefont {A.~K.}\ \bibnamefont
  {Paravastu}}, \bibinfo {author} {\bibfnamefont {R.~D.}\ \bibnamefont
  {Leapman}}, \bibinfo {author} {\bibfnamefont {W.-M.}\ \bibnamefont {Yau}}, \
  and\ \bibinfo {author} {\bibfnamefont {R.}~\bibnamefont {Tycko}},\ }\href
  {\doibase 10.1073/pnas.0806270105} {\bibfield  {journal} {\bibinfo  {journal}
  {Proc. Natl. Acad. Sci.}\ }\textbf {\bibinfo {volume} {105}},\ \bibinfo
  {pages} {18349} (\bibinfo {year} {2008})}\BibitemShut {NoStop}%
\bibitem [{\citenamefont {Yu}\ \emph {et~al.}(2010)\citenamefont {Yu},
  \citenamefont {Wang}, \citenamefont {Yang}, \citenamefont {Wang},
  \citenamefont {Cheng}, \citenamefont {Nussinov},\ and\ \citenamefont
  {Zheng}}]{Yu2010}%
  \BibitemOpen
  \bibfield  {author} {\bibinfo {author} {\bibfnamefont {X.}~\bibnamefont
  {Yu}}, \bibinfo {author} {\bibfnamefont {J.}~\bibnamefont {Wang}}, \bibinfo
  {author} {\bibfnamefont {J.~C.}\ \bibnamefont {Yang}}, \bibinfo {author}
  {\bibfnamefont {Q.}~\bibnamefont {Wang}}, \bibinfo {author} {\bibfnamefont
  {S.~Z.~D.}\ \bibnamefont {Cheng}}, \bibinfo {author} {\bibfnamefont
  {R.}~\bibnamefont {Nussinov}}, \ and\ \bibinfo {author} {\bibfnamefont
  {J.}~\bibnamefont {Zheng}},\ }\href {\doibase 10.1016/j.bpj.2009.10.003}
  {\bibfield  {journal} {\bibinfo  {journal} {Biophys. J.}\ }\textbf {\bibinfo
  {volume} {98}},\ \bibinfo {pages} {27} (\bibinfo {year} {2010})}\BibitemShut
  {NoStop}%
\bibitem [{\citenamefont {Knowles}\ \emph {et~al.}(2012)\citenamefont
  {Knowles}, \citenamefont {{De Simone}}, \citenamefont {Fitzpatrick},
  \citenamefont {Baldwin}, \citenamefont {Meehan}, \citenamefont {Rajah},
  \citenamefont {Vendruscolo}, \citenamefont {Welland}, \citenamefont
  {Dobson},\ and\ \citenamefont {Terentjev}}]{Knowles2012}%
  \BibitemOpen
  \bibfield  {author} {\bibinfo {author} {\bibfnamefont {T.~P.~J.}\
  \bibnamefont {Knowles}}, \bibinfo {author} {\bibfnamefont {A.}~\bibnamefont
  {{De Simone}}}, \bibinfo {author} {\bibfnamefont {A.~W.}\ \bibnamefont
  {Fitzpatrick}}, \bibinfo {author} {\bibfnamefont {A.}~\bibnamefont
  {Baldwin}}, \bibinfo {author} {\bibfnamefont {S.}~\bibnamefont {Meehan}},
  \bibinfo {author} {\bibfnamefont {L.}~\bibnamefont {Rajah}}, \bibinfo
  {author} {\bibfnamefont {M.}~\bibnamefont {Vendruscolo}}, \bibinfo {author}
  {\bibfnamefont {M.~E.}\ \bibnamefont {Welland}}, \bibinfo {author}
  {\bibfnamefont {C.~M.}\ \bibnamefont {Dobson}}, \ and\ \bibinfo {author}
  {\bibfnamefont {E.~M.}\ \bibnamefont {Terentjev}},\ }\href {\doibase
  10.1103/PhysRevLett.109.158101} {\bibfield  {journal} {\bibinfo  {journal}
  {Phys. Rev. Lett.}\ }\textbf {\bibinfo {volume} {109}},\ \bibinfo {pages} {1}
  (\bibinfo {year} {2012})}\BibitemShut {NoStop}%
\bibitem [{\citenamefont {L{\"{u}}hrs}\ \emph {et~al.}(2005)\citenamefont
  {L{\"{u}}hrs}, \citenamefont {Ritter}, \citenamefont {Adrian}, \citenamefont
  {Riek-Loher}, \citenamefont {Bohrmann}, \citenamefont {D{\"{o}}beli},
  \citenamefont {Schubert},\ and\ \citenamefont {Riek}}]{Luhrs2005}%
  \BibitemOpen
  \bibfield  {author} {\bibinfo {author} {\bibfnamefont {T.}~\bibnamefont
  {L{\"{u}}hrs}}, \bibinfo {author} {\bibfnamefont {C.}~\bibnamefont {Ritter}},
  \bibinfo {author} {\bibfnamefont {M.}~\bibnamefont {Adrian}}, \bibinfo
  {author} {\bibfnamefont {D.}~\bibnamefont {Riek-Loher}}, \bibinfo {author}
  {\bibfnamefont {B.}~\bibnamefont {Bohrmann}}, \bibinfo {author}
  {\bibfnamefont {H.}~\bibnamefont {D{\"{o}}beli}}, \bibinfo {author}
  {\bibfnamefont {D.}~\bibnamefont {Schubert}}, \ and\ \bibinfo {author}
  {\bibfnamefont {R.}~\bibnamefont {Riek}},\ }\href {\doibase
  10.1073/pnas.0506723102} {\bibfield  {journal} {\bibinfo  {journal} {Proc.
  Natl. Acad. Sci. U. S. A.}\ }\textbf {\bibinfo {volume} {102}},\ \bibinfo
  {pages} {17342} (\bibinfo {year} {2005})}\BibitemShut {NoStop}%
\bibitem [{\citenamefont {Otzen}(2013)}]{Otzen2013}%
  \BibitemOpen
  \bibfield  {author} {\bibinfo {author} {\bibfnamefont {D.~E.}\ \bibnamefont
  {Otzen}},\ }\href@noop {} {\emph {\bibinfo {title} {Amyloid fibrils and
  prefibrillar aggregates: molecular and biological properties}}}\ (\bibinfo
  {publisher} {Wiley-VCH, Weinheim, Germany},\ \bibinfo {year}
  {2013})\BibitemShut {NoStop}%
\bibitem [{\citenamefont {Correia}\ \emph {et~al.}(2006)\citenamefont
  {Correia}, \citenamefont {Loureiro-Ferreira}, \citenamefont {Rodrigues},\
  and\ \citenamefont {Brito}}]{Correia2006}%
  \BibitemOpen
  \bibfield  {author} {\bibinfo {author} {\bibfnamefont {B.~E.}\ \bibnamefont
  {Correia}}, \bibinfo {author} {\bibfnamefont {N.}~\bibnamefont
  {Loureiro-Ferreira}}, \bibinfo {author} {\bibfnamefont {J.~R.}\ \bibnamefont
  {Rodrigues}}, \ and\ \bibinfo {author} {\bibfnamefont {R.~M.~M.}\
  \bibnamefont {Brito}},\ }\href {\doibase 10.1110/ps.051787106} {\bibfield
  {journal} {\bibinfo  {journal} {Protein Sci.}\ }\textbf {\bibinfo {volume}
  {15}},\ \bibinfo {pages} {28} (\bibinfo {year} {2006})}\BibitemShut {NoStop}%
\bibitem [{\citenamefont {Davis}\ and\ \citenamefont
  {Berkowitz}(2010)}]{Davis2010}%
  \BibitemOpen
  \bibfield  {author} {\bibinfo {author} {\bibfnamefont {C.~H.}\ \bibnamefont
  {Davis}}\ and\ \bibinfo {author} {\bibfnamefont {M.~L.}\ \bibnamefont
  {Berkowitz}},\ }\href {\doibase 10.1002/prot.22763} {\bibfield  {journal}
  {\bibinfo  {journal} {Proteins: Struct., Funct., Bioinf.}\ }\textbf {\bibinfo
  {volume} {78}},\ \bibinfo {pages} {2533} (\bibinfo {year}
  {2010})}\BibitemShut {NoStop}%
\bibitem [{\citenamefont {{Garc{\'{i}} Cuesta}}\ and\ \citenamefont {{J
  S{\'{a}} nchez de Mer{\'{a}}}}(2014)}]{GarciCuesta2014}%
  \BibitemOpen
  \bibfield  {author} {\bibinfo {author} {\bibfnamefont {I.}~\bibnamefont
  {{Garc{\'{i}} Cuesta}}}\ and\ \bibinfo {author} {\bibfnamefont {A.~M.}\
  \bibnamefont {{J S{\'{a}} nchez de Mer{\'{a}}}}},\ }\href {\doibase
  10.1039/c3cp53551g} {\bibfield  {journal} {\bibinfo  {journal} {Phys. Chem.
  Chem. Phys. Phys. Chem. Chem. Phys}\ }\textbf {\bibinfo {volume} {16}},\
  \bibinfo {pages} {4369} (\bibinfo {year} {2014})}\BibitemShut {NoStop}%
\bibitem [{\citenamefont {Green}\ \emph {et~al.}(2004)\citenamefont {Green},
  \citenamefont {Goldsbury}, \citenamefont {Kistler}, \citenamefont {Cooper},\
  and\ \citenamefont {Aebi}}]{Green2004}%
  \BibitemOpen
  \bibfield  {author} {\bibinfo {author} {\bibfnamefont {J.~D.}\ \bibnamefont
  {Green}}, \bibinfo {author} {\bibfnamefont {C.}~\bibnamefont {Goldsbury}},
  \bibinfo {author} {\bibfnamefont {J.}~\bibnamefont {Kistler}}, \bibinfo
  {author} {\bibfnamefont {G.~J.~S.}\ \bibnamefont {Cooper}}, \ and\ \bibinfo
  {author} {\bibfnamefont {U.}~\bibnamefont {Aebi}},\ }\href {\doibase
  10.1074/jbc.M312452200} {\bibfield  {journal} {\bibinfo  {journal} {J. Biol.
  Chem.}\ }\textbf {\bibinfo {volume} {279}},\ \bibinfo {pages} {12206}
  (\bibinfo {year} {2004})}\BibitemShut {NoStop}%
\bibitem [{\citenamefont {Meisl}\ \emph {et~al.}(2014)\citenamefont {Meisl},
  \citenamefont {Yang}, \citenamefont {Hellstrand}, \citenamefont {Frohm},
  \citenamefont {Kirkegaard}, \citenamefont {Cohen}, \citenamefont {Dobson},
  \citenamefont {Linse},\ and\ \citenamefont {Knowles}}]{Meisl2014}%
  \BibitemOpen
  \bibfield  {author} {\bibinfo {author} {\bibfnamefont {G.}~\bibnamefont
  {Meisl}}, \bibinfo {author} {\bibfnamefont {X.}~\bibnamefont {Yang}},
  \bibinfo {author} {\bibfnamefont {E.}~\bibnamefont {Hellstrand}}, \bibinfo
  {author} {\bibfnamefont {B.}~\bibnamefont {Frohm}}, \bibinfo {author}
  {\bibfnamefont {J.~B.}\ \bibnamefont {Kirkegaard}}, \bibinfo {author}
  {\bibfnamefont {S.~I.~A.}\ \bibnamefont {Cohen}}, \bibinfo {author}
  {\bibfnamefont {C.~M.}\ \bibnamefont {Dobson}}, \bibinfo {author}
  {\bibfnamefont {S.}~\bibnamefont {Linse}}, \ and\ \bibinfo {author}
  {\bibfnamefont {T.~P.~J.}\ \bibnamefont {Knowles}},\ }\href {\doibase
  10.1073/pnas.1401564111} {\bibfield  {journal} {\bibinfo  {journal} {Proc.
  Natl. Acad. Sci. U.S.A.}\ }\textbf {\bibinfo {volume} {111}},\ \bibinfo
  {pages} {9384} (\bibinfo {year} {2014})}\BibitemShut {NoStop}%
\bibitem [{\citenamefont {Buell}\ \emph {et~al.}(2010)\citenamefont {Buell},
  \citenamefont {Blundell}, \citenamefont {Dobson}, \citenamefont {Welland},
  \citenamefont {Terentjev},\ and\ \citenamefont {Knowles}}]{Buell2010}%
  \BibitemOpen
  \bibfield  {author} {\bibinfo {author} {\bibfnamefont {A.~K.}\ \bibnamefont
  {Buell}}, \bibinfo {author} {\bibfnamefont {J.~R.}\ \bibnamefont {Blundell}},
  \bibinfo {author} {\bibfnamefont {C.~M.}\ \bibnamefont {Dobson}}, \bibinfo
  {author} {\bibfnamefont {M.~E.}\ \bibnamefont {Welland}}, \bibinfo {author}
  {\bibfnamefont {E.~M.}\ \bibnamefont {Terentjev}}, \ and\ \bibinfo {author}
  {\bibfnamefont {T.~P.~J.}\ \bibnamefont {Knowles}},\ }\href@noop {}
  {\bibfield  {journal} {\bibinfo  {journal} {Phys. Rev. Lett.}\ }\textbf
  {\bibinfo {volume} {104}},\ \bibinfo {pages} {1} (\bibinfo {year}
  {2010})}\BibitemShut {NoStop}%
\bibitem [{\citenamefont {Buell}\ \emph {et~al.}(2012)\citenamefont {Buell},
  \citenamefont {Dhulesia}, \citenamefont {White}, \citenamefont {Knowles},
  \citenamefont {Dobson},\ and\ \citenamefont {Welland}}]{Buell2012}%
  \BibitemOpen
  \bibfield  {author} {\bibinfo {author} {\bibfnamefont {A.~K.}\ \bibnamefont
  {Buell}}, \bibinfo {author} {\bibfnamefont {A.}~\bibnamefont {Dhulesia}},
  \bibinfo {author} {\bibfnamefont {D.~A.}\ \bibnamefont {White}}, \bibinfo
  {author} {\bibfnamefont {T.~P.~J.}\ \bibnamefont {Knowles}}, \bibinfo
  {author} {\bibfnamefont {C.~M.}\ \bibnamefont {Dobson}}, \ and\ \bibinfo
  {author} {\bibfnamefont {M.~E.}\ \bibnamefont {Welland}},\ }\href {\doibase
  10.1002/anie.201108040} {\bibfield  {journal} {\bibinfo  {journal} {Angew.
  Chem., Int. Ed. Engl.}\ }\textbf {\bibinfo {volume} {51}},\ \bibinfo {pages}
  {5247} (\bibinfo {year} {2012})}\BibitemShut {NoStop}%
\bibitem [{\citenamefont {Auer}\ \emph {et~al.}(2007)\citenamefont {Auer},
  \citenamefont {Dobson},\ and\ \citenamefont {Vendruscolo}}]{Vendruscolo2007}%
  \BibitemOpen
  \bibfield  {author} {\bibinfo {author} {\bibfnamefont {S.}~\bibnamefont
  {Auer}}, \bibinfo {author} {\bibfnamefont {C.~M.}\ \bibnamefont {Dobson}}, \
  and\ \bibinfo {author} {\bibfnamefont {M.}~\bibnamefont {Vendruscolo}},\
  }\href@noop {} {\bibfield  {journal} {\bibinfo  {journal} {HFSP Journal}\
  }\textbf {\bibinfo {volume} {1}},\ \bibinfo {pages} {137} (\bibinfo {year}
  {2007})}\BibitemShut {NoStop}%
\bibitem [{\citenamefont {Richter}\ and\ \citenamefont
  {Eigen}(1974)}]{Richter1974}%
  \BibitemOpen
  \bibfield  {author} {\bibinfo {author} {\bibfnamefont {P.~H.}\ \bibnamefont
  {Richter}}\ and\ \bibinfo {author} {\bibfnamefont {M.}~\bibnamefont
  {Eigen}},\ }\href {\doibase 10.1016/0301-4622(74)80050-5} {\bibfield
  {journal} {\bibinfo  {journal} {Biophys. Chem.}\ }\textbf {\bibinfo {volume}
  {2}},\ \bibinfo {pages} {255} (\bibinfo {year} {1974})}\BibitemShut {NoStop}%
\bibitem [{\citenamefont {Bleiholder}\ \emph {et~al.}(2011)\citenamefont
  {Bleiholder}, \citenamefont {Dupuis}, \citenamefont {Wyttenbach},\ and\
  \citenamefont {Bowers}}]{Bleiholder2011}%
  \BibitemOpen
  \bibfield  {author} {\bibinfo {author} {\bibfnamefont {C.}~\bibnamefont
  {Bleiholder}}, \bibinfo {author} {\bibfnamefont {N.~F.}\ \bibnamefont
  {Dupuis}}, \bibinfo {author} {\bibfnamefont {T.}~\bibnamefont {Wyttenbach}},
  \ and\ \bibinfo {author} {\bibfnamefont {M.~T.}\ \bibnamefont {Bowers}},\
  }\href {\doibase 10.1038/nchem.945} {\bibfield  {journal} {\bibinfo
  {journal} {Nat Chem}\ }\textbf {\bibinfo {volume} {3}},\ \bibinfo {pages}
  {172} (\bibinfo {year} {2011})}\BibitemShut {NoStop}%
\bibitem [{\citenamefont {Yong}\ \emph {et~al.}(2002)\citenamefont {Yong},
  \citenamefont {Lomakin}, \citenamefont {Kirkitadze}, \citenamefont {Teplow},
  \citenamefont {Chen},\ and\ \citenamefont {Benedek}}]{Yong2002}%
  \BibitemOpen
  \bibfield  {author} {\bibinfo {author} {\bibfnamefont {W.}~\bibnamefont
  {Yong}}, \bibinfo {author} {\bibfnamefont {A.}~\bibnamefont {Lomakin}},
  \bibinfo {author} {\bibfnamefont {M.}~\bibnamefont {Kirkitadze}}, \bibinfo
  {author} {\bibfnamefont {D.}~\bibnamefont {Teplow}}, \bibinfo {author}
  {\bibfnamefont {S.-H.}\ \bibnamefont {Chen}}, \ and\ \bibinfo {author}
  {\bibfnamefont {G.}~\bibnamefont {Benedek}},\ }\href@noop {} {\bibfield
  {journal} {\bibinfo  {journal} {Proc. Natl. Acad. Sci. U. S. A.}\ }\textbf
  {\bibinfo {volume} {99}},\ \bibinfo {pages} {150} (\bibinfo {year}
  {2002})}\BibitemShut {NoStop}%
\bibitem [{\citenamefont {Sabat{\'{e}}}\ and\ \citenamefont
  {Estelrich}(2005)}]{Sabate2005}%
  \BibitemOpen
  \bibfield  {author} {\bibinfo {author} {\bibfnamefont {R.}~\bibnamefont
  {Sabat{\'{e}}}}\ and\ \bibinfo {author} {\bibfnamefont {J.}~\bibnamefont
  {Estelrich}},\ }\href@noop {} {\bibfield  {journal} {\bibinfo  {journal} {J.
  Phys. Chem. B}\ }\textbf {\bibinfo {volume} {109}},\ \bibinfo {pages} {11027}
  (\bibinfo {year} {2005})}\BibitemShut {NoStop}%
\bibitem [{\citenamefont {Frare}\ \emph {et~al.}(2009)\citenamefont {Frare},
  \citenamefont {Mossuto}, \citenamefont {{de Laureto}}, \citenamefont {Tolin},
  \citenamefont {Menzer}, \citenamefont {Dumoulin}, \citenamefont {Dobson},\
  and\ \citenamefont {Fontana}}]{Frare2009}%
  \BibitemOpen
  \bibfield  {author} {\bibinfo {author} {\bibfnamefont {E.}~\bibnamefont
  {Frare}}, \bibinfo {author} {\bibfnamefont {M.~F.}\ \bibnamefont {Mossuto}},
  \bibinfo {author} {\bibfnamefont {P.~P.}\ \bibnamefont {{de Laureto}}},
  \bibinfo {author} {\bibfnamefont {S.}~\bibnamefont {Tolin}}, \bibinfo
  {author} {\bibfnamefont {L.}~\bibnamefont {Menzer}}, \bibinfo {author}
  {\bibfnamefont {M.}~\bibnamefont {Dumoulin}}, \bibinfo {author}
  {\bibfnamefont {C.~M.}\ \bibnamefont {Dobson}}, \ and\ \bibinfo {author}
  {\bibfnamefont {A.}~\bibnamefont {Fontana}},\ }\href {\doibase
  10.1016/j.jmb.2009.01.049} {\bibfield  {journal} {\bibinfo  {journal} {J.
  Mol. Biol.}\ }\textbf {\bibinfo {volume} {387}},\ \bibinfo {pages} {17}
  (\bibinfo {year} {2009})}\BibitemShut {NoStop}%
\bibitem [{\citenamefont {Serio}\ \emph {et~al.}(2000)\citenamefont {Serio},
  \citenamefont {Cashikar}, \citenamefont {Kowal}, \citenamefont {Sawicki},
  \citenamefont {Moslehi}, \citenamefont {Serpell}, \citenamefont {Arnsdorf},\
  and\ \citenamefont {Lindquist}}]{Serio2000a}%
  \BibitemOpen
  \bibfield  {author} {\bibinfo {author} {\bibfnamefont {T.~R.}\ \bibnamefont
  {Serio}}, \bibinfo {author} {\bibfnamefont {A.~G.}\ \bibnamefont {Cashikar}},
  \bibinfo {author} {\bibfnamefont {A.~S.}\ \bibnamefont {Kowal}}, \bibinfo
  {author} {\bibfnamefont {G.~J.}\ \bibnamefont {Sawicki}}, \bibinfo {author}
  {\bibfnamefont {J.~J.}\ \bibnamefont {Moslehi}}, \bibinfo {author}
  {\bibfnamefont {L.}~\bibnamefont {Serpell}}, \bibinfo {author} {\bibfnamefont
  {M.~F.}\ \bibnamefont {Arnsdorf}}, \ and\ \bibinfo {author} {\bibfnamefont
  {S.~L.}\ \bibnamefont {Lindquist}},\ }\href {\doibase
  10.1126/science.289.5483.1317} {\bibfield  {journal} {\bibinfo  {journal}
  {Science}\ }\textbf {\bibinfo {volume} {289}},\ \bibinfo {pages} {1317}
  (\bibinfo {year} {2000})}\BibitemShut {NoStop}%
\bibitem [{\citenamefont {Auer}\ \emph {et~al.}(2008)\citenamefont {Auer},
  \citenamefont {Meersman}, \citenamefont {Dobson},\ and\ \citenamefont
  {Vendruscolo}}]{Auer2008a}%
  \BibitemOpen
  \bibfield  {author} {\bibinfo {author} {\bibfnamefont {S.}~\bibnamefont
  {Auer}}, \bibinfo {author} {\bibfnamefont {F.}~\bibnamefont {Meersman}},
  \bibinfo {author} {\bibfnamefont {C.~M.}\ \bibnamefont {Dobson}}, \ and\
  \bibinfo {author} {\bibfnamefont {M.}~\bibnamefont {Vendruscolo}},\ }\href
  {\doibase 10.1371/journal.pcbi.1000222} {\bibfield  {journal} {\bibinfo
  {journal} {PLoS Comput. Biol.}\ }\textbf {\bibinfo {volume} {4}},\ \bibinfo
  {pages} {e1000222} (\bibinfo {year} {2008})}\BibitemShut {NoStop}%
\bibitem [{\citenamefont {Uversky}(2010)}]{Uversky2010}%
  \BibitemOpen
  \bibfield  {author} {\bibinfo {author} {\bibfnamefont {V.~N.}\ \bibnamefont
  {Uversky}},\ }\href {\doibase 10.1111/j.1742-4658.2010.07721.x} {\bibfield
  {journal} {\bibinfo  {journal} {FEBS J.}\ }\textbf {\bibinfo {volume}
  {277}},\ \bibinfo {pages} {2940} (\bibinfo {year} {2010})}\BibitemShut
  {NoStop}%
\bibitem [{\citenamefont {Urbanc}\ \emph {et~al.}(2010)\citenamefont {Urbanc},
  \citenamefont {Betnel}, \citenamefont {Cruz}, \citenamefont {Bltan},\ and\
  \citenamefont {Teplow}}]{Urbanc2010}%
  \BibitemOpen
  \bibfield  {author} {\bibinfo {author} {\bibfnamefont {B.}~\bibnamefont
  {Urbanc}}, \bibinfo {author} {\bibfnamefont {M.}~\bibnamefont {Betnel}},
  \bibinfo {author} {\bibfnamefont {L.}~\bibnamefont {Cruz}}, \bibinfo {author}
  {\bibfnamefont {G.}~\bibnamefont {Bltan}}, \ and\ \bibinfo {author}
  {\bibfnamefont {D.~B.}\ \bibnamefont {Teplow}},\ }\href {\doibase
  10.1021/ja9096303} {\bibfield  {journal} {\bibinfo  {journal} {J. Am. Chem.
  Soc.}\ }\textbf {\bibinfo {volume} {132}},\ \bibinfo {pages} {4266} (\bibinfo
  {year} {2010})}\BibitemShut {NoStop}%
\bibitem [{\citenamefont {Guo}\ \emph {et~al.}(2005)\citenamefont {Guo},
  \citenamefont {Gorman}, \citenamefont {Rico}, \citenamefont {Chakrabartty},\
  and\ \citenamefont {Laurents}}]{Guo2005}%
  \BibitemOpen
  \bibfield  {author} {\bibinfo {author} {\bibfnamefont {M.}~\bibnamefont
  {Guo}}, \bibinfo {author} {\bibfnamefont {P.~M.}\ \bibnamefont {Gorman}},
  \bibinfo {author} {\bibfnamefont {M.}~\bibnamefont {Rico}}, \bibinfo {author}
  {\bibfnamefont {A.}~\bibnamefont {Chakrabartty}}, \ and\ \bibinfo {author}
  {\bibfnamefont {D.~V.}\ \bibnamefont {Laurents}},\ }\href {\doibase
  10.1016/j.febslet.2005.05.036} {\bibfield  {journal} {\bibinfo  {journal}
  {FEBS Letters}\ }\textbf {\bibinfo {volume} {579}},\ \bibinfo {pages} {3574}
  (\bibinfo {year} {2005})}\BibitemShut {NoStop}%
\bibitem [{\citenamefont {Maibaum}\ \emph {et~al.}(2004)\citenamefont
  {Maibaum}, \citenamefont {Dinner},\ and\ \citenamefont
  {Chandler}}]{Maibaum2004}%
  \BibitemOpen
  \bibfield  {author} {\bibinfo {author} {\bibfnamefont {L.}~\bibnamefont
  {Maibaum}}, \bibinfo {author} {\bibfnamefont {A.~R.}\ \bibnamefont {Dinner}},
  \ and\ \bibinfo {author} {\bibfnamefont {D.}~\bibnamefont {Chandler}},\
  }\href {\doibase 10.1021/jp037487t} {\bibfield  {journal} {\bibinfo
  {journal} {J. Phys. Chem. B}\ }\textbf {\bibinfo {volume} {108}},\ \bibinfo
  {pages} {6778} (\bibinfo {year} {2004})}\BibitemShut {NoStop}%
\bibitem [{\citenamefont {Groenewold}\ and\ \citenamefont
  {Kegel}(2001)}]{Groenewold2001}%
  \BibitemOpen
  \bibfield  {author} {\bibinfo {author} {\bibfnamefont {J.}~\bibnamefont
  {Groenewold}}\ and\ \bibinfo {author} {\bibfnamefont {W.~K.}\ \bibnamefont
  {Kegel}},\ }\href {\doibase 10.1021/jp011646w} {\bibfield  {journal}
  {\bibinfo  {journal} {J. Phys. Chem. B}\ }\textbf {\bibinfo {volume} {105}},\
  \bibinfo {pages} {11702} (\bibinfo {year} {2001})}\BibitemShut {NoStop}%
\bibitem [{\citenamefont {Woo}\ \emph {et~al.}(1996)\citenamefont {Woo},
  \citenamefont {Carraro},\ and\ \citenamefont {Chandler}}]{Woo1996}%
  \BibitemOpen
  \bibfield  {author} {\bibinfo {author} {\bibfnamefont {H.}~\bibnamefont
  {Woo}}, \bibinfo {author} {\bibfnamefont {C.}~\bibnamefont {Carraro}}, \ and\
  \bibinfo {author} {\bibfnamefont {D.}~\bibnamefont {Chandler}},\ }\href@noop
  {} {\bibfield  {journal} {\bibinfo  {journal} {Faraday Discuss.}\ }\textbf
  {\bibinfo {volume} {104}},\ \bibinfo {pages} {183} (\bibinfo {year}
  {1996})}\BibitemShut {NoStop}%
\bibitem [{\citenamefont {Hills}\ and\ \citenamefont
  {Brooks}(2007)}]{Hills2007}%
  \BibitemOpen
  \bibfield  {author} {\bibinfo {author} {\bibfnamefont {R.~D.}\ \bibnamefont
  {Hills}}\ and\ \bibinfo {author} {\bibfnamefont {C.~L.}\ \bibnamefont
  {Brooks}},\ }\href@noop {} {\bibfield  {journal} {\bibinfo  {journal} {J.
  Mol. Biol.}\ }\textbf {\bibinfo {volume} {368}},\ \bibinfo {pages} {894}
  (\bibinfo {year} {2007})}\BibitemShut {NoStop}%
\bibitem [{\citenamefont {Soreghan}\ \emph {et~al.}(1994)\citenamefont
  {Soreghan}, \citenamefont {Kosmoski},\ and\ \citenamefont
  {Glabe}}]{Soreghan1994}%
  \BibitemOpen
  \bibfield  {author} {\bibinfo {author} {\bibfnamefont {B.}~\bibnamefont
  {Soreghan}}, \bibinfo {author} {\bibfnamefont {J.}~\bibnamefont {Kosmoski}},
  \ and\ \bibinfo {author} {\bibfnamefont {C.}~\bibnamefont {Glabe}},\ }\href
  {\doibase citeulike-article-id:3576076} {\bibfield  {journal} {\bibinfo
  {journal} {J. Biol. Chem.}\ }\textbf {\bibinfo {volume} {269}},\ \bibinfo
  {pages} {28551} (\bibinfo {year} {1994})}\BibitemShut {NoStop}%
\bibitem [{\citenamefont {Nuallain}\ \emph {et~al.}(2005)\citenamefont
  {Nuallain}, \citenamefont {Shivaprasad}, \citenamefont {Kheterpal},\ and\
  \citenamefont {Wetzel}}]{Nuallain2005}%
  \BibitemOpen
  \bibfield  {author} {\bibinfo {author} {\bibfnamefont {B.~O.}\ \bibnamefont
  {Nuallain}}, \bibinfo {author} {\bibfnamefont {S.}~\bibnamefont
  {Shivaprasad}}, \bibinfo {author} {\bibfnamefont {I.}~\bibnamefont
  {Kheterpal}}, \ and\ \bibinfo {author} {\bibfnamefont {R.}~\bibnamefont
  {Wetzel}},\ }\href@noop {} {\bibfield  {journal} {\bibinfo  {journal}
  {Biochemistry}\ }\textbf {\bibinfo {volume} {44}},\ \bibinfo {pages} {12709}
  (\bibinfo {year} {2005})}\BibitemShut {NoStop}%
\bibitem [{\citenamefont {Kim}\ and\ \citenamefont {Lee}(2004)}]{Kim2004}%
  \BibitemOpen
  \bibfield  {author} {\bibinfo {author} {\bibfnamefont {J.}~\bibnamefont
  {Kim}}\ and\ \bibinfo {author} {\bibfnamefont {M.}~\bibnamefont {Lee}},\
  }\href {\doibase 10.1016/j.bbrc.2004.02.059} {\bibfield  {journal} {\bibinfo
  {journal} {Biochem. Biophys. Res. Commun.}\ }\textbf {\bibinfo {volume}
  {316}},\ \bibinfo {pages} {393} (\bibinfo {year} {2004})}\BibitemShut
  {NoStop}%
\bibitem [{\citenamefont {Kramers}(1940)}]{Kramers1940}%
  \BibitemOpen
  \bibfield  {author} {\bibinfo {author} {\bibfnamefont {H.}~\bibnamefont
  {Kramers}},\ }\href {\doibase 10.1016/S0031-8914(40)90098-2} {\bibfield
  {journal} {\bibinfo  {journal} {Physica}\ }\textbf {\bibinfo {volume} {7}},\
  \bibinfo {pages} {284} (\bibinfo {year} {1940})}\BibitemShut {NoStop}%
\bibitem [{\citenamefont {H{\"{a}}nggi}\ \emph {et~al.}(1990)\citenamefont
  {H{\"{a}}nggi}, \citenamefont {Talkner},\ and\ \citenamefont
  {Borkovec}}]{Hanggi1990}%
  \BibitemOpen
  \bibfield  {author} {\bibinfo {author} {\bibfnamefont {P.}~\bibnamefont
  {H{\"{a}}nggi}}, \bibinfo {author} {\bibfnamefont {P.}~\bibnamefont
  {Talkner}}, \ and\ \bibinfo {author} {\bibfnamefont {M.}~\bibnamefont
  {Borkovec}},\ }\href@noop {} {\bibfield  {journal} {\bibinfo  {journal} {Rev.
  Mod. Phys.}\ }\textbf {\bibinfo {volume} {62}},\ \bibinfo {pages} {251}
  (\bibinfo {year} {1990})}\BibitemShut {NoStop}%
\bibitem [{\citenamefont {Michel}\ and\ \citenamefont
  {Ruelle}(2013)}]{Michel2013}%
  \BibitemOpen
  \bibfield  {author} {\bibinfo {author} {\bibfnamefont {D.}~\bibnamefont
  {Michel}}\ and\ \bibinfo {author} {\bibfnamefont {P.}~\bibnamefont
  {Ruelle}},\ }\href@noop {} {\bibfield  {journal} {\bibinfo  {journal} {J.
  Math. Chem.}\ }\textbf {\bibinfo {volume} {51}},\ \bibinfo {pages} {2271}
  (\bibinfo {year} {2013})}\BibitemShut {NoStop}%
\bibitem [{\citenamefont {Raines}(1988)}]{Raines1988}%
  \BibitemOpen
  \bibfield  {author} {\bibinfo {author} {\bibfnamefont {R.~T.}\ \bibnamefont
  {Raines}},\ }\href@noop {} {\bibfield  {journal} {\bibinfo  {journal} {J.
  Chem. Educ.}\ }\textbf {\bibinfo {volume} {65}},\ \bibinfo {pages} {757}
  (\bibinfo {year} {1988})}\BibitemShut {NoStop}%
\bibitem [{\citenamefont {Berg}\ and\ \citenamefont {von
  Hippel}(1985)}]{Berg1985a}%
  \BibitemOpen
  \bibfield  {author} {\bibinfo {author} {\bibfnamefont {O.~G.}\ \bibnamefont
  {Berg}}\ and\ \bibinfo {author} {\bibfnamefont {P.~H.}\ \bibnamefont {von
  Hippel}},\ }\href
  {http://www.ncbi.nlm.nih.gov/entrez/query.fcgi?cmd=Retrieve&db=PubMed&dopt=Citation&list_uids=3890878}
  {\bibfield  {journal} {\bibinfo  {journal} {Annu. Rev. Biophys. Biophys.
  Chem.}\ }\textbf {\bibinfo {volume} {14}},\ \bibinfo {pages} {131} (\bibinfo
  {year} {1985})}\BibitemShut {NoStop}%
\bibitem [{\citenamefont {Sakono}\ and\ \citenamefont
  {Zako}(2010)}]{Sakono2010}%
  \BibitemOpen
  \bibfield  {author} {\bibinfo {author} {\bibfnamefont {M.}~\bibnamefont
  {Sakono}}\ and\ \bibinfo {author} {\bibfnamefont {T.}~\bibnamefont {Zako}},\
  }\href@noop {} {\bibfield  {journal} {\bibinfo  {journal} {FEBS J.}\ }\textbf
  {\bibinfo {volume} {277}},\ \bibinfo {pages} {1348} (\bibinfo {year}
  {2010})}\BibitemShut {NoStop}%
\bibitem [{\citenamefont {Zheng}\ \emph {et~al.}(2007)\citenamefont {Zheng},
  \citenamefont {Jang}, \citenamefont {Ma}, \citenamefont {Tsai},\ and\
  \citenamefont {Nussinov}}]{Zheng2007}%
  \BibitemOpen
  \bibfield  {author} {\bibinfo {author} {\bibfnamefont {J.}~\bibnamefont
  {Zheng}}, \bibinfo {author} {\bibfnamefont {H.}~\bibnamefont {Jang}},
  \bibinfo {author} {\bibfnamefont {B.}~\bibnamefont {Ma}}, \bibinfo {author}
  {\bibfnamefont {C.-J.}\ \bibnamefont {Tsai}}, \ and\ \bibinfo {author}
  {\bibfnamefont {R.}~\bibnamefont {Nussinov}},\ }\href {\doibase
  10.1529/biophysj.107.110700} {\bibfield  {journal} {\bibinfo  {journal}
  {Biophys. J.}\ }\textbf {\bibinfo {volume} {93}},\ \bibinfo {pages} {3046}
  (\bibinfo {year} {2007})}\BibitemShut {NoStop}%
\bibitem [{\citenamefont {Williams}\ \emph {et~al.}(2006)\citenamefont
  {Williams}, \citenamefont {Shivaprasad},\ and\ \citenamefont
  {Wetzel}}]{Williams2006}%
  \BibitemOpen
  \bibfield  {author} {\bibinfo {author} {\bibfnamefont {A.~D.}\ \bibnamefont
  {Williams}}, \bibinfo {author} {\bibfnamefont {S.}~\bibnamefont
  {Shivaprasad}}, \ and\ \bibinfo {author} {\bibfnamefont {R.}~\bibnamefont
  {Wetzel}},\ }\href {\doibase 10.1016/j.jmb.2006.01.041} {\bibfield  {journal}
  {\bibinfo  {journal} {J. Mol. Biol.}\ }\textbf {\bibinfo {volume} {357}},\
  \bibinfo {pages} {1283} (\bibinfo {year} {2006})}\BibitemShut {NoStop}%
\bibitem [{\citenamefont {Nag}\ \emph {et~al.}(2011)\citenamefont {Nag},
  \citenamefont {Sarkar}, \citenamefont {Bandyopadhyay}, \citenamefont {Sahoo},
  \citenamefont {Sreenivasan}, \citenamefont {Kombrabail}, \citenamefont
  {Muralidharan},\ and\ \citenamefont {Maiti}}]{Nag2011}%
  \BibitemOpen
  \bibfield  {author} {\bibinfo {author} {\bibfnamefont {S.}~\bibnamefont
  {Nag}}, \bibinfo {author} {\bibfnamefont {B.}~\bibnamefont {Sarkar}},
  \bibinfo {author} {\bibfnamefont {A.}~\bibnamefont {Bandyopadhyay}}, \bibinfo
  {author} {\bibfnamefont {B.}~\bibnamefont {Sahoo}}, \bibinfo {author}
  {\bibfnamefont {V.~K.~A.}\ \bibnamefont {Sreenivasan}}, \bibinfo {author}
  {\bibfnamefont {M.}~\bibnamefont {Kombrabail}}, \bibinfo {author}
  {\bibfnamefont {C.}~\bibnamefont {Muralidharan}}, \ and\ \bibinfo {author}
  {\bibfnamefont {S.}~\bibnamefont {Maiti}},\ }\href {\doibase
  10.1074/jbc.M110.199885} {\bibfield  {journal} {\bibinfo  {journal} {J. Biol.
  Chem.}\ }\textbf {\bibinfo {volume} {286}},\ \bibinfo {pages} {13827}
  (\bibinfo {year} {2011})}\BibitemShut {NoStop}%
\bibitem [{\citenamefont {Stroud}\ \emph {et~al.}(2012)\citenamefont {Stroud},
  \citenamefont {Liu}, \citenamefont {Teng},\ and\ \citenamefont
  {Eisenberg}}]{Stroud2012a}%
  \BibitemOpen
  \bibfield  {author} {\bibinfo {author} {\bibfnamefont {J.~C.}\ \bibnamefont
  {Stroud}}, \bibinfo {author} {\bibfnamefont {C.}~\bibnamefont {Liu}},
  \bibinfo {author} {\bibfnamefont {P.~K.}\ \bibnamefont {Teng}}, \ and\
  \bibinfo {author} {\bibfnamefont {D.}~\bibnamefont {Eisenberg}},\ }\href
  {\doibase 10.1073/pnas.1203193109} {\bibfield  {journal} {\bibinfo  {journal}
  {Proc. Natl. Acad. Sci.}\ }\textbf {\bibinfo {volume} {109}},\ \bibinfo
  {pages} {7717} (\bibinfo {year} {2012})}\BibitemShut {NoStop}%
\end{thebibliography}
%

\vspace{2cm}

\begin{appendix}
\section{Values of A\mathinhead{\beta_{1-42}}{\beta1-42} parameters}  \label{app:abpara}

The values of all parameters are summarized in the Table 1. 
An approximate value of $\Delta_c$ is suggested to be approximately 20 $k_BT$ ($T$ refers to the room temperature 25 $^{\circ}$C) in the practice of coarse-grained molecular simulations \cite{Bieler2012}. It has to be noted that this value is not a reliable result of rigorous all-atom simulations, as one would have wished and as is the case for many other energy parameters we use. In the coarse-grained simulation \cite{Bieler2012} the authors merely assumed this value to be reasonable, based on the fact that single $\beta$-mers are not spontaneously formed in solution. 

All-atom simulations were carried out to investigate the interactions between two protofilaments in the model of A$\beta_{17-42}$, which is the fraction of peptide sequence from the 17th to the 42nd amino acid residues of the native A$\beta_{1-42}$, and is believed to be the main cause of the universal structural motif in A$\beta_{17-42}$ fibrils. This result suggested a steric zipper formed on the interface of two protofilaments \cite{Zheng2007}. Thus, the $\Delta_{s}$ value can be roughly estimated to be -22 $k_BT$, the typical interaction free energy of a steric zipper \cite{GarciCuesta2014}. 

To estimate the value of $\mu^{\circ}$, we connect the theoretical elongation free energy formula, namely $\Delta_{\beta}+\Delta_s-\mu^{\circ}+\Delta_c+ k_B T\ln{(C_1/C_1^{\circ})}$ from \eqref{eq:d-proto}, with the experimentally measured magnitude of elongation free energy of fibrils. Although this experimentally measured vlaue for A$\beta_{1-42}$is not available, a similar (shorter) amyloidogenic peptide, A$\beta_{1-40}$, was used to measure its elongation free energy through evaluating the monomer concentration in thermal equilibrium with its fibrils \cite{Williams2006}. The result has produced a value -11.6 $k_B T$ in the A$\beta_{1-40}$ system. We shall assume that A$\beta_{1-42}$ should resemble it closely, and have roughly the same elongation free energy. By inserting $\Delta_{\beta}$ of -44 $k_B T$, $\Delta_s$ of -22 $k_B T$, $\Delta_c$ of 20 $k_B T$, and the elongation free energy of -11.6 $k_B T$ in the theoretical elongation free energy formula, $\mu^{\circ}$ is evaluated as -34.5 $k_B T$. 

The value of $\Delta_{\alpha}$ cannot be determined from experiments and can only be investigated by molecular simulations. Nevertheless, different results have been obtained depending on the constraints on the conformation of peptides that are used in simulations. In ref.~\cite{Saric2014} and~\cite{Bieler2012}, $\Delta_{\alpha}$ has the value of -8.4 $k_B T$ and -6 $k_B T$ respectively, but these values may be under-estimated since each $\alpha$-mer is assumed to still hold the same structure as before o micellation, and can only interact through one hydrophobic patch on the surface of the native monomeric state. This scene is unrealistic because micellation usually undergoes structural change to become a more stable state, like collapsed amorphous micelles. On the other hand, ref.~\cite{Hills2007} does not have this pre-assumption and allows peptides to reorganize its structure to form an amorphous micelle. Even though it only models the peptides with the core sequence that forms $\beta$-sheet in the aggregation and may thus also undervalues $\Delta_{\alpha}$ term. Despite this defect, we will use the values given in ref.~\cite{Hills2007} to estimate the value of $\Delta_{\alpha}$. The bond interaction is approximately -19.2 kcal/mol, which is taken from the amorphous dimer case of Fig.~3 in ref.~\cite{Hills2007}. The free energy change due to the entropic loss is evaluated as 9.2 kcal/mol under the assumption that the peptide loses all its conformational entropy upon bonding in the supplement of it. Together with these two values, $\Delta_{\alpha}$ is roughly -17 $k_B T$. 

$\tau_I$ is the time-scale for internal $\alpha$ to $\beta$ rearrangement of amyloidogenic proteins, which cannot be theoretically evaluated, as no simple theoretical models are proposed to include internal friction in a dense protein structure. Although $\tau_I$ for A$\beta_{1-42}$ peptide is still unknown, it can be approximated from the known value for insulin (in ref.~\cite{Buell2010}) that has the closest protein length to A$\beta_{1-42}$ peptide, giving $\tau_I$ as $10^{-5} s$.

For the $\Delta_{\alpha\beta}$ value, although all-atomic simulations on the free energy of one $\alpha\beta$ bond in ref.~\cite{Saric2014} was implemented, it had the pre-assumption/constrains on the conformation of $\alpha$-mers, not allowed to change from its native monomeric structure in solution, and the resultant $\Delta_{\alpha\beta}$ value is doubtable to be used in the micelle model of this work where $\alpha$-mers within a micelle may not have the same conformation as in the monomeric state. Unfortunately, no other specific experiments or all-atomic molecular simulations considering this structural change of monomers into a micelle have been conducted to determine the value of this coupling energy. 

We set the lower limit of $\Delta_{\alpha\beta}$ by considering its role to facilitate the conversion process. \eqref{eq:mic} can be used to obtain the free energy change from an $N$-size micelle into the conversion intermediate with only one $\alpha$-mer converted. If we first ignore the electrostatic and entropic repulsion and focus on the role of the $\alpha\beta$ bond free energy, the result gives: $\Delta_c+(\Delta_{\alpha\beta}-\Delta_{\alpha})\left[N_{\alpha} (N)-N_{\alpha} (N-1)\right]$. If $\Delta_{\alpha\beta}$ equals to $\Delta_{\alpha}$, it means that conversion of one $\alpha$-mer on the surface of an micelle or conversion of one $\alpha$-mer in solution are equally likely to happen because the free energy change is the same, i.e. $\Delta_c$, in both cases. In other words, micellation cannot help facilitate the conversion of $\alpha$-mers and lower the free energy barrier along the conversion process. The lower limit of $\Delta_{\alpha\beta}$ is therefore set to be -18 $k_B T$ in our model, a little higher than $\Delta_{\alpha}$. On the other hand, we consider the final conversion step for intermediates with one $\alpha$-mer left to fully become the $\beta$-aggregate. This step gives the same free energy states involved in the elongation process: an $\alpha$-mer first attaches to the end of the pre-existing $\beta$-aggregate, and then converts into the $\beta$-mer. The free energy of the intermediate at $x=N-1$ shall be higher than the final free energy of the $\beta$ -aggregate ($x=N$), so that the final $\beta$-aggregate is more favored than partially converted intermediates, and the elongation process will not get thermodynamically trapped in the intermediate state. We thus have the constraint on the upper limit of $\Delta_{\alpha\beta}$: $0>\Delta_{\beta}+\Delta_{s}+\Delta_{c}-2\Delta_{\alpha\beta}$. Therefore, $\Delta_{\alpha\beta}$ should be weaker than -23 $k_B T$ and has its lower limit of -17 $k_B T$ as discussed earlier. 

\begin{table}\label{tab:para}
\caption{Summary of parameters. The values of bond energies are given in units of $k_BT$ at room temperature, with the main reference to the source where available. }
\begin{tabular}{cccccccc}
&$\Delta_{\beta}$&$\Delta_{c}$&$\Delta_{s}$&$\mu^{\circ}$&$\Delta_{\alpha}$&$\Delta_{\alpha\beta}$&$\tau_{I}$\\
\midrule
value& -44&20&-22&-34.5&-17&-(17$-$23) & $10^{-5}$ s                \\
ref.&\cite{Davis2010}&\cite{Saric2014}&\cite{GarciCuesta2014}&\cite{Williams2006}&\cite{Hills2007}&(see below)&\cite{Buell2010}\\
\bottomrule
\end{tabular}
\end{table}

\section{Derivation of \mathinhead{k_{1}}{k1}}  \label{app:k1}
The shape of monomers can be approximated by a sphere with the radius of 1 nm from the hydrodynamic radius experiment of A$\beta_{1-42}$ \cite{Nag2011}, while the trimer of the paired protofilament (the critical nucleus in the NP mechanism) takes roughly the shape of a cuboid with dimensions of 46$\mathrm{\AA}$, 46$\mathrm{\AA}$ and 8.4$\mathrm{\AA}$ based on the structure of a single $\beta$-mer given in ref.~\cite{Stroud2012a}. In this scenario, it is clear that the collision between monomers and the critical nuclei is the non-spherical case, and in fact, no theoretical formula of the collision rate is available for such a complicated shape. Yet, a convenient way to estimate this structural effect is to approximate it as in the elongated ellipsoid case, as illustrated in Fig. \ref{fig:appB}(a), where as long as the center of mass of one monomer enters the surface of this ellipsoid (with the trimer being its center), the collision is considered effective. 

In this case, the inverse of the time required for one monomer to diffuse to or hit the surface of the trimer, $1/\tau_D$, is written as \cite{Richter1974}:  
\begin{align} \label{eq:ellip}
\frac{1}{\tau_D}=\frac{4\pi D_m C_1A_{x1}\sqrt[2]{1-(A_{x2}/A_{x1})^2}}{\pi/2-\arctan{[A_{x2}/A_{x1}\sqrt[2]{1-(A_{x2}/A_{x1})^2}]}}
\end{align}
Here, $A_{x1}$ and $A_{x2}$ are the major and minor semi-axes of the ellipsoid, respectively. We further let $A_{x1}$ equal to half the sum of the length of the trimer and the monomer diameter, which is $(46+20)/2$ $\mathrm{\AA}$ and refers to the case where the monomer contacts the top surface of trimer. Similarly, $A_{x2}$ equals to half the sum of the heigh and the monomer diameter, giving $(8.4+20)/2$ $\mathrm{\AA}$ and corresponding to the case of side attachment of one monomer to this trimer. A comparison with the Smoluchowski equation and \eqref{eq:ellip} gives the geometric factor $f_{geo}$ used in obtaining the $k_1$ value as:
\begin{align} \label{eq:fgeo}
f_{geo}=\frac{A_{x1}\sqrt[2]{1-(A_{x2}/A_{x1})^2}}{\pi/2-\arctan{[A_{x2}/A_{x1}\sqrt[2]{1-(A_{x2}/A_{x1})^2}]}}
\end{align}
With $A_{x1}$ of 33 $\mathrm{\AA}$ and $A_{x2}$ of 14.2 $\mathrm{\AA}$,
$f_{geo}$ is estimated roughly as 26.4 $\mathrm{\AA}$.

The constant $D_m$ is assumed to be roughly twice the diffusion coefficient of the spherical monomer: although we know that the trimer has a different volume, the difference in the diffusion constant between it and the monomer is minor -- while no simple calculation for an accurate mutual diffusion coefficient is available. From the Stoke-Einstein equation, $D_m$ is expressed in terms of the monomer size $r_1$ and solvent viscosity $\eta$ as $k_B T/3\pi \eta r_1$. Putting this relation and  \eqref{eq:fgeo} into \eqref{eq:ellip}, we obtain:  
\begin{align} \label{eq:tD}
\frac{1}{\tau_D}=\frac{4k_BT C_1 f_{geo}}{3\eta r_1}
\end{align}

Since we take the rate of linear aggregation as the rate of progressing from $N=3$ to $N=4$ (the process repeats with the same parameters further on, due to the linear $N$-dependence of the free energy, Fig. \ref{fig:benerg}). Therefore, this rate is determined by the barrier to reach the critical nucleation size $N=3$ of a paired protofilament, plus the small free energy barrier for monomer aggregation on regular filament elongation, see Fig. \ref{fig:appB}(b) and the detailed simulations \cite{Vendruscolo2007}. 
Putting \eqref{eq:tD} into the \eqref{eq:elong} we finally obtain \eqref{eq:k1}.

Why do the filaments not effectively depolymerize, after reaching and exceeding the critical nucleus size? The rate constant of depolymerization $k_d$ (that is, the transition from $N=4$ back to $N=3$) is controlled by the activation energy $F_{\beta , 2}(3)-F_{\beta , 2}(4)+\Delta F_{el}$ and its ratio to $k_p$, the further polymerization rate constant, is proportional to the exponential factor 
\begin{equation}
\frac{k_d}{k_p} \propto \exp \left(- \frac{F_{\beta , 2}(3)-F_{\beta , 2}(4)}{k_BT} \right) \approx  \exp \left(-11.5 - \ln C_1 \right),
\end{equation}
using the \eqref{eq:d-proto} with $N\geq 3$, and the values of energy parameters in Table 1. At the lowest limit of our examined concentrations, $\ln C_1 =-6.5$ and the ratio of reverse-forward rates is $\propto \exp{-5}$: even this upper bound is low enough to consider only the forward rate of filament polymerization.

\begin{figure} 
\centering	\includegraphics[width=0.9\columnwidth]{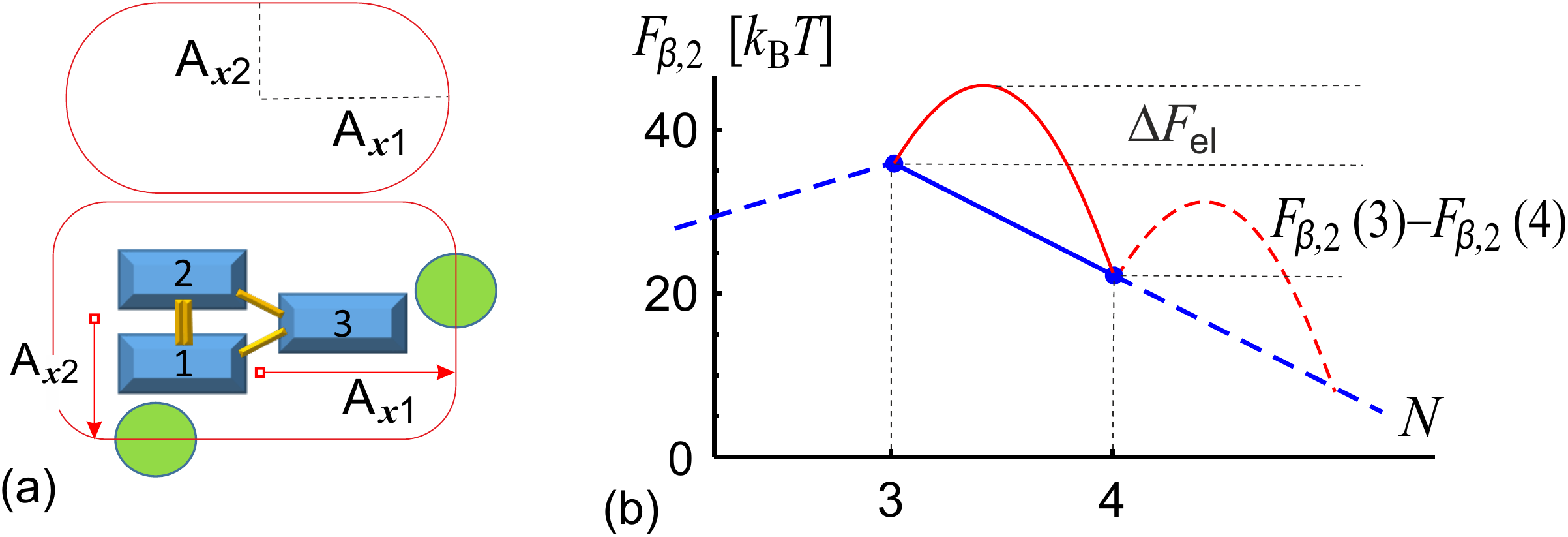}
	\caption{(a) Sketch of the non-spherical shape of the trimer nucleus, with the semi-axes $A_{x1}=(2.3+1)$\,nm, and $A_{x2}=(0.42+1)$\,nm, defined following the notations of \cite{Richter1974}. (b) A sketch of a fragment of the free energy $F_{\beta,2}(N)$ illustrating the additional (small) free energy barriers between the consecutive states of an elongating filament pair \cite{Vendruscolo2007}. }
	\label{fig:appB}
\end{figure} 

\section{Derivation of \mathinhead{k_{2}}{k2}}  \label{app:k2}
We consider a system of only three states: $S_{mono}$, monomers, $S_\mathrm{int}$, the intermediate species, and $S_{aggr}$, the final $\beta$-aggregate. This system can go from $S_{mono}$ to $S_\mathrm{int}$ with the micellation rate constant ($k_+$), $S_\mathrm{int}$ to $S_{aggr}$ with the conversion rate constant ($k_c$), $S_\mathrm{int}$ to $S_{mono}$ with the dissociation rate constant ($k_-$). At $S_\mathrm{int}$, the system can either dissociate back into monomers or fully convert into the $\beta$-aggregate. The inverse conversion from $S_{aggr}$ is not considered based on the fact that the elongation rate of any $\beta$-aggregate that exceeds the critical nucleus size is rather fast, leaving almost no chance to reverse the aggregation. The set of kinetic equations for each state are written as:
\begin{align} \label{eq:3kset}
&\frac{dC_{1}}{dt}=k_-C_\mathrm{int}-k_+ C_{1} \notag \\
&\frac{dC_\mathrm{int}}{dt}=-(k_-+k_c)C_\mathrm{int}+k_+C_{1} \notag \\
&\frac{dC_{aggr}}{dt}=k_cC_\mathrm{int}
\end{align}
, where $C_{1}$, $C_\mathrm{int}$ and $C_{aggr}$ are defined as the concentrations of the $S_{mono}$ (monomers), $S_\mathrm{int}$ (intermediate) and $S_{aggr}$ (the final $\beta$-aggregate) states, respectively. Our aim is to find the average time for the system to first reach the final state $S_{aggr}$ considering all the possible paths. That is to find the average nucleation rate constant $k_2$ defined below :
\begin{equation} \label{eq:k2}
\frac{dC_{aggr}}{dt}=k_2C_{1} 
\end{equation}
All the rate constants defined in Eqs.~\ref{eq:3kset}~and~\ref{eq:k2} have the dimension of the inverse of time, which then can be used to find the time scale involved in this nucleation process. The time required to travel from $S_{mono}$ to $S_\mathrm{int}$ is defined as $\tau_+$, which is simply $1/k_+$. Similarly, $\tau_-$, the time from $S_\mathrm{int}$ to $S_{mono}$, is $1/k_-$, while $\tau_c$, the time from $S_\mathrm{int}$ to $S_{aggr}$, equals to $1/k_c$. 

To find the average time for nucleation, we need to find out the time required for each path. For example, the first path is $S_{mono} \rightarrow S_\mathrm{int} \rightarrow S_{aggr}$ and the second path corresponds to $S_{mono}\rightarrow S_\mathrm{int}\rightarrow S_{mono}\rightarrow S_\mathrm{int}\rightarrow S_{aggr}$. All other paths will have more than two times of dissociating back and associating into a micelle again. For the first path, the total travel time is $(\tau_+ + \tau_c)$, whereas the second path requires $[(\tau_+ + \tau_- )+1*(\tau_+ + \tau_c)]$. A general form for the travel time of the $(n+1)$th path, i.e. $\tau_{n+1}$, shall be
\begin{equation} \label{eq:ttra}
\tau_{n+1}=(\tau_+ + \tau_- )+n(\tau_+ + \tau_c) 
\end{equation}

In this system, only at $S_\mathrm{int}$ the decision is made to fully convert to the aggregate or to dissociate back into monomers. The probability to convert is denoted as $P_c$, which is proportional to $k_c$, while the probability to dissociate back is  $P_-$ and is proportional to $k_-$. We then have the normalized probability of conversion and dissociation:
\begin{equation} \label{eq:PcP-}
P_c=\frac{k_c}{k_c+k_-};\ \ P_-= \frac{k_-}{k_c+k_-}
\end{equation}

For the $(n+1)$th path, the system needs to dissociate back $n$ times and finally a conversion follows, and the probability for this to happen, $P_{n+1}$, is readily written as
\begin{equation} \label{eq:Pn}
P_{n+1}=P_-^n P_c
\end{equation}
The average time to reach $S_{aggr}$, denoted as $\tau_{ave}$, is the summation of products of the time required for different paths and their probability to happen:
\begin{equation} \label{eq:tave}
\tau_{ave}=\Sigma_0^{\infty} \tau_{n+1} P_{n+1}
\end{equation}
Inserting Eqs.~\ref{eq:ttra},~\ref{eq:PcP-}~and~\ref{eq:Pn} into \eqref{eq:tave} and  taking the inverse of it, we arrive at the formula of \eqref{eq:3-state}.

\section{Derivation of \mathinhead{k_{+}}{k+}} \label{app:k+}
The rate collisions between a monomer and a micelle of size $(N-1)$ is expressed by the Smoluchowski relation using the mutual diffusion coefficient $D_m = D_1 + D_{N-1}$: $4\pi D_m (r_1+r_{N-1})C_1 C_{N-1}$. Here $r_1$ is the radius of a monomer, while $r_{N-1}$ is the radius of an $(N-1)$-size micelle. $C_{N-1}$ is the concentration of the $(N-1)$-size micelle, which is between the  $N_h$ and $N_l$ size in the three-state kinetic model (in order to have a metastable/intermediate state). This rate, by definition, is equal to $k_+C_1$, which is how we shall determine this rate constant. 

Due to the micellation free energy barrier for monomers to cross to aggregate into this $(N-1)$-size micelle, we assume a pre-thermal equilibrium for the concentrations of the micelles whose size is smaller than $N_h$, the micelle size that gives the micellation free energy barrier. Accordingly, we can easily write the $C_{N_h}$ term as: 
\begin{equation} \label{eq:CNh}
C_{N_h}=C_1e^{\frac{-\Delta F_\mathrm{mic}}{k_BT}}, 
\end{equation}
where $\Delta F_\mathrm{mic}$ refers to the micellation free energy and is simply \eqref{eq:mic} at $N=N_h$.

In order for this $N_h$-size micelle to aggregate into $(N-1)$-size, it has to adsorb additional $(N-N_h-1)$ monomers. We may further assume that this $N_h$-size micelle acts as a deep absorbing sink. Monomers are adsorbed as soon as they contact the surface of the $N_h$-size micelle, which requires that the center of mass of one monomer to fall inside the spherical volume of $ 4\pi\left(r_{N_h}+r_1\right)^3/3$, a region that is bound by the radii of one monomer and the pre-existing micelle of $N_h$ size. The volume conservation is implemented to estimate $r_{N_h}$ as $r_1\sqrt[3]{N_h}$. In this way, the probability of adsorption of $n$ monomers to this pre-micelle, $P_{a,n}$, is:
\begin{equation} \label{eq:Pan}
P_{\mathrm{ads},n}=\left[4\pi r_1^3 \left(1+\sqrt[3]{N_h}\right)^3 C_1/3\right]^n
\end{equation}

With \eqref{eq:CNh} and \eqref{eq:Pan}, $C_{N-1}$ is given as
\begin{align} \label{eq:C_N-1}
C_{N-1}=& P_{\mathrm{ads},N-N_h-1} C_{N_h}   \\
&= \left[\frac{4}{3}\pi r_1^3 (1+\sqrt[3]{N_h})^3 \right]^{N-N_h-1}
e^{-\Delta F_\mathrm{mic}/k_B T}C_1^{N-N_h}.  \notag
\end{align}

For the estimation of the mutual diffusion coefficient $D_m$, the Stokes-Einstein equation is used to express $D_m$ in terms of the viscosity of solvents $\eta$, temperature $T$ and the size of the micelle and monomer. Together with the $r_{N-1}$ expressed in $r_1$, it gives:
\begin{equation} 
D_m= D_1+D_{N-1}=\frac{k_BT}{6\pi\eta r_1} \left(1+\frac{1}{\sqrt[3]{N-1}} \right).
\end{equation}

Inserting \eqref{eq:CNh} and \eqref{eq:C_N-1} into $4\pi D_m C_{N-1}$, we obtain $k_+$ as written in \eqref{eq:k+}.	
\end{appendix} 

\end{document}